\documentclass[usenatbib]{mn2e}
\usepackage{amsmath}
\usepackage[pdftex]{graphicx}
\usepackage{mn2e-breakabs}
\usepackage{times}
\usepackage{url}

\newcommand{\hMpc}{\mbox{$h^{-1}{\rm Mpc}$}}
\newcommand{\volunit}{\mbox{$h^{-3}{\rm Mpc}^{3}$}}
\newcommand{\denunit}{\mbox{$h^{3}{\rm Mpc}^{-3}$}}
\newcommand{\ldenunit}{\mbox{$L_\odot h {\rm Mpc}^{-3}$}}
\newcommand{\sbunit}{\mbox{mag arcsec$^{-2}$}}
\newcommand{\Vmax}{\mbox{$V_{\rm max}$}}

\title[GAMA luminosity functions]
{Galaxy and Mass Assembly (GAMA): $ugriz$ galaxy luminosity functions}

\author[J.~Loveday et al.]
{{\parbox{\textwidth}{\raggedright J.~Loveday,$^{1}$\thanks{E-mail:~J.Loveday@sussex.ac.uk}
P.~Norberg,$^{2,3}$
I.K.~Baldry,$^{4}$
S.P.~Driver,$^{5,6}$
A.M.~Hopkins,$^{7}$
J.A.~Peacock,$^{2}$
S.P.~Bamford,$^{8}$
J.~Liske,$^{9}$
J.~Bland-Hawthorn,$^{10}$
S.~Brough,$^{7}$
M.J.I.~Brown,$^{11}$
E.~Cameron,$^{12}$
C.J.~Conselice,$^{8}$
S.M.~Croom,$^{10}$
C.S.~Frenk,$^{3}$
M.~Gunawardhana,$^{10}$
D.T.~Hill,$^{5}$
D.H.~Jones,$^{11}$
L.S.~Kelvin,$^{5}$
K.~Kuijken,$^{13}$
R.C.~Nichol,$^{14}$
H.R.~Parkinson,$^{2}$
S.~Phillipps,$^{15}$
K.A.~Pimbblet,$^{11}$
C.C.~Popescu,$^{16}$
M.~Prescott,$^{4}$
A.S.G.~Robotham,$^{5}$
R.G.~Sharp,$^{17}$
W.J.~Sutherland,$^{18}$
E.N.~Taylor,$^{10}$
D.~Thomas,$^{14}$
R.J.~Tuffs,$^{19}$
E.~van~Kampen,$^{9}$
D.~Wijesinghe$^{10}$
}}\\
\vspace{0.4cm}\\
{\parbox{\textwidth}{\raggedright 
$^{1}$Astronomy Centre, University of Sussex, Falmer, Brighton BN1 9QH, UK
\\
$^{2}$Institute for Astronomy, University of Edinburgh, Royal
Observatory, Blackford Hill, Edinburgh EH9 3HJ, UK
\\
$^{3}$Institute for Computational Cosmology, Department of Physics,
Durham University, South Road, Durham DH1 3LE, UK
\\
$^{4}$Astrophysics Research Institute, Liverpool John Moores University,
Twelve Quays House, Egerton Wharf, Birkenhead, CH41 1LD, UK
\\
$^{5}$School of Physics \& Astronomy, University of St Andrews, North
Haugh, St Andrews, KY16 9SS, UK
\\
$^{6}$International Centre for Radio Astronomy Research (ICRAR),
    The University of Western Australia, 35 Stirling Highway, Crawley,
    WA6009, Australia
\\
$^{7}$Australian Astronomical Observatory, PO Box 296, Epping, NSW 1710, Australia
\\
$^{8}$Centre for Astronomy and Particle Theory, University of
Nottingham, University Park, Nottingham NG7 2RD, UK
\\
$^{9}$European Southern Observatory, Karl-Schwarzschild-Str.~2, 85748
Garching, Germany
\\
$^{10}$Sydney Institute for Astronomy, School of Physics, University of
Sydney, NSW 2006, Australia
\\
$^{11}$School of Physics, Monash University, Clayton, Victoria 3800, Australia
\\
$^{12}$Department of Physics, Swiss Federal Institute of Technology
(ETH-Z{\" u}rich), 8093 Z{\" u}rich, Switzerland
\\
$^{13}$Leiden University, P.O.~Box 9500, 2300 RA Leiden, The Netherlands
\\
$^{14}$Institute of Cosmology and Gravitation (ICG), University of
Portsmouth, Dennis Sciama Building, Burnaby Road, Portsmouth PO1 3FX, UK
\\
$^{15}$Astrophysics Group, HH Wills Physics Laboratory,
University of Bristol, Tyndall Avenue, Bristol BS8 1TL
\\
$^{16}$Jeremiah Horrocks Institute, University of Central Lancashire,
Preston PR1 2HE, UK
\\
$^{17}$Research School of Astronomy \& Astrophysics,
The Australian National University,
Cotter Road,
Weston Creek, ACT 2611,
Australia
\\
$^{18}$Astronomy Unit, Queen Mary University London, Mile End Rd, London
E1 4NS, UK
\\
$^{19}$Max Planck Institute for Nuclear Physics (MPIK), Saupfercheckweg
1, 69117 Heidelberg, Germany
\\
}}}

\begin{document}

\maketitle

\begin{abstract}
Galaxy and Mass Assembly (GAMA) is a project to study galaxy 
formation and evolution, combining imaging data from
ultraviolet to radio with spectroscopic data from the AAOmega spectrograph
on the Anglo-Australian Telescope.
Using data from phase 1 of GAMA, taken over three observing seasons,
and correcting for various minor sources of incompleteness, 
we calculate galaxy luminosity functions (LFs) and their evolution 
in the $ugriz$ passbands.

At low redshift, $z < 0.1$,
we find that blue galaxies, defined according to a magnitude-dependent
but non-evolving colour cut, are reasonably well fitted over a range
of more than ten magnitudes by simple Schechter functions in all bands.
Red galaxies, and the combined blue-plus-red sample, 
require double power-law Schechter functions to fit a dip in their LF
faintwards of the characteristic magnitude $M^*$ before a steepening faint end.
This upturn is at least partly due to dust-reddened disc galaxies.

We measure evolution of the galaxy LF over the redshift range 
$0.002 < z < 0.5$ both by using a parametric fit and by measuring binned
LFs in redshift slices.
The characteristic luminosity $L^*$ is found to increase with redshift 
in all bands, 
with red galaxies showing stronger luminosity evolution than blue galaxies.
The comoving number density of blue galaxies increases with redshift,
while that of red galaxies decreases, consistent with prevailing movement
from blue cloud to red sequence.
As well as being more numerous at higher redshift, blue galaxies also dominate
the overall luminosity density beyond redshifts $z \simeq 0.2$.
At lower redshifts, the luminosity density is dominated by red galaxies in 
the $riz$ bands, by blue galaxies in $u$ and $g$.
\end{abstract}


\section{Introduction}

Measurements of the galaxy luminosity function (LF) and its evolution
provide important constraints on theories of galaxy formation and evolution,
(see e.g. \citealt{2003ApJ...599...38B}).
It is currently believed that galaxies formed hierarchically
from the merger of sub-clumps.
Looking back in time with increasing redshift, 
the star formation rate appears to peak at 
redshift $z \simeq 1$, above which it plateaus and slowly declines towards 
$z \simeq 6$ \citep{clbf2000,2004ApJ...615..209H,2006ApJ...651..142H,2008ApJ...683L...5Y,2009ApJ...705L.104K}.
Since $z \simeq 1$, galaxies are thought to have evolved mostly passively
as their stellar populations age, with occasional activity
triggered by accretion and interactions with other galaxies.
\citet{2007ApJ...660L..47N} have suggested that the first major burst of 
star formation is delayed to later times for low mass galaxies, contributing 
to the downsizing phenomenon.

There has long been a discrepancy between the measured number density
of low-luminosity galaxies (the `faint end' of the LF)
and predictions from cold dark matter (CDM) hierarchical simulations, 
in the sense that fewer low-luminosity galaxies than predicted by most models
are observed \citep{2002MNRAS.335..712T}.
Of course, interpretation of these simulation results is subject to 
uncertainties in the baryon physics.
In particular, more effective feedback in low mass halos might act to
suppress the faint end of the LF.
However, it is also possible that many surveys have underestimated the number
of dwarf galaxies due to the correlation between luminosity and surface 
brightness which makes them hard to detect 
\citep{1999ApJ...526L..69D,2002MNRAS.329..579C,2007MNRAS.377..523C,2009A&A...493..489C}.
\citet{arXiv:1107.2930} have recently demonstrated that the LF faint-end
slope steepens with decreasing surface brightness.

Galaxy LFs have previously been measured in the
$ugriz$ bands from the
Sloan Digital Sky Survey \citep[SDSS,][]{york2000} by \citet{blan2003L},
\citet{2004MNRAS.347..601L}, \citet{blan2005D}, \citet{2009MNRAS.399.1106M},
and \citet{2010MNRAS.404.1215H}.
\citet{blan2003L} analysed a sample of 147,986 galaxies,
roughly equivalent to SDSS Data Release 1 \citep[DR1,][]{abazajian2003}.
They fit the LF with a series of overlapping Gaussian functions, allowing
the amplitude of each Gaussian to vary, along with two parameters $Q$ and
$P$ describing, respectively, luminosity and density evolution.
They maximized the joint likelihood of absolute magnitude and redshift,
rather than the likelihood of absolute magnitude given redshift,
making this estimator more sensitive to evolution, as well as to density
fluctuations due to large-scale structure.
They found luminosity densities at $z = 0.1$ to increase systematically with
effective wavelength of survey band, and for luminosity evolution to
decline systematically with wavelength.
Allowing for LF evolution enabled reconciliation of previously
discrepant luminosity densities obtained from SDSS commissioning data
\citep{blan2001} and the Two-degree field Galaxy Redshift Survey
\citep{folkes1999,2002MNRAS.336..907N}.

\citet{2004MNRAS.347..601L} measured the $r$-band LF in redshift slices
from SDSS DR1 and found that the comoving  number density of galaxies 
brighter than $M_r - 5 \lg h = -21.5$ mag was a factor $\simeq 3$ higher
at redshift $z = 0.3$ than today, due to luminosity and/or density evolution.

\citet{blan2005D} focused on the faint end of the LF of low-redshift
galaxies from SDSS DR2 \citep{abazajian2004}, and found that a double-power-law 
Schechter function was required to fit an upturn in the LF at 
$M_r - 5 \lg h \ga -18$ mag with faint-end slope $\alpha_2 \simeq -1.5$
after correcting for low surface-brightness incompleteness.

\citet{2009MNRAS.399.1106M} have analysed SDSS DR6 \citep{sdss-dr6}
which is roughly five times larger than the sample analysed by 
\citet{blan2003L}.
Their results are generally consistent with those of Blanton et al., although
they do point out a bright-end excess above Schechter function fits,
particularly in the $u$ and $g$ bands, due primarily to 
active galactic nuclei (AGNs).
A bright-end excess above a best-fitting Schechter function has also been observed 
in near-IR passbands by \citet{2006MNRAS.369...25J}.

\citet{2010MNRAS.404.1215H} analysed combined datasets from the
Millennium Galaxy Catalogue \citep{liske2003}, SDSS and the
UKIDSS Large Area Survey \citep{2007MNRAS.379.1599L} over a common volume
of $\simeq 71,000 \volunit$ within redshift $z = 0.1$ to obtain LFs
in the $ugrizYJHK$ bands.
They found that LFs in all bands were reasonably well fitted by Schechter
functions, apart from tentative upturns at the faint ends of the
$i$- and $z$-band LFs.
Hill et al. provided the first homogeneous measurement of the luminosity
density (LD) over the optical--near-IR regimes, finding a smooth spectral
energy distribution (SED).

Here we present an estimate of $ugriz$ galaxy LFs
from the Galaxy and Mass Assembly 
\citep[GAMA,][]{2009A&G....50e..12D,2011MNRAS.413..971D} survey.
GAMA provides an ideal sample with which to constrain the galaxy LF at low
to moderate redshifts due to its combination of moderately deep spectroscopic
magnitude limit ($r < 19.4$ or $r < 19.8$) and wide-area sky coverage 
(three $4 \times 12$ deg$^2$ regions).

We describe the input galaxy sample and incompleteness, velocity and 
$K$-corrections in Section~\ref{sec:data}.
Our LF estimation procedure is described in Section~\ref{sec:lf} and
tested using simulations in Appendix~\ref{sec:test}.
We present our results and a discussion of luminosity and density evolution
in Section~\ref{sec:results}, with our conclusions summarized in 
Section~\ref{sec:concs}.

Unless otherwise stated, we assume a Hubble constant of $H_0 = 100 h$
km/s/Mpc and an $\Omega_M = 0.3, \Omega_\Lambda = 0.7$ cosmology in
calculating distances, co-moving volumes and luminosities.

\section{Data and observations} \label{sec:data}

\subsection{Input catalogue}

The input catalogue for GAMA is described in detail by 
\citet{2010MNRAS.404...86B}.
In brief, it consists of three $4 \times 12$ deg$^2$ regions centred
approximately on the equator and at right ascensions of 9, 12 and 14.5 hours.
These fields are known as G09, G12 and G15 respectively.
Primary galaxy targets were selected from Data Release 6 
\citep[DR6,][]{sdss-dr6} of the
Sloan Digital Sky Survey \citep[SDSS,][]{york2000} to extinction-corrected,
Petrosian magnitude limits of $r < 19.4$ mag in
the G09 and G15 fields and $r < 19.8$ mag in the G12 field. 

We require Petrosian and model magnitudes and their errors in all five 
SDSS passbands in order to determine $K$-corrections (Section~\ref{sec:kcorr}),
and so we match objects in the GAMA team catalogue TilingCatv16
to objects in the SDSS DR6 PhotoObj table on SDSS ObjID using the SDSS
{\sc CasJobs}\footnote{\url{http://casjobs.sdss.org/CasJobs/}} service.
We use only objects with GAMA {\sc survey\_class} $ \ge 3$ in order to exclude
additional filler targets from the sample.
We exclude objects, which, upon visual inspection, showed no evidence of
galaxy light, were not the main part of a deblended galaxy, or had compromised
photometry ({\sc vis\_class} = 2, 3 or 4 respectively).
See \citet{2010MNRAS.404...86B} for further details of these target flags
and Section~\ref{sec:outliers} for a discussion of additional 
visual inspection of extreme luminosity objects.

In estimating LFs, we use Petrosian magnitudes corrected for 
Galactic extinction according to the dust maps of \cite{sfd98}.
We make no attempt here to correct for intrinsic dust extinction within
each galaxy, as was done by \cite{2007MNRAS.379.1022D}, nor to extrapolate
the Petrosian magnitudes to total, as done, for example, by \citet{2005AJ....130.1535G}
and \citet{2011MNRAS.412..765H}.
These systematic corrections to SDSS photometry, much more significant than 
any small random errors, will be considered in a subsequent paper.

An exception to our use of Petrosian magnitudes is for $u$-band data, 
where we instead use a pseudo-Petrosian magnitude defined by 
\begin{equation} \label{eqn:umag}
u_{\rm pseudo-Petro} = u_{\rm model} - r_{\rm model} + r_{\rm petro}.
\end{equation}
The reason for this is that the Petrosian $u$-band quantities are noisy
and suffer from systematic sky-subtraction errors \citep{baldry2005}.
The pseudo-Petrosian $u$-band magnitude defined above,
(using the SDSS $r$ band since it has highest signal-to-noise),
and referred to
as $u_{\rm select}$ by \cite{baldry2005}, is much better behaved
at faint magnitudes.

For colour selection (see Section~\ref{sec:colour}), we use SDSS model
magnitudes in defining $(g-r)$ colour, as recommended by the SDSS 
website\footnote{\url{http://www.sdss.org/dr7/algorithms/photometry.html}}.

\subsection{Spectroscopic observations}

GAMA spectroscopic observations are described in the first GAMA data release
paper \citep{2011MNRAS.413..971D}.
Observations for the GAMA Phase 1 campaign
were made over 100 nights between 2008 February and 2010 May,
comprising 493 overlapping two-degree fields.
Redshifts were assigned in a semi-automated fashion by the observers
at the telescope.
A subsequent re-redshifting exercise (Liske et al. in prep.) was used to assign
a normalised quality $nQ$ to each redshift, 
according to each particular observer and their assigned quality $Q$.
Here we use reliable ($nQ > 2$) redshifts from all three years of the 
GAMA Phase 1 campaign.
In addition to pre-existing redshifts and those obtained with the 
Anglo-Australian Telescope, twenty redshifts of brighter galaxies were
obtained with the Liverpool Telescope.
The GAMA-II campaign, extending the survey to additional southern fields,
began in 2011, but only GAMA-I redshifts are used here.

\subsection{Completeness}

Although GAMA has a very high spectroscopic completeness ($>98$  per cent;
\citealt{2011MNRAS.413..971D}), the small level of incompleteness is likely to
preferentially affect low surface brightness, low luminosity galaxies,
or galaxies lacking distinctive spectral features.
We have identified three sources of incompleteness that potentially affect 
the survey: the completeness of
the input catalogue (imaging completeness), completeness of the targets
for which spectra have been obtained (target completeness) and the success 
rate of obtaining spectroscopic redshifts (spectroscopic success rate).
These three sources of incompleteness, and how we correct for them,
are now considered in turn.

\subsubsection{Imaging completeness}  \label{sec:imcomp}

\begin{figure}
\includegraphics[width=\linewidth]{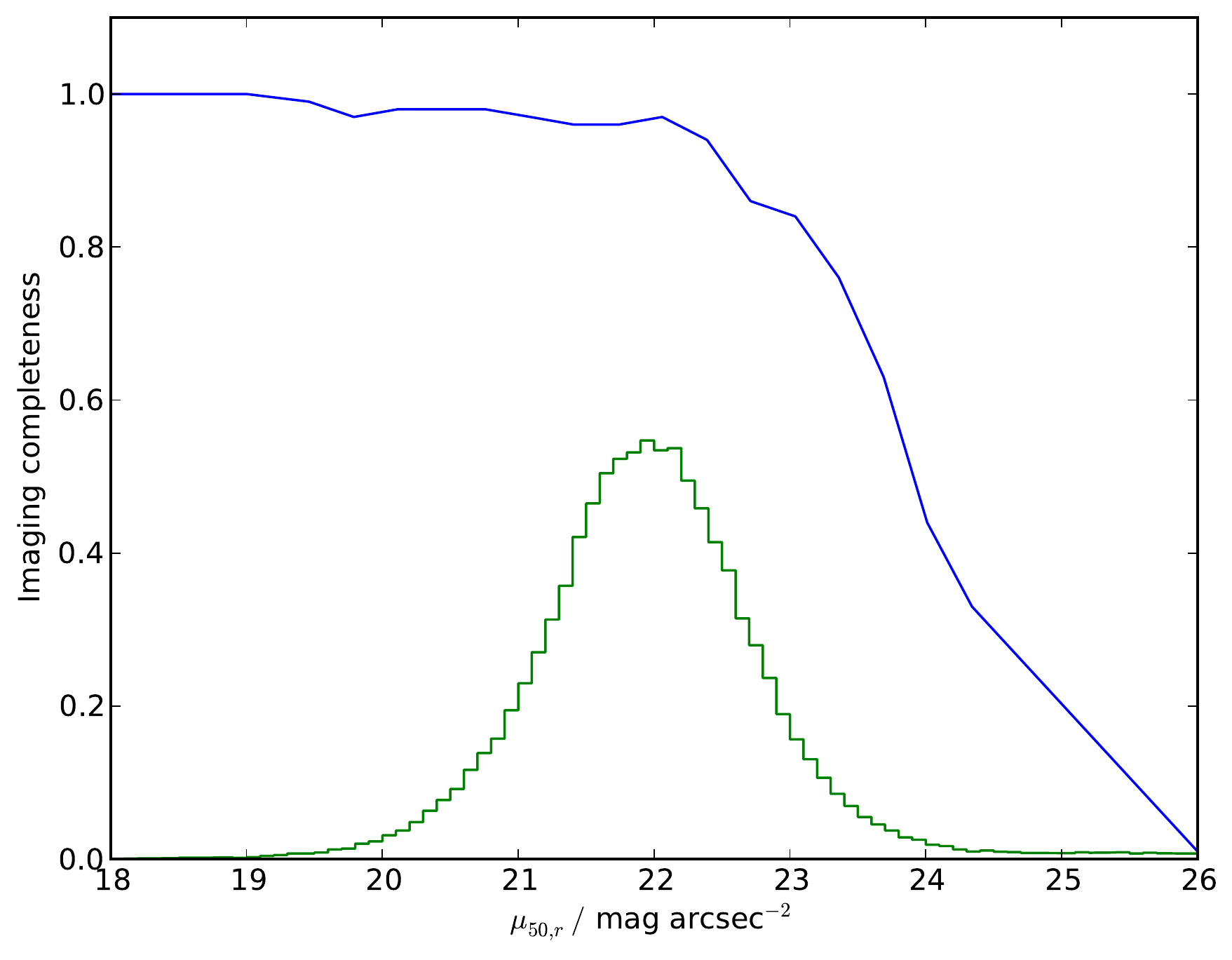}
\caption{Top line shows imaging completeness as a function of $r$-band
half-light surface brightness, $\mu_{50,r}$, from \citet{blan2005D},
modified as described in the text.
The histogram shows the normalised counts of $\mu_{50,r}$ for galaxies
in the GAMA sample.
}
\label{fig:imcomp}
\end{figure}

Imaging completeness has been estimated for the SDSS main galaxy sample
by \citet{blan2005D}, who passed fake galaxy images through the SDSS
photometric pipeline.
Blanton et al. found
that imaging completeness is nearly independent of apparent magnitude
(at least down to $m_r \approx 18$ mag), depending mostly on
$r$-band half-light surface brightness, $\mu_{50,r}$ (their Fig.~2).
Thus, while GAMA goes about 2 mag fainter than the SDSS main galaxy sample, 
the Blanton et al.
imaging completeness should still be approximately applicable.
We have used their imaging completeness estimates modified
in the following ways\footnote{An alternative way of estimating 
imaging completeness is to determine
what fraction of galaxies detected in 
the much deeper co-added data from SDSS Stripe 82 \citep{abazajian2009}
are detected in regular, single-epoch SDSS imaging.
However, one needs to allow for the large number 
of bright star or noise images misclassified
as low surface-brightness galaxies in the SDSS co-added catalogue,
and so this approach was abandoned.
It will be re-explored once high-quality VST KIDS imaging 
of the GAMA regions becomes available.}:
\begin{enumerate}
\item Blanton et al. determine imaging completeness over the surface brightness
range $18 < \mu_{50,r} < 24.5$ \sbunit.
Extrapolating their completeness as faint as $\mu_{50,r} = 26$ \sbunit\
results in negative completeness values.
We therefore arbitrarily assume 1 per cent imaging completeness at 
$\mu_{50,r} = 26$ \sbunit\ and linearly interpolate from the faintest
tabulated Blanton et al. completeness point at 
($\mu_{50,r}$, $f_{\rm ph}$) = (24.34, 0.33).
\item The Blanton et al. imaging completeness decreases at the bright end,
$\mu_{50,r} \la 19$ \sbunit, due to a lower angular size limit of
$\theta_{50} > 2$ arcsec and a star-galaxy separation criterion
$\Delta_{\rm sg} = r_{\rm psf} - r_{\rm model} > 0.24$ for the SDSS
main galaxy sample which excludes some compact, high surface-brightness (HSB)
galaxies.
GAMA target selection uses a far less stringent $\Delta_{\rm sg} > 0.05$,
backed up by $J-K$ colour selection, and so is much more complete in HSB
galaxies.
We therefore omit the Blanton et al. completeness points at 
$\mu_{50,r} < 19.2$ \sbunit\ and instead assume 100 per cent completeness at 
$\mu_{50,r} = 19.0$ \sbunit\ and brighter.
\end{enumerate}
Our revised imaging completeness curve, along with a histogram of $\mu_{50,r}$
values for GAMA galaxies, is given in Fig.\ref{fig:imcomp}.

\subsubsection{Target completeness}

\begin{figure*}
\includegraphics[width=\linewidth]{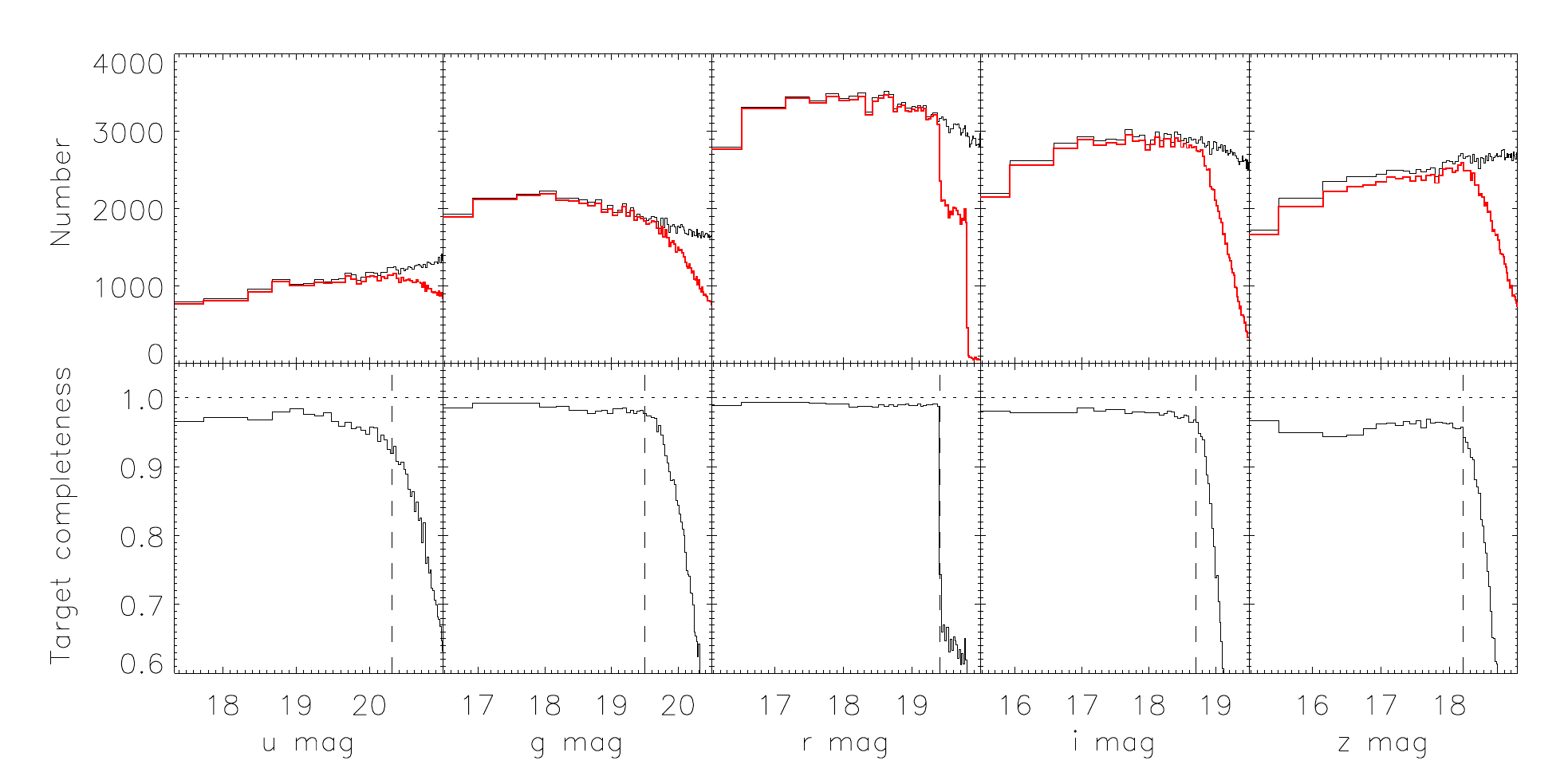}
\caption{GAMA target completeness as a function of magnitude
(pseudo-Petrosian for $u$, Petrosian for $griz$).
The upper panels show galaxy counts in varying-width magnitude bins,
chosen to give roughly equal numbers of galaxies per bin, for GAMA targets 
(thin black histogram) and counts for 
galaxies that have been spectroscopically observed (thick red histogram).
The lower panels show target completeness, ie. the ratio of observed to
target counts, with the horizontal dotted line indicating 100 per cent target 
completeness.
Vertical dashed lines indicate our chosen magnitude limits in each band:
20.3, 19.5, 19.4, 18.7, 18.2 in $ugriz$ respectively.
}
\label{fig:targcomp}
\end{figure*}

Target completeness in the $r$ band may be assessed relative to the 
GAMA tiling catalogue, which contains all galaxies to $r = 19.8$ mag in the
GAMA regions.
In the $ugiz$ bands, however, there is no well-defined magnitude limit.
We therefore re-implement the GAMA target selection criteria detailed by
\citet{2010MNRAS.404...86B} on samples of objects selected from
the SDSS DR6 {\sc photoObj} table.
We replace the \citet{2010MNRAS.404...86B} magnitude limits (their equation~6)
with the following:
$u < 21.0$, $g < 20.5$, $r < 20.0$, $i < 19.5$ or $z < 19.0$.

Target completeness in each band is then simply defined as the fraction of
target galaxies that have been spectroscopically observed,
either by GAMA or by another redshift survey, as a function of 
apparent magnitude in that band.
This is shown in Fig.~\ref{fig:targcomp}, where we have used magnitude bins 
which are equally spaced in $m' = 10^{0.52(m - m_{\rm min})}$.
This binning is chosen to give a roughly equal number of galaxies per bin,
thus avoiding large Poisson uncertainties at bright magnitudes.
In the $r$ band, target completeness is around 98-99 per cent 
brighter than $r = 19.4$ mag corresponding to the magnitude limit of the
GAMA G09 and G15 fields.

In the other four bands, the drop in completeness at faint magnitudes
is more gradual due to the spread in galaxy colours.
Magnitude limits in each band are set to the faintest magnitude 
at which target completeness is at least 92 per cent ($u$ band) or where completeness
starts to drop rapidly.
These magnitude limits are 20.3, 19.5, 19.4, 18.7, 18.2 in $ugriz$ respectively
and are indicated by the vertical dashed lines in Fig.~\ref{fig:targcomp}.

An alternative approach to estimating the LF in bands other than that
of target selection is to perform a multivariate LF, e.g. \citet{love2000},
or to use a $1/\Vmax$ estimator where $\Vmax$ is calculated using the
selection-band magnitude, e.g. \citet{2009MNRAS.397..868S}.
While using more data, these estimators suffer from a colour bias as one 
approaches the sample flux limit, and so we prefer the more conservative 
approach adopted here.

\subsubsection{Redshift success rate} \label{sec:zsuccess}

\begin{figure}
\includegraphics[width=\linewidth]{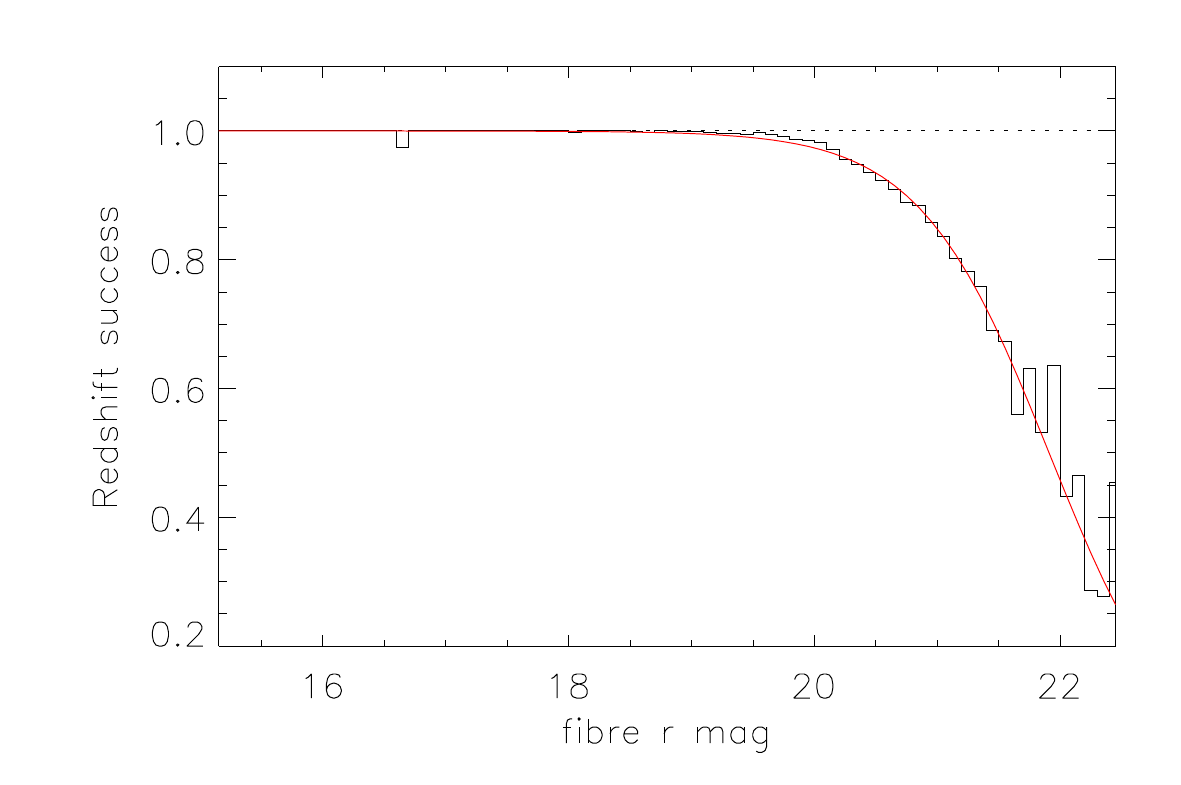}
\caption{GAMA redshift success rate as a function of fibre $r$-band
magnitude plotted as a black histogram, 
with the horizontal dotted line indicating 100 per cent success.
The red curve shows the best-fit sigmoid function.
}
\label{fig:speccomp}
\end{figure}

Redshift success rate is most likely to depend on the flux that
goes down a 2dF fibre, that is a seeing-convolved 2-arcsec-diameter
aperture.
The closest quantity available in the SDSS database is \verb|fiberMag_r|, 
hereinafter $r_{\rm fibre}$,
corresponding to the flux contained within a 3-arcsec-diameter aperture
centred on the galaxy.
We therefore determine histograms of $r_{\rm fibre}$ (uncorrected for
Galactic extinction) for all objects with high-quality redshifts ($nQ > 2$)
and for all objects with spectra.
The ratio of these two histograms then gives redshift success
as a function of $r_{\rm fibre}$, and is shown in Fig.~\ref{fig:speccomp}.
Note that some spectra observed in poor conditions have been re-observed
at a later date in order to obtain this high success rate.

We see that redshift success rate is essentially 100 per cent for 
$r_{\rm fibre} < 19.5$, declines gently to around 98 per cent by $r_{\rm fibre} = 20$
and then declines steeply at fainter magnitudes.
We have fitted a sigmoid function $f = 1/(1 + e^{a(r_{\rm fibre} - b)})$
to the binned success rate.
Sigmoid functions have previously been used to model survey completeness,
e.g. by \citet{2007MNRAS.377..815E}.
Our best-fit sigmoid function has parameters 
$a = 1.89$~mag$^{-1}$, $b = 21.91$~mag, 
shown by the red line in Fig.~\ref{fig:speccomp}, 
and we use this fit in determining redshift success rate.

\subsubsection{Galaxy weights}

\begin{figure*}
\includegraphics[width=\linewidth]{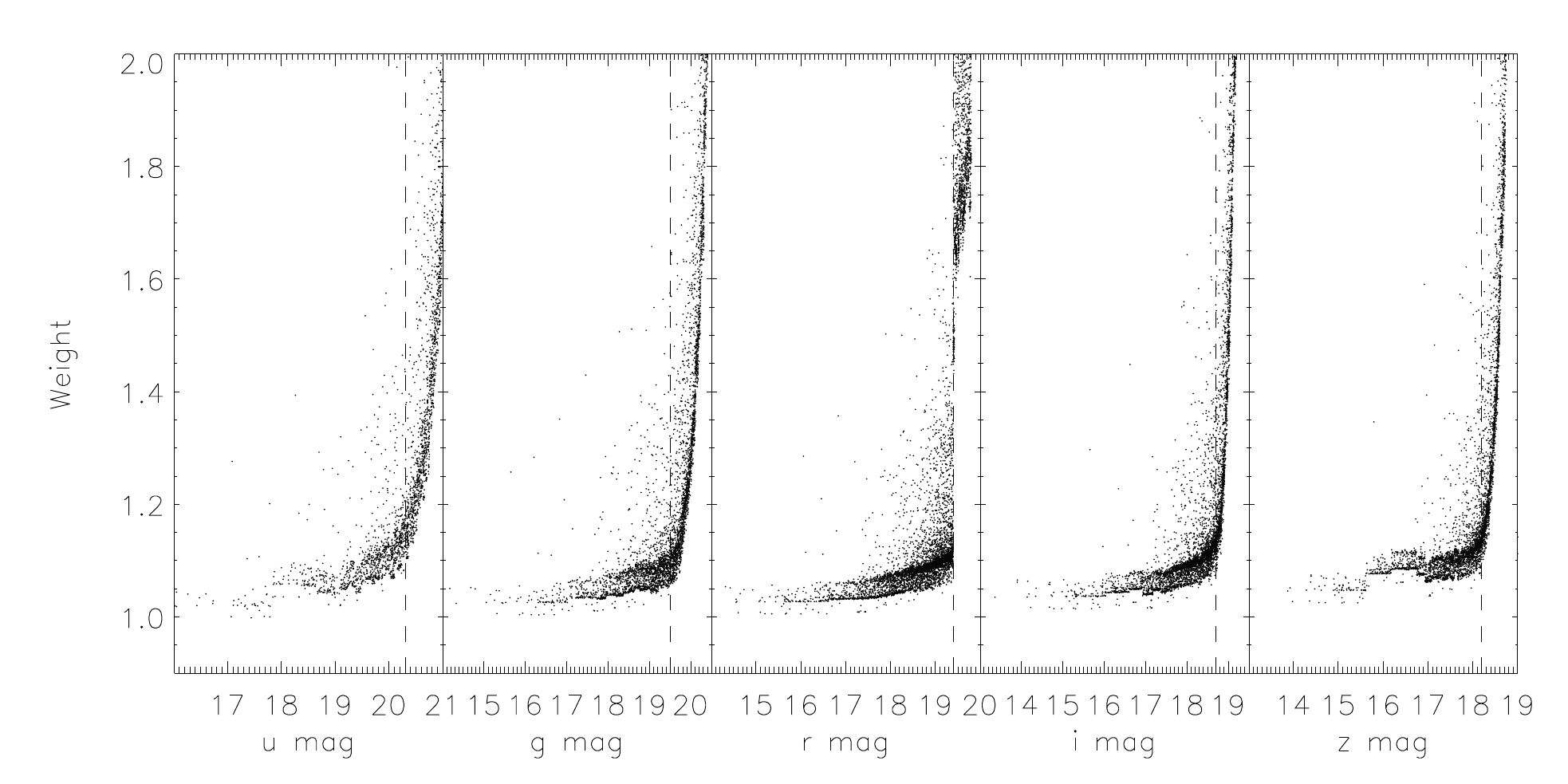}
\caption{Completeness-correction weights 
as a function of magnitude for a random 5 per cent subset of GAMA galaxies.
The vertical dashed lines show the magnitude limits applied in the LF
analysis of each band.}
\label{fig:weights}
\end{figure*}

Each galaxy is given a weight which is equal to the reciprocal of the
product of the imaging completeness, target completeness 
and redshift success rate.
Imaging completeness $C_{\rm im}$ is determined from the galaxy's apparent 
$r$-band half-light surface brightness, $\mu_{50,r}$ 
by linear interpolation of the curve in Fig.\ref{fig:imcomp}.
Target completeness $C_{\rm targ}$ is determined separately in each band from 
the galaxy's magnitude according to Fig.~\ref{fig:targcomp} and
the spectroscopic success rate $C_{\rm spec}$ is determined from the 
sigmoid function fit described in section~\ref{sec:zsuccess}.

The weight $W_i$ assigned to galaxy $i$ is then 
\begin{equation} \label{eqn:weight}
W_i = 1/(C_{\rm im} C_{\rm targ} C_{\rm spec}).
\end{equation}
These weights, as a function of magnitude in each band,
are shown for a randomly selected 5 per cent of galaxies in Fig.~\ref{fig:weights}.
The majority of galaxies brighter than our magnitude limits
have weight $W_i < 1.1$, 
with a small fraction extending to $W_i \simeq 2$.

\subsection{Velocity corrections} \label{sec:flowcorr}

The redshifting software {\sc runz} \citep{2004AAONw.106...16S}
provides heliocentric redshifts.
Before converting heliocentric redshifts to any other velocity reference frame,
we first eliminate likely stars from our sample by rejecting objects
with a heliocentric redshift $z_{\rm helio} < 0.002$ ($cz < 600$ km/s).
This lower redshift cut is conservatively chosen, 
as the {\em 2nd Catalogue of Radial Velocities with Astrometric Data}
\citep{2007AN....328..889K} includes only one star with radial velocity 
$RV > 500$ km/s, and the vast majority of
Galactic stars have $RV < 200$ km/s.
Furthermore, Fig.~7 of \citet{2011MNRAS.413..971D} shows that the 
redshift-error distribution for GAMA is essentially zero by
$cz = 600$ km/s.

Having eliminated 2111 likely stars from our sample, heliocentric redshifts
are converted to the CMB rest frame $z_{\rm CMB}$ according to the dipole of
\cite{1996ApJ...470...38L}.
For nearby galaxies ($z_{\rm CMB} < 0.03$), we apply the multiattractor 
flow model of \cite{tbad2000}.
Note there are triple-valued solutions of 
$z_{\rm CMB} \rightarrow z_{\rm Tonry}$
over a small range in G12 (near the Virgo cluster), here the average
distance is used. The solution is tapered to $z_{\rm CMB}$ from 
$z_{\rm CMB} = 0.02$ to $z_{\rm CMB} = 0.03$ (see \citealt{Baldry2011} for details).
We will see later that the Tonry et al. flow correction affects only
the very faintest end of the LF.

\subsection{$K$-corrections} \label{sec:kcorr}

\begin{figure}
\includegraphics[width=\linewidth]{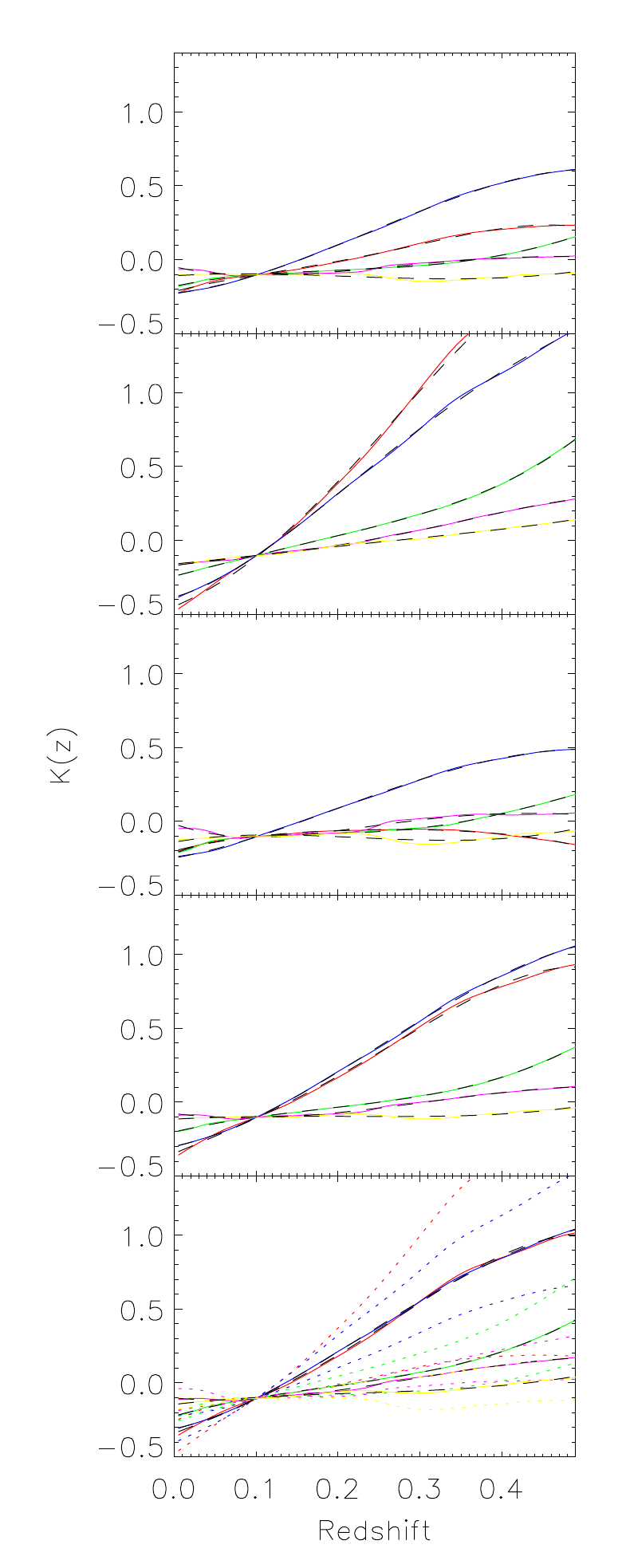}
\caption{Top four panels: $K$-corrections as a function of redshift
(red, blue, green, magenta and yellow respectively for $ugriz$)
for the first four galaxies in our sample.
Black dashed lines show fourth-order polynomial fits to each band.
Bottom panel: median $K$-corrections (coloured continuous lines) 
and 5 and 95 percentile ranges (dotted lines) for the entire GAMA sample.
Black dashed lines show fourth-order polynomial fits to the medians.
}
\label{fig:kcorrz}
\end{figure}

When estimating intrinsic galaxy luminosities, it is necessary to correct
for the fact that a fixed observed passband corresponds to a different
range of wavelengths in the rest frames of galaxies at different redshifts,
the so-called $K$-correction \citep{1956AJ.....61...97H}.
The $K$-correction depends on the passband used, the 
redshift of the galaxy and its spectral energy distribution (SED).
Here we use {\sc kcorrect v4\_2} \citep{blan2003K,br2007} 
in order to estimate and apply these corrections.
Briefly, this code estimates an SED for each galaxy by finding the
non-negative, linear combination of five template spectra that gives 
the best fit to the five SDSS model magnitudes of that galaxy.
{\sc kcorrect} can then calculate the $K$-correction in any given passband 
at any redshift.
Before calling {\sc kcorrect} itself, we use {\sc k\_sdssfix} to 
convert SDSS asinh magnitudes to the AB system and to add in quadrature 
to the random $ugriz$ mag errors given in the SDSS database typical
systematic errors of (0.05, 0.02, 0.02, 0.02, 0.03) mag respectively.

We determine $K$-corrections in a passband blueshifted by $z_0 = 0.1$.
These magnitudes are indicated with a superscript
prefix of 0.1, e.g. $^{0.1}M_r$.
This choice of rest frame allows direct comparison with most
previously published LFs based on SDSS data.

Our LF estimators require $K$-corrections to be calculated for each galaxy
at many different redshifts in order to work out visibility limits.
To make this calculation more efficient, we fit a fourth-order 
polynomial $P_k(z) = \sum_{i=0}^4 a_i (z - z_0)^i$, with $z_0 = 0.1$,
to the $K(z)$ distribution for each galaxy and use this polynomial
fit to determine $K$-corrections as a function of redshift.
Using a polynomial of this order, the rms difference between the {\sc kcorrect}
prediction and the polynomial fit is 0.01 mag or less in all five bands.
Calculated $K$-corrections and their polynomial fits are shown
for the first four galaxies in our sample, along with the median 
$K$-corrections and the 5 and 95 percentile ranges for the full sample, 
in Fig.~\ref{fig:kcorrz}.

Strictly, one should use heliocentric redshift when calculating
$K$-corrections, since they depend on the {\em observed} passband.
However, for consistency with finding the minimum and maximum redshifts
at which a galaxy would still be visible when using the $1/V_{\rm max}$
LF estimator, we use the flow-corrected redshift as described in 
section~\ref{sec:flowcorr}.
The difference in $K$-correction when using heliocentric or flow-corrected 
redshift is entirely negligible.

\subsection{Colour sub-samples} \label{sec:colour}

\begin{figure}
\includegraphics[width=\linewidth]{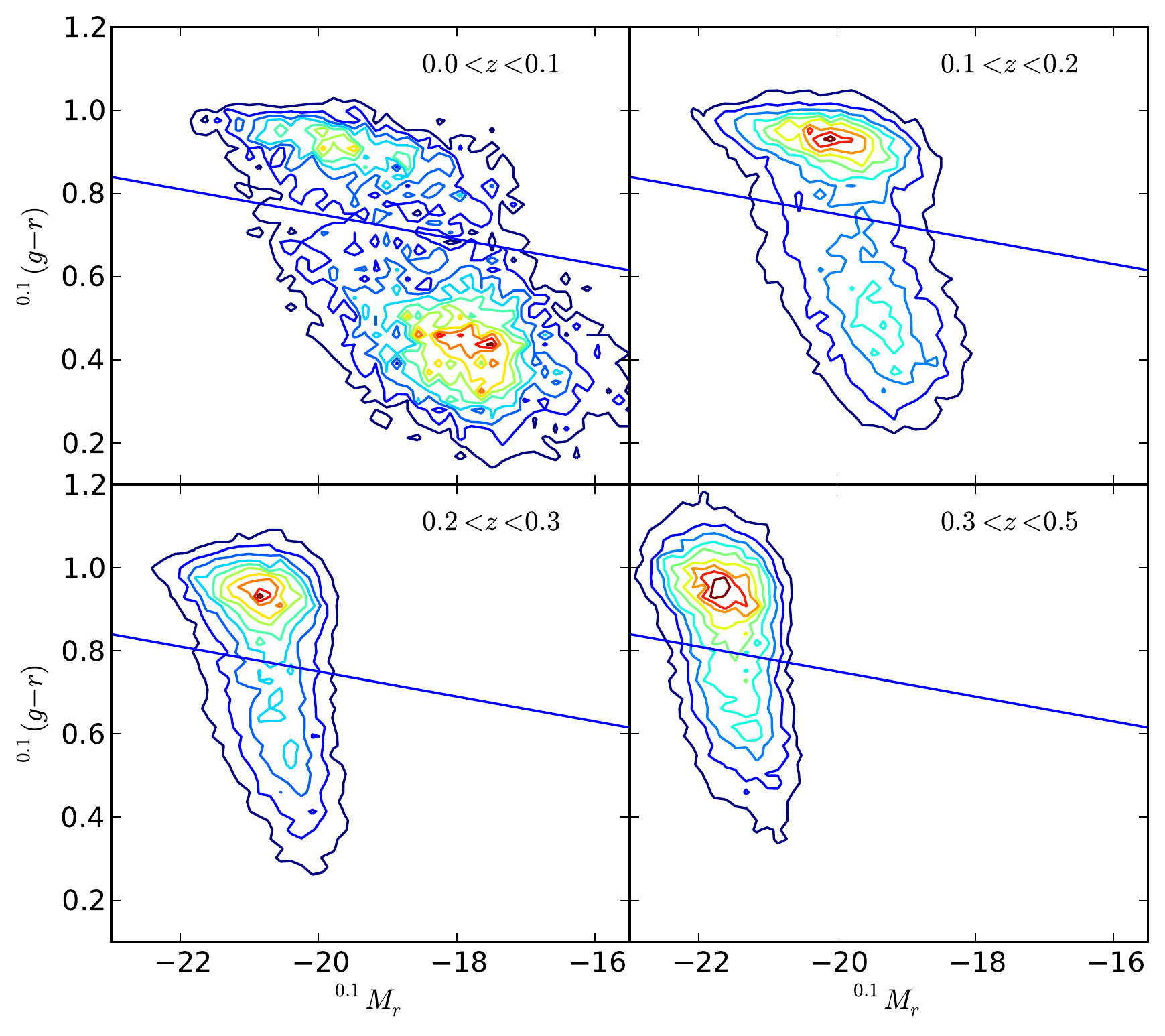}
\caption{$^{0.1}(g-r)$ colour versus $^{0.1}M_r$ $r$-band absolute magnitude 
contour plots for GAMA galaxies in four redshift ranges as
labelled.
Ten contours, spaced linearly in density, are colour-coded from black to red
in order of increasing density.
The straight line shows the magnitude-dependent colour cut separating 
blue and red galaxies given by equation~(\ref{eqn:colourcut}).
}
\label{fig:colourMag}
\end{figure}

As well as analysing flux-limited samples of galaxies in the $ugriz$ bands
(hereafter the combined sample),
we separate the galaxies into blue and red sub-samples.
Following \citet{2011ApJ...736...59Z}, we use a colour cut 
based on $K$-corrected $(g-r)$ model colour and absolute $r$-band magnitude 
that is insensitive to redshift:
\begin{equation}
^{0.1}(g-r)_{\rm model} = 0.15 - 0.03 ^{0.1}M_r,
\label{eqn:colourcut}
\end{equation}
We have adjusted the zero-point of 0.21 mag in \citet{2011ApJ...736...59Z}
to 0.15 mag
in order to better follow the `green valley' and to get more equal-sized 
samples of blue and red galaxies.
This colour cut works well at all redshifts (Fig.~\ref{fig:colourMag}),
although we see that the colour bimodality becomes far less obvious
beyond redshift $z = 0.2$ due to the lack of low-luminosity, blue galaxies
at these high redshifts.

Although colour bimodality is more pronounced in $(u-r)$ colour,
e.g. \citet{2001AJ....122.1861S}, \citet{2004ApJ...600..681B},
$u$-band photometry, even after forming a `pseudo-Petrosian' magnitude 
(equation~\ref{eqn:umag}) is rather noisy, and so we prefer to base our
colour cuts on the more robust $(g-r)$ colour.
This colour cut (in the original form of Zehavi et al.) has also been used
to investigate the angular clustering of galaxies by \citet{Christodoulou2011}.
One should also note that colour is not a proxy for galaxy morphology:
many red galaxies are in fact dust-obscured disc galaxies
(Driver et al. in prep.).

\subsection{Outlier inspection} \label{sec:outliers}

\begin{table}
\caption{Classification of extreme high- and low-luminosity objects.}
\label{tab:post_class}
\begin{tabular}{lrr}
\hline
{\sc post\_class} & $^{0.1}M_u < -20$ mag & $^{0.1}M_r > -15$ mag\\
1 OK & 4,743 & 299\\
2 QSO & 18 & 0\\
3 Major shred & 68 & 62\\
4 Minor shred & 0 & 7\\
5 Problem deblend & 151 & 16\\
6 Bad sky background & 246 & 14\\
\hline
\end{tabular}
\end{table}

\begin{figure}
\includegraphics[width=\linewidth]{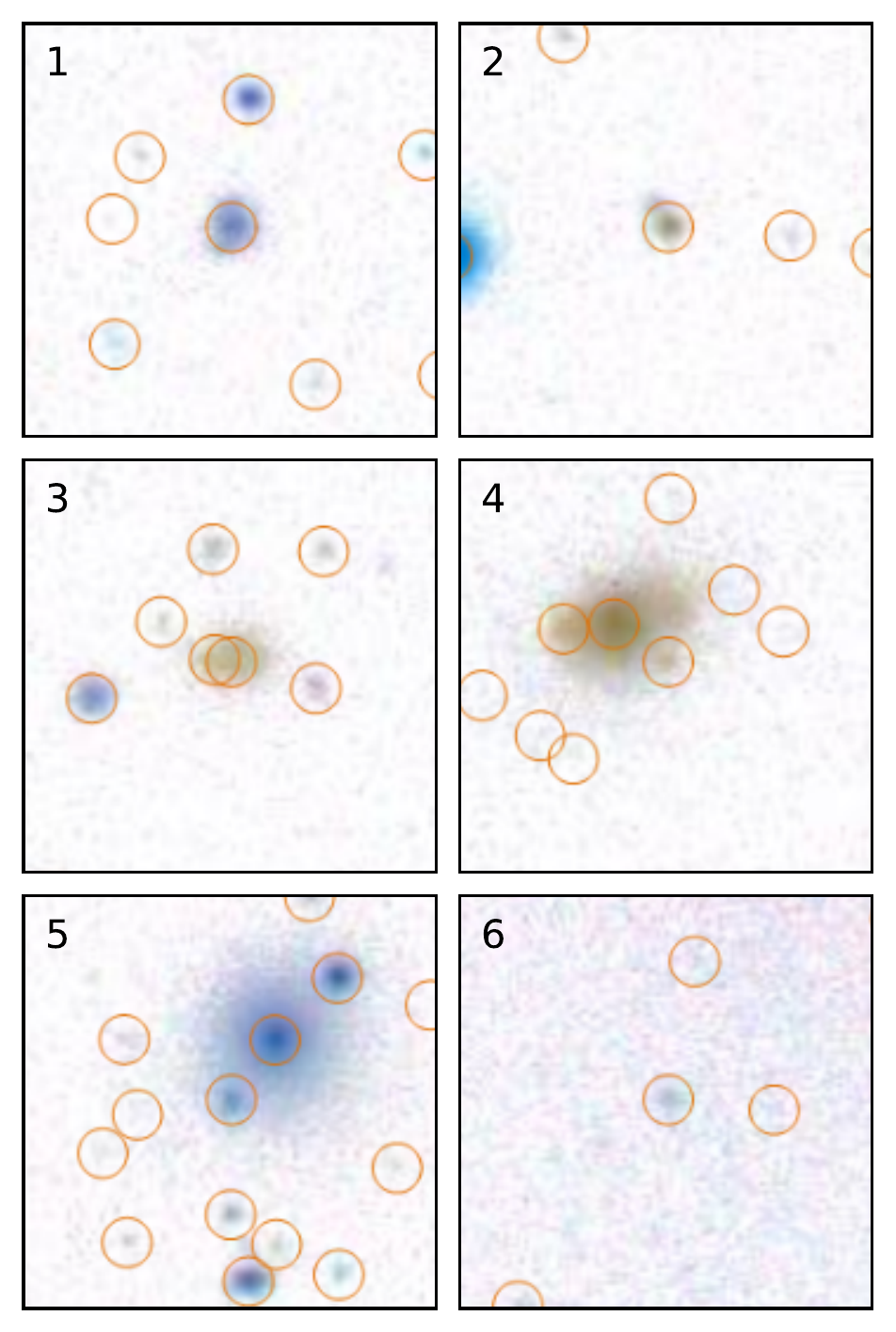}
\caption{Examples of objects classified from 1 to 6.
The GAMA targets are at the centre of each image,
which are 40 arcsec on each side.
The colour table has been inverted, so that red objects appear blue and
vice-versa, in order to obtain a light background.
Circles denote SDSS image detections: multiple circles on a 
single object (example classifications 3 and 4)
suggest that it has been over-deblended.
{\sc post\_class} classifications are shown in the top-left corner 
of each image, their meaning is given in Table~\ref{tab:post_class}.
}
\label{fig:post_class}
\end{figure}

\begin{figure}
\includegraphics[width=\linewidth]{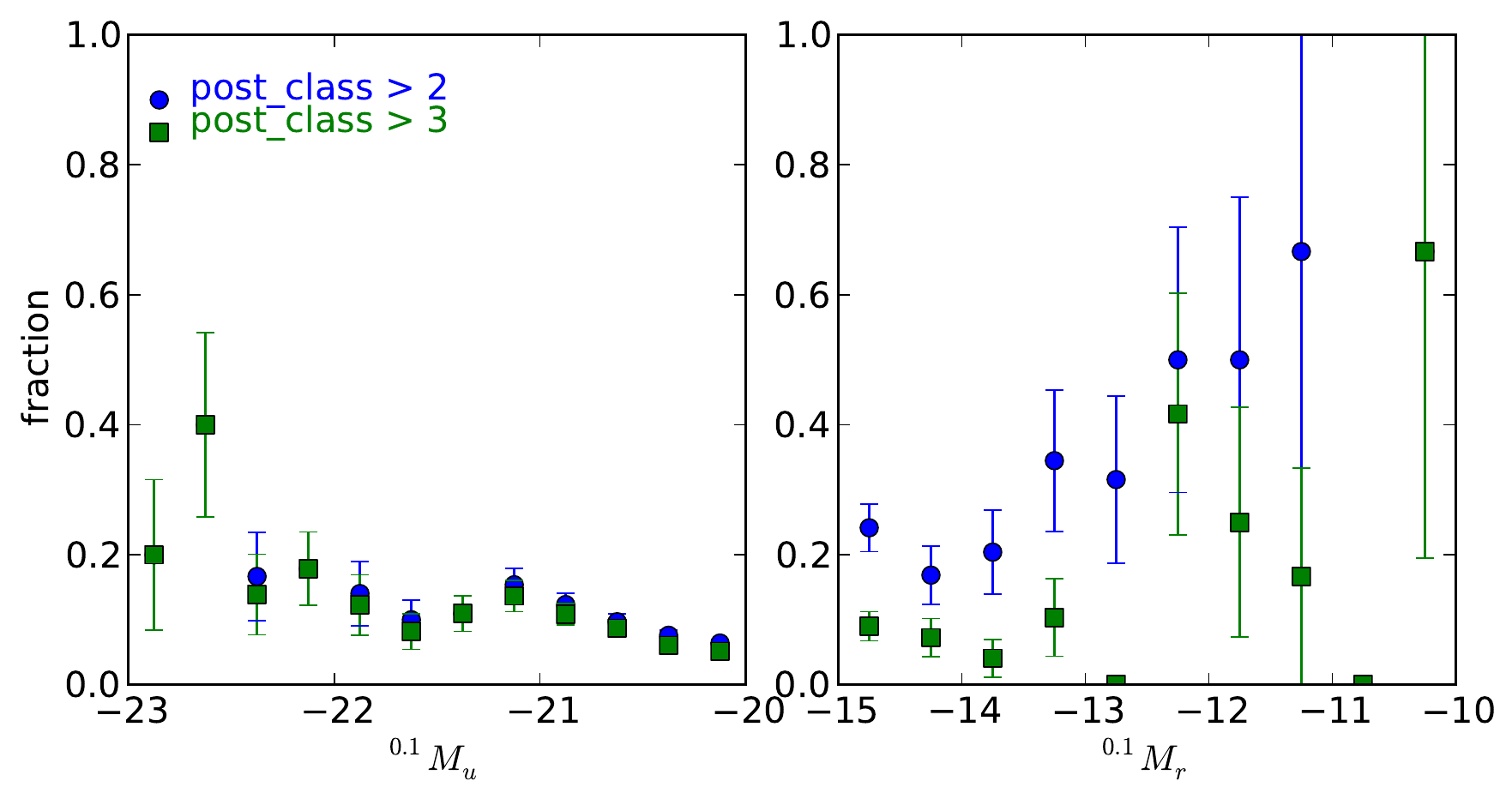}
\caption{Fraction of objects with {\sc post\_class} $> 2$ (blue circles)
and $> 3$ (green squares) as a function of $^{0.1}M_u$ (left panel)
and $^{0.1}M_r$ (right panel).
Error bars show Poisson errors on the counts.}
\label{fig:exclude_frac}
\end{figure}

\begin{figure}
\includegraphics[width=\linewidth]{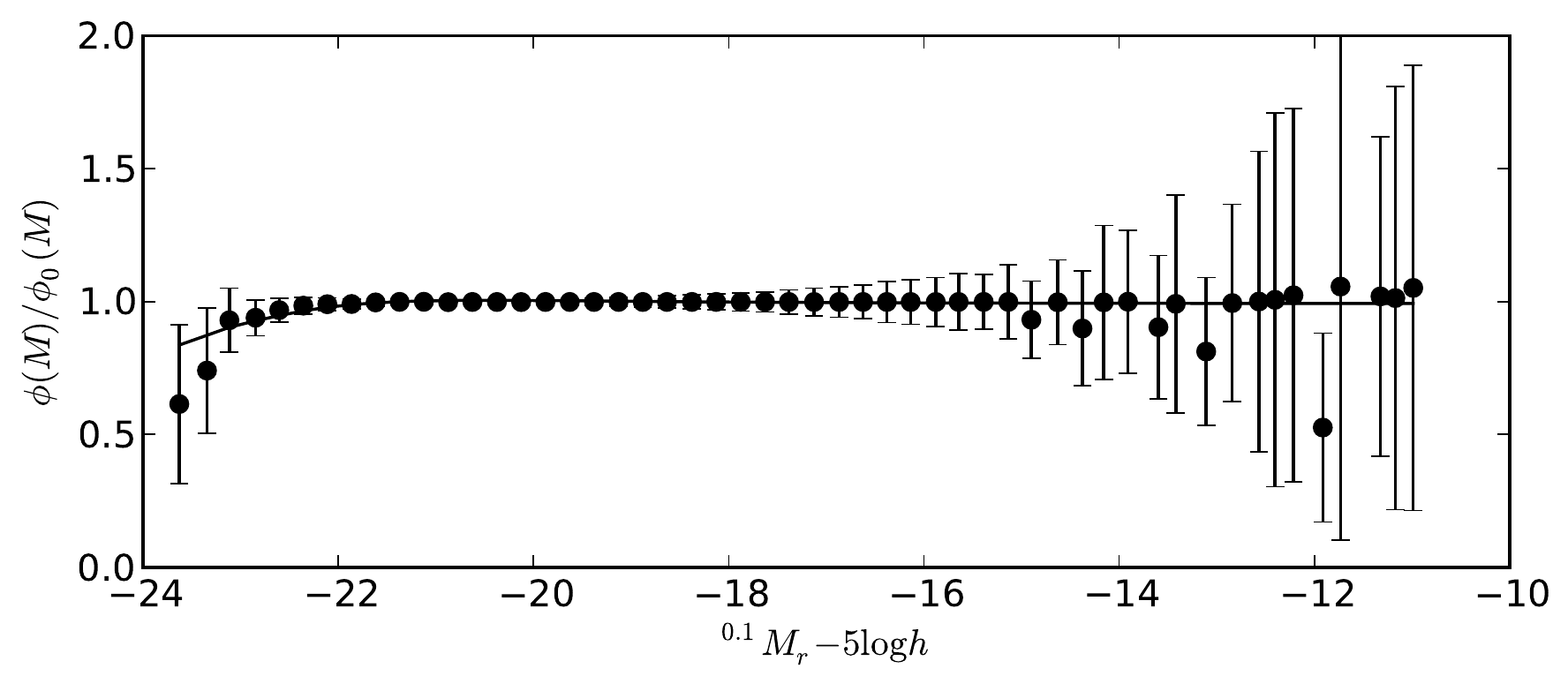}
\includegraphics[width=\linewidth]{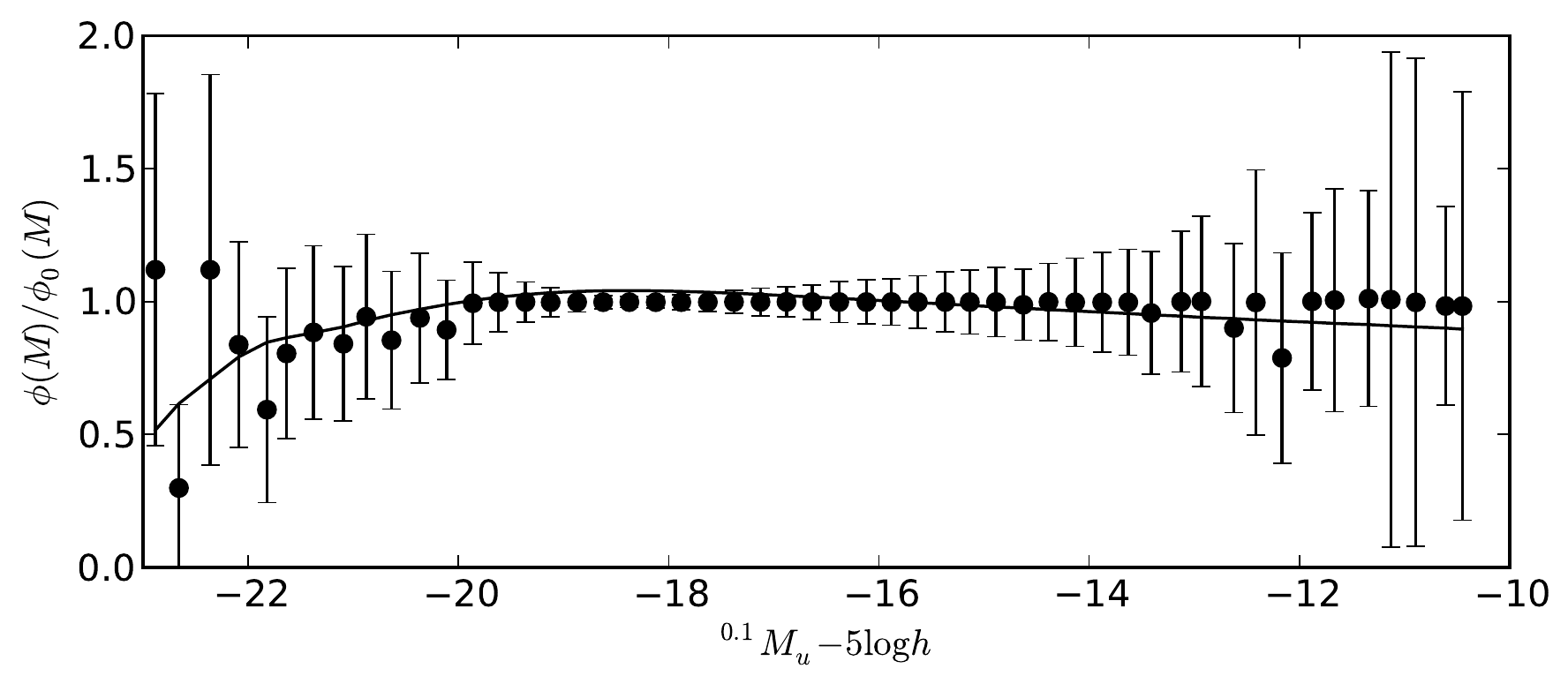}
\caption{Ratio of LFs determined using {\sc post\_class} 
$\le 3$ objects to that using all objects in $r$ band (top) and $u$ band 
(bottom).
Symbols show ratio of SWML estimates and their uncertainties,
the continuous lines shows the ratio of parametric fits to the two samples.
}
\label{fig:pc_test}
\end{figure}

We measure the LF over a very wide range of luminosities, $-23$ to $-11$ mag
in the $r$ band.
Galaxies at the extremes of this luminosity range are very rare in a 
flux-limited survey, due either to their intrinsic low number density at
high luminosity or small detection volume at low luminosity, and thus it is
likely that a significant fraction of these putative extreme objects are
in fact artifacts due to incorrectly assigned redshifts or magnitudes.
The first author has therefore inspected image cutouts showing 
SDSS image detections of 5,226 very luminous GAMA targets with
$^{0.1}M_u < -20$ mag and 398 very faint targets with $^{0.0}M_r > -15$ mag.
We choose the $u$ band to select luminous outliers since the $u$-band
LF shows the largest bright-end excess.

Table~\ref{tab:post_class} shows how the inspected images were classified.
The classification codes, which we call {\sc post\_class} in order
to distinguish them from the pre-target-selection {\sc vis\_class}
have the following meanings.
\begin{enumerate}
\renewcommand{\theenumi}{(\arabic{enumi})}
\item OK --- nothing from the image would lead one to expect poor 
photometry for that object.
\item The object looks like a QSO, i.e. blue and point-like.
This classification is ignored in the analysis (treated as OK) due to the
difficulty of distinguishing QSOs and compact blue galaxies from the 
imaging data alone.
\item The central part of a galaxy which has been shredded into multiple
detections.
It is likely that the luminosity is somewhat underestimated in these cases.
\item The target is a minor part of a shredded galaxy.  
Luminosity is likely to be greatly underestimated.
\item The galaxy is very close to a second object which either has not
been deblended, or is likely to significantly affect the estimated luminosity
in either direction.
\item Photometry is likely severely compromised by rapidly varying sky 
background due typically to the presence of a nearby saturated star.
\end{enumerate}
Examples of objects with these classifications are shown in 
Fig.~\ref{fig:post_class}.
In practice, there is some ambiguity in assigning a galaxy with classification
4 or 5, but as far as the LF analysis is concerned, it makes no difference.

These inspections were based on version 10 of the GAMA tiling catalogue,
excluding objects with {\sc vis\_class} = 2--4.
The major and minor shreds in Table~\ref{tab:post_class} have also been
inspected by IKB.
In the case of major shreds, 
we have summed the flux from the components of the shredded
galaxy to derive a `deblend-fixed' magnitude.
In total, 281 GAMA-I galaxies have had their magnitudes fixed in this manner.

In the case of six minor shreds which both JL and IKB agreed on, 
the value of {\sc vis\_class}
has been set equal to 3 in the latest version (v16) of the tiling catalogue.

The fractions of galaxies with {\sc post\_class} of 3 or 4 or higher
as a function of $M_u$ and $M_r$ are shown in Fig.~\ref{fig:exclude_frac}.
We see in the left panel that by magnitudes of $M_u = -20$,
less than 10 per cent of objects have suspect photometry.
Once we allow for fixing of over-deblended galaxies,
a similar fraction of objects with $M_r \simeq -15$ have suspect photometry
(right panel).

For our analysis, we have chosen to exclude any galaxies with {\sc post\_class}
of 4 or higher, i.e. we include major shreds with fixed fluxes but exclude
minor shreds, problem deblends and bad sky objects.
Fig.~\ref{fig:pc_test} shows the ratio of the $r$ and $u$ band LFs
using {\sc post\_class} $< 4$ galaxies to that determined using all galaxies.
We see that excluding objects with suspect photometry has a relatively minor 
effect on the determined LF: the very bright end and some faint-end bins are
systematically lower by up to 50 per cent; 
these changes are comparable to the size of the error bars.

Finally, we note that \citet{2011MNRAS.413.1236B} have independently 
checked a sample of
GAMA galaxies with the lowest detected H$\alpha$ luminosity.
Our four faintest $r$-band luminosity galaxies are also in the
Brough et al. sample.

\section{Estimating the luminosity function and its evolution} \label{sec:lf}

\subsection{Parameterizing the evolution}

In order to parametrize the evolution of the galaxy LF,
we follow \cite{lin99} in assuming a \cite{schec76} function in
which the characteristic magnitude $M^*$ and galaxy density $\phi^*$ 
are allowed to vary with redshift, but where the faint-end slope $\alpha$ is 
assumed to be non-evolving.\footnote{Evolution in the LF faint-end slope 
$\alpha$ with redshift is still rather poorly constrained.
\citet{ecbhg96} claim that $\alpha$ steepens with redshift, due to an 
increase in the number of faint, star-forming galaxies at $z \simeq 0.5$.
\citet{2005A&A...439..863I} also measure a possible steepening of
$\alpha$ with redshift.
In contrast, \citet{2008ApJ...672..198L} find that $\alpha$ gets shallower 
at higher redshifts.
Our assumption of fixed $\alpha$ is largely based on practical necessity,
since $\alpha$ can only be reliably measured at redshifts $z \la 0.2$ from the 
GAMA data.}

Specifically, in magnitudes, the Schechter function is given by
\begin{equation} \label{eqn:schec}
\phi(M) = 0.4 \ln 10 \phi^*(10^{0.4(M^* - M)})^{1 + \alpha}
\exp(-10^{0.4(M^* - M)}),
\end{equation}
where the Schechter parameters $\alpha$, $M^*$ and $\phi^*$ vary with 
redshift as:
\begin{align}
\alpha(z) &= \alpha(z_0),\nonumber \\
M^*(z) &= M^*(z_0) - Q(z - z_0),\label{eqn:evol}\\
\phi^*(z) &= \phi^*(0) 10^{0.4 P z}.\nonumber
\end{align}
Here the fiducial redshift $z_0$ is the same redshift to which magnitudes
are $K$-corrected, in our case $z_0 = 0.1$.
The Schechter parameters $\alpha$, $M^*(z_0)$ and $\phi^*(0)$ and
evolution parameters $Q$ and $P$ are determined via the maximum-likelihood
methods described by \citet{lin99}.

First, the shape parameters $\alpha$, $M^*(z_0)$ and luminosity evolution
parameter $Q$ are fit simultaneously and independently of the other parameters
by maximising the log likelihood 
\begin{equation} \label{eqn:styLike}
\ln {\cal L} = \sum_{i=1}^{N_{\rm gal}} W_i \ln p_i.
\end{equation}
Here, $W_i$ is the incompleteness correction weighting 
(equation~\ref{eqn:weight})
and the probability of galaxy $i$ having absolute magnitude $M_i$
given its redshift $z_i$ is
\begin{equation} \label{eqn:sty_prob}
p_i \equiv p(M_i|z_i) = \phi(M_i)\left/
\int_{{\rm max}[M_{\rm min}(z_i), M_1]}^{{\rm min}[M_{\rm max}(z_i), M_2]}
\phi(M) dM\right.,
\end{equation}
where $M_1$, $M_2$ are the absolute magnitude limits of the sample, 
$M_{\rm min}(z_i)$, $M_{\rm max}(z_i)$ are the minimum and maximum absolute 
magnitudes visible at redshift $z_i$,
and $\phi(M)$ is the differential LF given by (\ref{eqn:schec}).

The density evolution parameter $P$ and normalization $\phi^*(0)$ cancel in
the ratio in (\ref{eqn:sty_prob}) and so must be determined separately.
Lin et al. show that the parameter $P$ may be determined by maximising the
second likelihood
\begin{equation} \label{eqn:pLike}
\ln {\cal L}' = \sum_{i=1}^{N_{\rm gal}} W_i \ln p'_i, 
\end{equation}
where
\begin{align} \label{eqn:p_prob}
p'_i & \equiv p[z_i|M_i(0), Q] \nonumber\\
& = 10^{0.4 P z_i}\left/
\int_{{\rm max}[z_{\rm min}[M_i(0)], z_1]}^{{\rm min}[z_{\rm max}[M_i(0), z_2]}
10^{0.4 P z} \frac{dV}{dz} dz\right.,
\end{align}
where $z_1$, $z_2$ are the redshift limits of the sample, 
$z_{\rm min}$, $z_{\rm max}$ are the redshift limits over which galaxy $i$
may be observed, given the survey's apparent magnitude limits and 
its absolute magnitude evolution-corrected to redshift zero,
$M_i(0) = M_i(z_i) + Q z_i$.
Note that the value of $P$ is independent of the fiducial redshift $z_0$.

Finally, we fit for the overall normalisation $\phi^*(0)$.
We depart slightly from the prescription of \cite{lin99} here.
In place of their equation~14, we use a minimum variance estimate
of the space density $\bar{n}$ of galaxies:
\begin{equation}
 \bar{n} = \sum_{i=1}^{N_{\rm gal}} \frac{W_i U(z_i)}{10^{0.4 P z_i}} \left/ 
            \int_{z_{\rm min}}^{z_{\rm max}} \frac{dV}{dz} S(z) U(z) dz\right.,
 \label{eqn:nbar}
\end{equation}
where $S(z)$ is the galaxy selection function,
$U(z)$ a redshift weighting function chosen to give minimum variance
and $dV/dz$ is the volume element at redshift $z$.
The selection function for galaxies with luminosities $L_1$ to $L_2$
is 
\begin{equation}
 S(z) = \int_{{\rm max}(L_{\rm min}(z),L_1)}^{{\rm min}(L_{\rm max}(z),L_2)}
        \phi(L,z) dL \left/ \int_{L_1}^{L_2} \phi(L,z) dL\right..
\label{eqn:selfn}
\end{equation}
Note that the integration limits in the numerator depend on the assumed
$K$-correction.
In this case, we use the median $K$-correction
of the galaxies in the sample under consideration:
see Fig.\ref{fig:kcorrz} for median $K$-corrections for the full sample
as a function of redshift.
Our results change by much less than the estimated 1-sigma errors 
(see Section~\ref{sec:errors}) if we use mean
instead of median $K$-corrections.

We adopt the redshift weighting function
\begin{equation}
 U(z) = \frac{1}{1 + 4 \pi (\bar{n}/\bar{W}) J_3(r_c) S(z)}, 
\quad J_3(r_c) = \int_0^{r_c} r^2 \xi(r) dr,
\label{eqn:zweight}
\end{equation}
where $\xi(r)$ is the two point galaxy correlation function
and $\bar{W}$ is the mean incompleteness weighting.
Provided $J_3(r_c)$ converges on a scale $r_c$ much smaller than the depth of
the survey, then this redshift weighting scheme (equation~\ref{eqn:zweight})
minimizes the variance in the estimate of $\bar {n}$ \citep{dh82}.
Larger values of $J_3$ weight galaxies at high redshift more highly;
we assume $4\pi J_3 = 30,000 h^{-3} {\rm Mpc}^3$.
This value comes from integrating the flux-limited
two-point galaxy correlation function
of \citet{zehavi2005}, $\xi(r) = (r/5.59)^{-1.84}$, to 
$r_c = 60 h^{-1} {\rm Mpc}$; 
at larger separations the value of $J_3$ becomes uncertain. 
However, the results are not too sensitive to the value
of $J_3$, the estimated densities changing by less than 8 per cent if $J_3$ 
is halved.
This possible systematic error is generally comparable to or less than the 
statistical uncertainty in $\phi^*$ (5-- 25  per cent).

We check our minimum variance normalisation by comparing,
in Tables~\ref{tab:faintfit}, \ref{tab:dpfit} and \ref{tab:evfit},
the observed number
of galaxies in each sample (within the specified apparent magnitude, 
absolute magnitude and redshift limits) with the prediction
\begin{equation} \label{eqn:Npred}
N_{\rm pred} = \frac{1}{\bar{W}} \int_{z_{\rm min}}^{z_{\rm max}} 
\int_{L_{\rm min}(z)}^{L_{\rm max}(z)}
\phi(L,z) \frac{dV}{dz} dz.
\end{equation}

\subsection{Luminosity density} \label{sec:ld}

Given our assumed evolutionary model, the predicted LD
is given by
\begin{equation} \label{eqn:ldfit}
{\rho_L}_{\rm fit} = {\rho_L}(0) 10^{0.4(P + Q)z},
\end{equation}
(\citealt{lin99} equation~11), where 
\begin{equation}\label{eqn:ld0}
{\rho_L}(0) = \int L \phi(L, z=0) dL = \phi^*(0) L^*(0) \Gamma(\alpha + 2),
\end{equation}
and $\Gamma(x)$ is the standard Gamma function.
In making this prediction, we are integrating over all possible luminosities,
and hence extrapolating our Schechter function fits.
This extrapolation introduces no more than 1 per cent in additional LD
beyond that contained within our luminosity limits.
We obtain luminosities in solar units using the following absolute magnitudes
for the Sun in SDSS bandpasses:
$^{0.1}{M_{\odot}}_{u,g,r,i,z} - 5 \lg h = 6.80, 5.45, 4.76, 4.58, 4.51$ mag
\citep{blan2003L}.

We also directly determine LD as a function of redshift
by summing the weighted luminosities of galaxies in a series of redshift bins:
\begin{equation} \label{eqn:ldsum}
{\rho_L}_j = \frac{1}{V_j} \sum_{i \in j} \frac{W_i L_i}{S_L(z_i)}.
\end{equation}
(Lin et al. equation~16).
Here $V_j$ is the volume of redshift bin $j$,
the sum is over each galaxy $i$ in bin $j$ and the factor
\begin{equation}
 S_L(z) = \int_{{\rm max}(L_{\rm min}(z),L_1)}^{{\rm min}(L_{\rm max}(z),L_2)}
        L \phi(L,z) dL \left/ \int_0^\infty L \phi(L,z) dL\right.,
\end{equation}
(Lin et al. equation~17) extrapolates 
for the luminosity of galaxies lying outside the accessible 
survey flux limits.

\subsection{Binned LF estimates} \label{sec:phiBin}

In order to assess how well the model (equation~\ref{eqn:evol}) parametrizes 
LF evolution, we also make non-parametric, binned, estimates of
the LF in independent redshift ranges using the $1/\Vmax$ 
\citep{schmidt68,eales1993} and stepwise maximum likelihood
(SWML, \citealt{eep88}) methods.
We use 60 magnitude bins from $M = -25$ to $M = -10$ with $\Delta M = 0.25$
and a series of redshift slices.

\begin{figure}
\includegraphics[width=\linewidth]{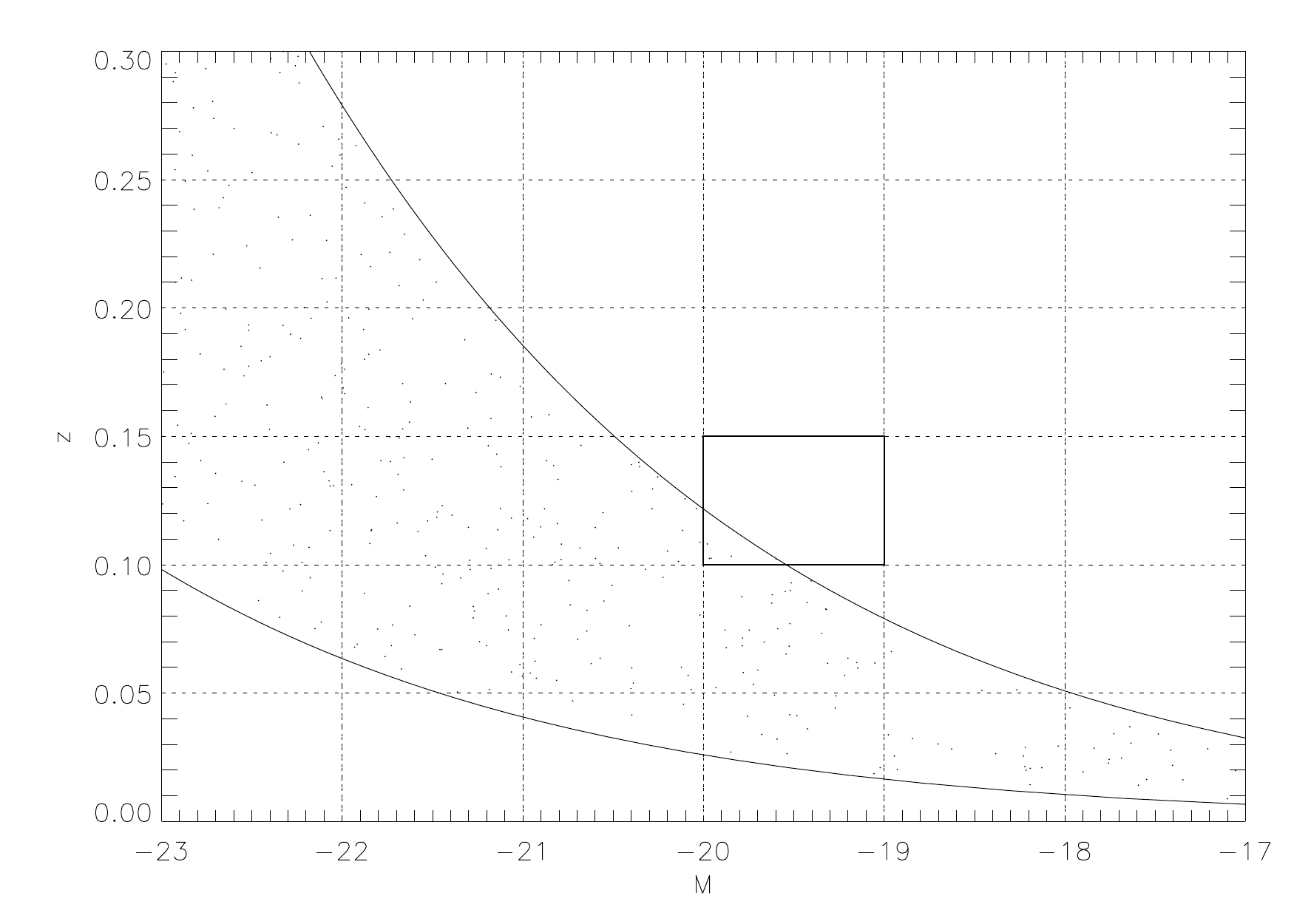}
\caption{Illustration of estimating $\phi(M,z)$ in bins of absolute magnitude
$M$ and redshift $z$ represented by dotted lines
for a fictitious survey with apparent magnitude limits
$m_{\rm bright} = 14.5$ and $m_{\rm faint} = 18$.
Galaxies (represented by points) are of course only found between these 
flux limits, corresponding to the lower and upper curved lines respectively.
Now consider the highlighted bin, centred on $M = -19.5$ and with
redshift limits $z_{\rm lo} = 0.10$, $z_{\rm hi} = 0.15$.
At the lower redshift limit, the absolute magnitude corresponding to
$m_{\rm faint}$ is $M \simeq -19.6$.
Since this is mid-bin, the LF estimated for this bin would be underestimated,
and therefore the bin should be excluded.
Thus for the redshift slice $0.10 < z < 0.15$, only magnitude bins 
brightward of $M = -20$ should be used.
(The fact that the magnitude bin centred on $M = -20.5$ is incomplete at 
redshifts $z > z_{\rm lo}$ will be compensated for by $1/\Vmax$ weighting.)
A similar incompleteness may arise for bins at low redshift and 
high luminosity.
For the redshift slice $0.00 < z < 0.05$, only magnitude bins {\em fainter} 
than $M = -21$ should be used.
}
\label{fig:phi_Mz}
\end{figure}

When estimating the LF over restricted redshift ranges, one has to be careful 
to only include magnitude bins that are fully sampled, since otherwise
the LF will
be underestimated in incompletely sampled bins, see Fig.~\ref{fig:phi_Mz}.
We therefore set the following magnitude limits for each redshift slice 
so that only complete bins are included:
\begin{align} 
M_{\rm faint} &< m_{\rm faint} - DM(z_{\rm lo}) - K(z_{\rm lo}), \nonumber \\
M_{\rm bright} &> m_{\rm bright} - DM(z_{\rm hi}) - K(z_{\rm hi}).
\label{eqn:philim}
\end{align}
Here, $m_{\rm faint}$ and $m_{\rm bright}$ are the flux limits of the survey,
$DM(z)$ is the distance modulus,
$K(z)$ is the $K$-correction for a galaxy with the median SED 
of those in the survey,
$z_{\rm lo}$ and $z_{\rm hi}$ are the limits of the redshift slice under
consideration, and $M_{\rm faint}$ and $M_{\rm bright}$ are the absolute
magnitude limits of each bin.
A bin should only be included if it satisfies both equations 
(\ref{eqn:philim}).

Again following \cite{lin99}, we incorporate the galaxy incompleteness
weights into the SWML maximum likelihood estimator by
multiplying each galaxy's log-probability by its weight 
before summing to form a log-likelihood (equation~\ref{eqn:styLike}).
In the $1/\Vmax$ estimate, we form a sum of the weight of each galaxy
divided by the volume within which it is observable.
We normalize the SWML estimates $\phi_{\rm SWML}$ in each redshift slice
to the $1/\Vmax$ estimates $\phi_{V_{\rm max}}$ by imposing the constraint
\begin{equation}
\sum_{k=1}^{N_{\rm bin}} \phi_{{\rm SWML}_k} V(M_k) = 
\sum_{k=1}^{N_{\rm bin}} \phi_{{V_{\rm max}}_k} V(M_k),
\end{equation}
where $V(M_k)$ is the volume (within the redshift limits of each slice)
over which a galaxy of absolute magnitude $M_k$, being the mean galaxy
absolute magnitude in bin $k$, is visible.
The predicted number of galaxies 
\begin{equation}
N_{\rm SWML} = \frac{1}{\bar{W}} \sum_{k=1}^{N_{\rm bin}} \phi_k V(M_k) \Delta M
\end{equation}
may also be compared with
the observed number of galaxies within each redshift range.

We can use our SWML LF estimates to assess the quality of the parametric fits
using a likelihood ratio test \citep{eep88}.
In this test, we compare the log-likelihoods $\ln {\cal L}_1$ and 
$\ln {\cal L}_2$ given by equation (\ref{eqn:styLike}) 
for the SWML and parametric estimates respectively.
The log likelihood ratio $-2 \ln({\cal L}_1/{\cal L}_2)$ is expected to 
follow a $\chi^2$ distribution with $\nu = N_p - 4$ degrees of freedom.
Here $N_p$ is the number of bins in the stepwise estimate and we subtract
1 degree of freedom for each of the fitted shape parameters 
$\alpha$, $M^*(0)$ and $Q$ and for the arbitrary normalisation.

To allow for the finite bin sizes and redshift ranges of the SWML estimates,
we calculate binned estimates of the parametric fits.
These are given by \citep{lin99}
\begin{equation} \label{eq:stybin}
\phi_k^{z_1 - z_2} = \frac{
\int_{M_k - \Delta M/2}^{M_k + \Delta M/2}
\int_{{\rm max}[z_{\rm min}(M), z_1]}^{{\rm min}[z_{\rm max}(M), z_2]}
\phi^2(M,z) \frac{dV}{dz} dz\, dM}
{
\int_{M_k - \Delta M/2}^{M_k + \Delta M/2}
\int_{{\rm max}[z_{\rm min}(M), z_1]}^{{\rm min}[z_{\rm max}(M), z_2]}
\phi(M,z) \frac{dV}{dz} dz\, dM}.
\end{equation}
Here, the parametric LF $\phi(M,z)$ is weighted by 
the number of galaxies at each magnitude and redshift, given by the factor
$\phi(M,z) \frac{dV}{dz}$ .
These binned versions of the parametric fits are also used when plotting
the LFs.
For absolute magnitudes in all plots,
we use the weighted mean magnitude of the galaxies
in each bin, rather than using the magnitude of the bin centre.
This helps to overcome the bias due to the finite width of magnitude bins.

\subsection{LF faint end}

It is now widely recognised that a Schechter function provides a poor fit
to galaxy LFs when measured over a wide range of magnitudes 
(e.g. \citealt{blan2005D}).
In order to parametrize the faint end, we separately analyse a
low redshift ($z < 0.1$) subset of the data and fit (non-evolving)
double power-law Schechter functions.

We use the parameterization of \cite{love97}, namely
\begin{equation}
\phi(L) = \phi^* \left(\frac{L}{L^*}\right)^\alpha 
	    \exp\left(\frac{-L}{L^*}\right)
	    \left[1 + \left(\frac{L}{L_t}\right)^\beta\right].
\label{eqn:dpschec}
\end{equation}
In this formulation, the standard Schechter function is multiplied by
the factor $[1 + (L/L_t)^\beta]$, where
$L_t < L^*$ is a transition luminosity between two power-laws of slope
$\alpha$ ($L \ll L_t$) and $\alpha + \beta$ ($L \gg L_t$).
It is fitted to unbinned data using an extension to the method of \cite{sty79}.
With this four-parameter fit (the normalisation $\phi^*$ is fitted separately), 
one has to be careful to choose sensible
starting values in order for the downhill simplex algorithm 
{\tt (scipy.optimize.fmin}) not to get stuck in local minima of $-\ln{\cal L}$.
(We also found that it helped to call the minimizer several times, 
using `best-fitting' parameters from one function call as starting parameters for
the next;
$-\ln{\cal L}$ was found to converge with 2--4 calls of the minimizer.)

Note that the double power-law Schechter function may
equivalently be written as the sum of two Schechter functions,
e.g. \citet{blan2005D,2008MNRAS.388..945B}, a fact which comes in useful
when integrating the LF.

When fitting a double power-law Schechter function, 
the likelihood ratio test has $\nu = N_p - 5$ degrees of freedom
(cf. section~\ref{sec:phiBin}).

\subsection{Error estimates}  \label{sec:errors}

Schechter and evolution parameter estimates are strongly correlated,
and so in Section~\ref{sec:results} we present 95 per cent likelihood contour
plots of shape parameters $\alpha$, $M^*$, $\beta$, $M_t$ 
and evolution parameters $Q$, $P$.
For uncertainties in tabulated measurements, we estimate errors using
the jackknife technique, as follows.
We divide the GAMA survey area into nine regions, 
each $4\times 4$ deg $^2$.
We then calculate the LF and LD using the methods discussed above, 
omitting each region in turn.
For any parameter $x$, we may then determine its variance using
\begin{equation}
{\rm Var}(x) = \frac{N-1}{N} \sum_{i=1}^N (x_i - \bar{x})^2,
\end{equation}
where $N=9$ is the number of jackknife regions and $\bar{x}$ is the mean of 
the parameters $x_i$ measured while excluding region $i$.
The jackknife method has the advantage of providing error estimates
which include both uncertainties in the fitted parameters as well as
sample variance.

The sample variance in galaxy density $n$ may also be determined using:
\begin{equation}
\left(\frac{\delta n}{n}\right)^2 = \frac{4 \pi J_3}{V},
\end{equation}
\citep{dh82,eep88}, where $J_3$ is defined in (\ref{eqn:zweight}) and 
$V$ is the volume of each sample between redshift limits.

For errors on binned LFs, we use Poisson errors for $1/\Vmax$ 
estimates and an inversion of the information matrix for SWML estimates
\citep{eep88}.

\section{Results}  \label{sec:results}

Before presenting our main results, we first check the effects of
correcting for imaging completeness and the choice of flow model
in converting redshifts to distances.

\subsection{Imaging completeness correction}

\begin{figure}
\includegraphics[width=\linewidth]{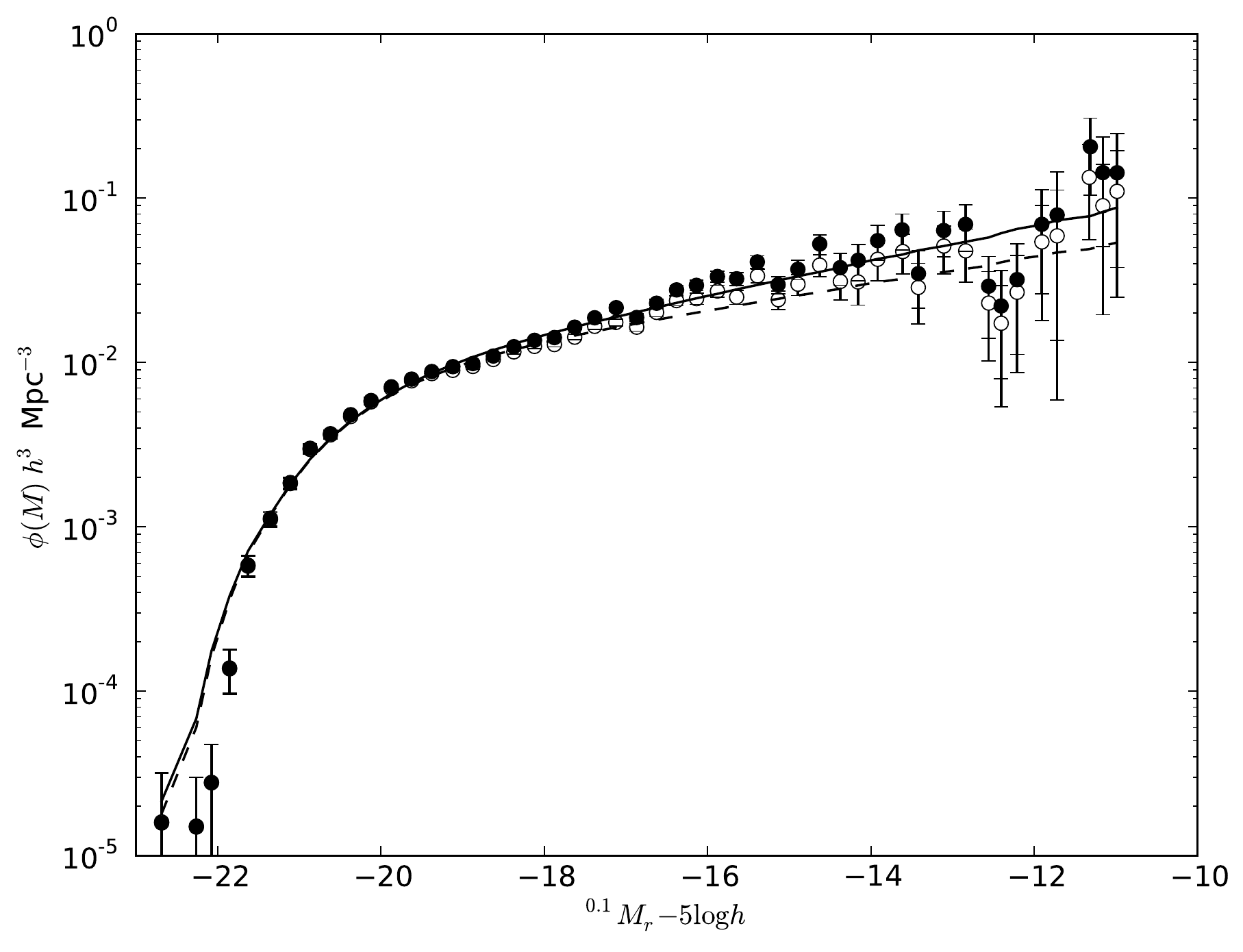}
\caption{LF estimates in the $r$ band for low-redshift
galaxies ($z < 0.1$) with (solid symbols and line) and without 
(open symbols, dashed line) applying a correction for imaging completeness.
Symbols show SWML estimates, lines show best-fitting Schechter functions.
}
\label{fig:lf_imcomp}
\end{figure}

\begin{table}
\caption{Change in fitted Schechter parameters for combined samples
when applying imaging completeness correction.}
\label{tab:lf_imcomp}
\begin{math}
\begin{array}{lrrr}
\hline
\mbox{Band} & \Delta \alpha & \Delta M^* /\mbox{mag} & \Delta \phi^*/\denunit\\
\hline
u & -0.05 & -0.03 & 0.00017\\
g & -0.05 & -0.04 & -0.00031\\
r & -0.06 & -0.07 & -0.00062\\
i & -0.05 & -0.06 & -0.00051\\
z & -0.03 & -0.03 & -0.00011\\
\hline
\end{array}
\end{math}
\end{table}

In Fig.~\ref{fig:lf_imcomp}, we compare $r$-band LFs calculated
for the combined sample, with distances calculated using the \cite{tbad2000} 
multiattractor flow model,
with and without the correction for imaging completeness described
in Section~\ref{sec:imcomp}.
As expected, we see that applying imaging completeness corrections 
boosts the LF faint end, while barely changing the bright end.
The changes in fitted Schechter parameters due to imaging completeness 
corrections are tabulated in Table~\ref{tab:lf_imcomp}.
Future plots and tabulated parameters will include imaging completeness 
corrections;
approximate uncorrected Schechter parameters may be obtained by subtracting
the appropriate quantities listed in Table~\ref{tab:lf_imcomp}.

\subsection{Effects of velocity flow model}

\begin{figure}
\includegraphics[width=\linewidth]{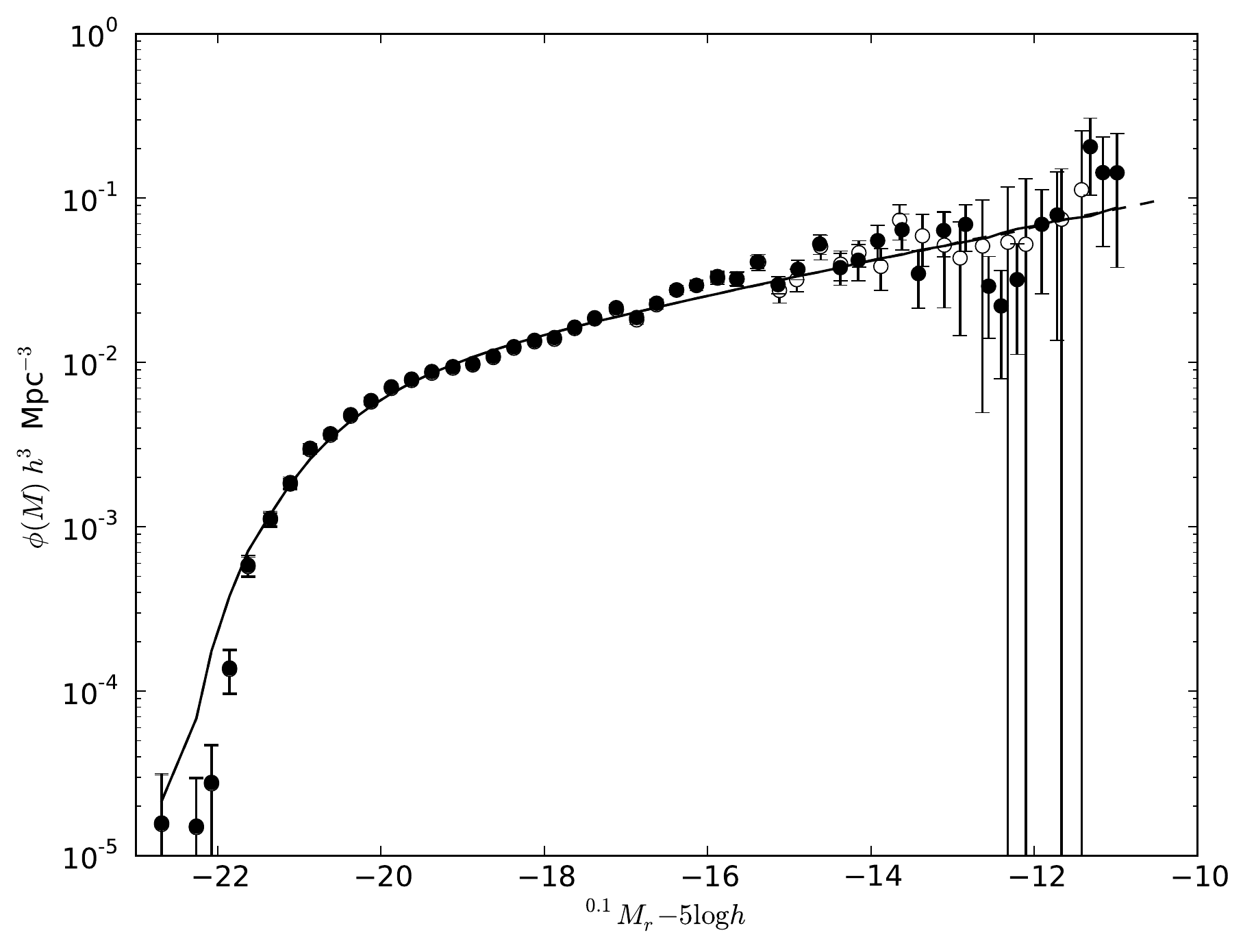}
\caption{LF estimates in the $r$ band for low-redshift
galaxies ($z < 0.1$) using
the CMB reference frame (open circles) and
the \protect\cite{tbad2000} multiattractor flow model (filled circles) using
the SWML estimator.
Solid and dashed lines show the best-fit Schechter functions which are
indistinguishable.
}
\label{fig:lf-Tonry}
\end{figure}

\begin{figure*}
\includegraphics[width=0.9\linewidth]{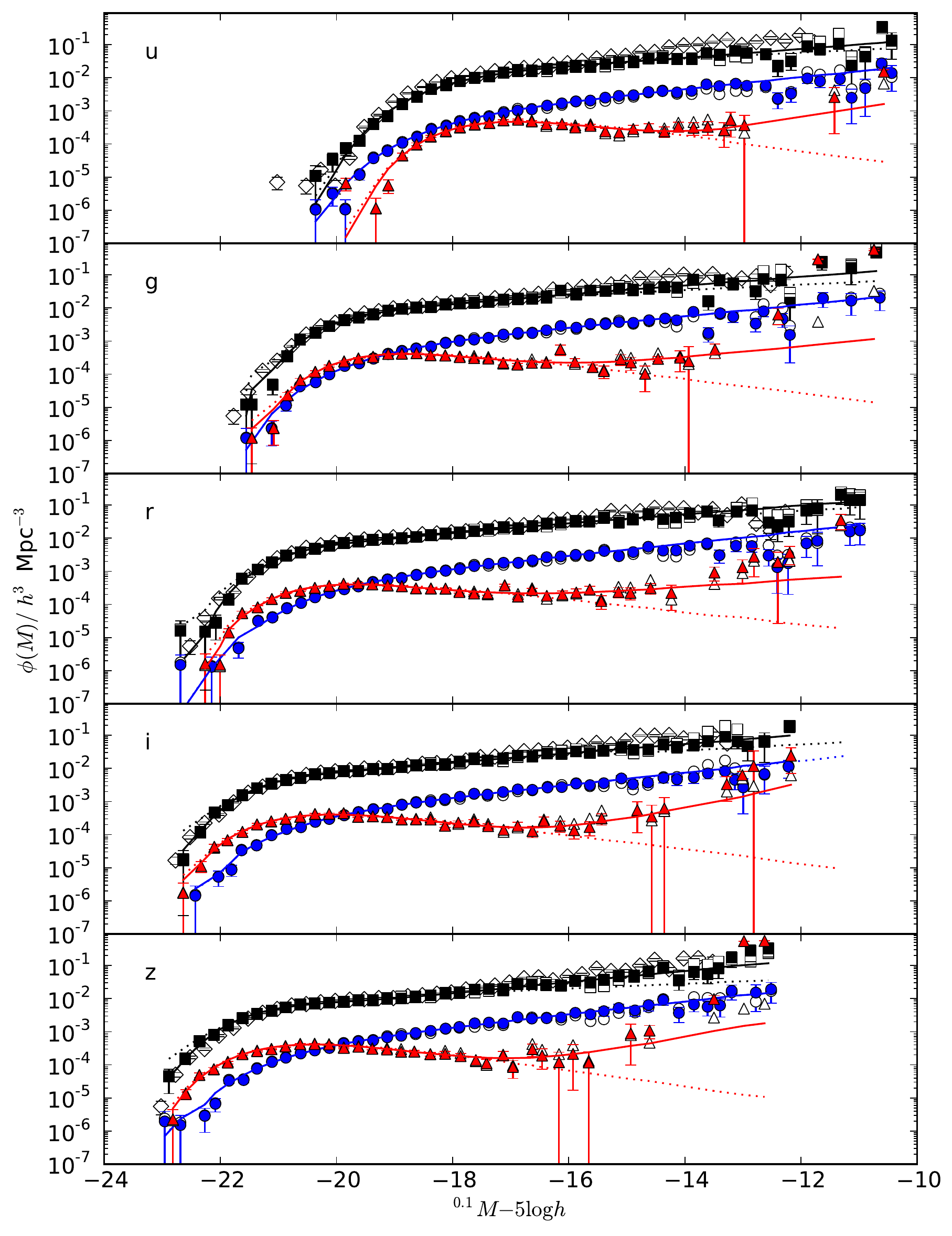}
\caption{$ugriz$ LFs at low redshift ($z < 0.1$).
Black squares show SWML estimates for combined red and blue samples, 
blue circles and red triangles show SWML LFs for the blue and red samples
respectively.
Open symbols of the same shapes show the corresponding $1/\Vmax$ estimates ---
these are hidden beneath the SWML estimates for all but the very 
faintest galaxies.
Continuous lines show the best-fit non-evolving double power-law Schechter
function fits, dotted lines show standard Schechter function fits.
LFs for the blue and red samples have been scaled by a factor of 0.1
to aid legibility.
Open diamonds show the `corrected' LF (without colour selection) 
from \protect\cite{blan2005D}.
}
\label{fig:lf_faint}
\end{figure*}

Luminosities of galaxies at the extreme faint end of the LF, being very
close by, will be sensitive to peculiar velocities.
In Fig.~\ref{fig:lf-Tonry}, we compare $r$-band LFs calculated
using the CMB reference frame \citep{1996ApJ...470...38L}
and the \cite{tbad2000} multiattractor flow model for galaxies at low redshift
($z < 0.1$).
We see that the CMB-frame and flow model 
LFs only begin to differ at the extreme faint end,
$M_r  - 5 \log h \ga -15$ mag, and even at these faint magnitudes the
differences are not large given the size of the error bars.
In particular, the recovered Schechter parameters are indistinguishable
between the two velocity frames.
Subsequent analysis will use the Tonry et al. 
flow model to determine luminosities.

\subsection{Luminosity function faint end}  \label{sec:lffaint}

\begin{table*}
\caption{Standard Schechter function fits for low-redshift galaxies.
Samples are as given in the first column.
$^{0.1}M_1$ and $^{0.1}M_2$ are the absolute magnitude limits, 
$N_{\rm gal}$ the number of galaxies in the sample and $N_{\rm pred}$
the predicted number of galaxies from integrating the LF 
(equation~\ref{eqn:Npred}).
$\alpha$, $^{0.1}M^*$ and $\phi^*$ are the usual Schechter parameters,
and $P_{\rm fit}$ gives the probability of the Schechter
function describing the observed LF determined from the likelihood ratio test
described in the text.
Luminosity densities ${\rho_L}_{\rm fit}$ and ${\rho_L}_{\rm sum}$
are calculated via equations (\ref{eqn:ld0}) and (\ref{eqn:ldsum}) 
respectively.
}
\label{tab:faintfit}

    \begin{math}
    \begin{array}{crrrrrrrrrr}
    \hline
    & ^{0.1}M_1 & ^{0.1}M_2 & 
    \multicolumn{1}{c}{N_{\rm gal}} & \multicolumn{1}{c}{N_{\rm pred}} &
    \multicolumn{1}{c}{\alpha} &
    \multicolumn{1}{c}{^{0.1}M^* - 5 \lg h} & \multicolumn{1}{c}{\phi^* \times 100} &
    \multicolumn{1}{c}{P_{\rm fit}} &
    \multicolumn{1}{c}{{\rho_L}_{\rm fit}} &
    \multicolumn{1}{c}{{\rho_L}_{\rm sum}}\\
    & \multicolumn{2}{c}{- 5 \lg h} & & & & & \multicolumn{1}{c}{/\denunit} & &
    \multicolumn{2}{c}{/10^8 \ldenunit}\\
    \hline
    \mbox{All}\\
        u & -21.0 & -10.0 &  9181 &  9402 \pm   766 &  -1.21 \pm   0.03 & -18.02 \pm   0.04 & 1.96 \pm 0.15 & 0.001 &  1.95 \pm 0.18 &  1.95 \pm 0.18\\
            g & -22.0 & -10.0 & 11158 & 11085 \pm   781 &  -1.20 \pm   0.01 & -19.71 \pm   0.02 & 1.33 \pm 0.12 & 0.000 &  1.79 \pm 0.14 &  1.79 \pm 0.15\\
            r & -23.0 & -10.0 & 12860 & 12789 \pm   956 &  -1.26 \pm   0.02 & -20.73 \pm   0.03 & 0.90 \pm 0.07 & 0.000 &  1.75 \pm 0.15 &  1.75 \pm 0.15\\
            i & -23.0 & -11.0 & 10438 & 10341 \pm   745 &  -1.22 \pm   0.01 & -21.13 \pm   0.02 & 0.90 \pm 0.08 & 0.000 &  2.06 \pm 0.18 &  2.06 \pm 0.18\\
            z & -24.0 & -12.0 &  8647 &  8535 \pm   658 &  -1.18 \pm   0.03 & -21.41 \pm   0.05 & 0.90 \pm 0.06 & 0.000 &  2.39 \pm 0.22 &  2.38 \pm 0.22\\
            \mbox{Blue}\\
        u & -21.0 & -10.0 &  6278 &  6664 \pm   509 &  -1.44 \pm   0.02 & -18.27 \pm   0.04 & 0.88 \pm 0.05 & 0.028 &  1.50 \pm 0.14 &  1.52 \pm 0.14\\
            g & -22.0 & -10.0 &  7356 &  7611 \pm   537 &  -1.42 \pm   0.02 & -19.58 \pm   0.05 & 0.71 \pm 0.03 & 0.002 &  1.12 \pm 0.10 &  1.12 \pm 0.10\\
            r & -23.0 & -10.0 &  8579 &  8893 \pm   680 &  -1.45 \pm   0.02 & -20.28 \pm   0.07 & 0.55 \pm 0.03 & 0.000 &  0.92 \pm 0.09 &  0.92 \pm 0.09\\
            i & -23.0 & -11.0 &  6432 &  6641 \pm   465 &  -1.45 \pm   0.02 & -20.68 \pm   0.06 & 0.50 \pm 0.03 & 0.381 &  1.02 \pm 0.09 &  1.02 \pm 0.09\\
            z & -24.0 & -12.0 &  4888 &  5089 \pm   400 &  -1.48 \pm   0.03 & -20.99 \pm   0.07 & 0.41 \pm 0.02 & 0.477 &  1.11 \pm 0.11 &  1.10 \pm 0.11\\
            \mbox{Red}\\
        u & -21.0 & -10.0 &  2903 &  2850 \pm   263 &  -0.40 \pm   0.08 & -17.34 \pm   0.06 & 1.29 \pm 0.12 & 0.000 &  0.52 \pm 0.05 &  0.53 \pm 0.05\\
            g & -22.0 & -10.0 &  3802 &  3758 \pm   307 &  -0.47 \pm   0.07 & -19.31 \pm   0.06 & 1.06 \pm 0.11 & 0.000 &  0.75 \pm 0.07 &  0.75 \pm 0.07\\
            r & -23.0 & -10.0 &  4281 &  4265 \pm   354 &  -0.53 \pm   0.04 & -20.28 \pm   0.06 & 0.98 \pm 0.09 & 0.000 &  0.90 \pm 0.08 &  0.89 \pm 0.08\\
            i & -23.0 & -11.0 &  4006 &  4014 \pm   332 &  -0.46 \pm   0.03 & -20.63 \pm   0.05 & 1.04 \pm 0.09 & 0.000 &  1.13 \pm 0.11 &  1.11 \pm 0.11\\
            z & -24.0 & -12.0 &  3759 &  3760 \pm   301 &  -0.40 \pm   0.05 & -20.87 \pm   0.06 & 1.10 \pm 0.07 & 0.000 &  1.39 \pm 0.13 &  1.38 \pm 0.13\\
            
    \hline
    \end{array}
    \end{math}
    
\end{table*}

\begin{table*}
\caption{Double power-law Schechter function fits for low-redshift galaxies.
Values for $^{0.1}M_1$, $^{0.1}M_2$ and $N_{\rm gal}$ are the same 
for each sample as in Table~\ref{tab:faintfit}.
Other columns have the same meaning as the previous Table and in addition
we tabulate the values of the double power-law Schechter parameters 
$\beta$ and $M_t$.
}
\label{tab:dpfit}

    \begin{math}
    \begin{array}{crrrrrrrrr}
    \hline
     & 
    \multicolumn{1}{c}{N_{\rm pred}} &
    \multicolumn{1}{c}{\alpha} & \multicolumn{1}{c}{\beta} &
    \multicolumn{1}{c}{^{0.1}M^* - 5 \lg h} &
    \multicolumn{1}{c}{^{0.1}M_t - 5 \lg h} &
    \multicolumn{1}{c}{\phi^* \times 100} & \multicolumn{1}{c}{P_{\rm fit}} &
    \multicolumn{1}{c}{{\rho_L}_{\rm fit}} &
    \multicolumn{1}{c}{{\rho_L}_{\rm sum}}\\
    & & & & & & /\denunit & & \multicolumn{2}{c}{/10^8 \ldenunit}\\
    \hline
    \mbox{All}\\
        u &  9397 \pm   761 & 
             -0.81 \pm   0.26 &
             -0.56 \pm   0.28 &
            -17.87 \pm   0.14 &
            -17.38 \pm   0.39 &
            1.32 \pm 0.26 &
            0.005 
            & 1.97 \pm 0.18 & 1.97 \pm 0.18\\g & 11199 \pm   798 & 
              0.09 \pm   0.10 &
             -1.41 \pm   0.10 &
            -19.05 \pm   0.05 &
            -18.99 \pm   0.06 &
            1.28 \pm 0.10 &
            0.000 
            & 1.83 \pm 0.15 & 1.83 \pm 0.15\\r & 12900 \pm   968 & 
              0.14 \pm   0.09 &
             -1.47 \pm   0.09 &
            -19.92 \pm   0.10 &
            -19.86 \pm   0.18 &
            1.02 \pm 0.13 &
            0.011 
            & 1.76 \pm 0.15 & 1.76 \pm 0.15\\i & 10447 \pm   759 & 
              0.10 \pm   0.01 &
             -1.44 \pm   0.03 &
            -20.32 \pm   0.04 &
            -20.10 \pm   0.12 &
            1.10 \pm 0.12 &
            0.606 
            & 2.07 \pm 0.18 & 2.07 \pm 0.18\\z &  8664 \pm   675 & 
             -0.07 \pm   0.35 &
             -1.35 \pm   0.27 &
            -20.63 \pm   0.17 &
            -19.99 \pm   0.33 &
            1.28 \pm 0.15 &
            0.729 
            & 2.41 \pm 0.23 & 2.41 \pm 0.23\\\mbox{Blue}\\
        u &  6663 \pm   508 & 
             -1.39 \pm   0.03 &
             -0.09 \pm   0.02 &
            -18.27 \pm   0.05 &
            -17.98 \pm   0.37 &
            0.45 \pm 0.02 &
            0.015 
            & 1.51 \pm 0.14 & 1.52 \pm 0.14\\g &  7610 \pm   527 & 
             -1.37 \pm   0.01 &
             -0.10 \pm   0.02 &
            -19.57 \pm   0.04 &
            -15.58 \pm   4.88 &
            0.42 \pm 0.08 &
            0.001 
            & 1.12 \pm 0.09 & 1.13 \pm 0.10\\r &  8898 \pm   674 & 
             -1.40 \pm   0.05 &
             -0.09 \pm   0.10 &
            -20.28 \pm   0.05 &
            -20.14 \pm   2.16 &
            0.28 \pm 0.03 &
            0.000 
            & 0.92 \pm 0.09 & 0.92 \pm 0.09\\i &  6650 \pm   468 & 
             -1.43 \pm   0.06 &
             -0.05 \pm   0.09 &
            -20.69 \pm   0.05 &
            -19.76 \pm   1.47 &
            0.25 \pm 0.03 &
            0.225 
            & 1.02 \pm 0.09 & 1.02 \pm 0.09\\z &  5081 \pm   403 & 
             -1.42 \pm   0.03 &
             -0.10 \pm   0.00 &
            -20.98 \pm   0.08 &
            -20.30 \pm   0.34 &
            0.22 \pm 0.01 &
            0.389 
            & 1.10 \pm 0.11 & 1.10 \pm 0.11\\\mbox{Red}\\
        u &  2845 \pm   264 & 
             -0.21 \pm   0.16 &
             -1.57 \pm   0.42 &
            -17.22 \pm   0.10 &
            -14.13 \pm   0.52 &
            1.34 \pm 0.14 &
            0.000 
            & 0.53 \pm 0.04 & 0.54 \pm 0.04\\g &  3751 \pm   302 & 
             -0.14 \pm   0.30 &
             -1.28 \pm   0.29 &
            -19.08 \pm   0.13 &
            -16.39 \pm   1.46 &
            1.14 \pm 0.17 &
            0.000 
            & 0.75 \pm 0.07 & 0.75 \pm 0.07\\r &  4245 \pm   349 & 
             -0.15 \pm   0.29 &
             -1.16 \pm   0.10 &
            -19.99 \pm   0.15 &
            -17.33 \pm   1.17 &
            1.09 \pm 0.15 &
            0.001 
            & 0.89 \pm 0.08 & 0.89 \pm 0.08\\i &  3974 \pm   327 & 
             -0.33 \pm   0.10 &
             -1.58 \pm   0.43 &
            -20.51 \pm   0.08 &
            -16.46 \pm   0.71 &
            1.12 \pm 0.10 &
            0.228 
            & 1.13 \pm 0.14 & 1.13 \pm 0.14\\z &  3730 \pm   302 & 
             -0.27 \pm   0.20 &
             -1.51 \pm   0.51 &
            -20.75 \pm   0.12 &
            -16.93 \pm   1.19 &
            1.16 \pm 0.09 &
            0.184 
            & 1.38 \pm 0.16 & 1.38 \pm 0.16\\
    \hline
    \end{array}
    \end{math}
    
\end{table*}

Having looked at the effects of incompleteness and flow corrections, 
we now study in detail the faint end of the LF 
for low redshift ($z < 0.1$) galaxies.
Fig.~\ref{fig:lf_faint} shows the LFs for our three (combined, blue and red)
samples in the $ugriz$ bands.
Also shown are LFs corrected for surface brightness
incompleteness by \citet{blan2005D} from the New York University 
Value-Added Galaxy Catalog (NYU-VAGC) low-redshift sample \citep{blan2005N}.
Since these Blanton et al. LFs were calculated using restframe $K$-corrections,
we apply an offset of $2.5 \lg(1 + z_0)$ to their absolute magnitudes in order
to convert to our $z_0 = 0.1$ band-shifted $K$-corrections.
Our faint-end LFs are systematically lower than those of 
Blanton et al., particularly in the $u$ band.
The difference can largely be explained by the different flow models used
by Blanton et al. and in the present analysis.
Re-analysing the Blanton et al. data using the Tonry et al. flow model
results in much better agreement \citep{Baldry2011} --- the extra 
1.7 mag depth of the GAMA versus the SDSS main galaxy sample 
means that uncertainties
due to the flow model affect the measured LF only at a correspondingly 
fainter magnitude.

Table~\ref{tab:faintfit} shows the number of galaxies and absolute
magnitude limits for each sample, along with the parameters of standard
Schechter function fits and luminosity densities.
Only for blue galaxies in the $u$, $i$ and $z$ bands does a standard
Schechter function provide a statistically acceptable fit to the data
at the 2 per cent level or better.
For red galaxies, we observe a decline in number density faintwards 
of the characteristic magnitude $M^*$ with a subsequent increase in faint-end
slope at $M_t \sim M^* + 3$.
For the red galaxies, and the combined sample, a double-power-law Schechter 
function (\ref{eqn:dpschec}) is required to fit the shape of the observed LFs.
These findings are in apparent agreement with the predictions of 
halo occupation distribution models, e.g. \citet{2008ApJ...682..937B}, 
in which luminous red galaxies are central galaxies, but fainter red galaxies
are increasingly more likely to be satellites in relatively massive halos.
An alternative perspective is provided by \citet{2010ApJ...721..193P},
who explain the change in faint-end slope of red galaxies 
via a simple picture for the quenching of star formation
by the distinct processes of `mass quenching' and `environment quenching'.
Our results provide the most precise demonstration of the changing faint-end
slope of red galaxies to date.

However, the observed upturn needs to be interpreted with caution, 
since, from a quick visual inspection, 
the 164 faint ($^{0.1}M_r - 5 \lg h > -16$ mag), red galaxies that
comprise the upturn include a significant fraction ($\simeq 50$ per cent)
of galaxies that appear to be disc like, as well as a number of artifacts.
It thus seems likely that dust-reddened disc systems, as well as dwarf
galaxies with intrinsically red stellar populations, contribute to the
faint-end upturn in the red galaxy LF.
Future work will investigate the LF dependence on morphology 
and dust reddening, utilising GAMA's multi-wavelength coverage.

Double-power-law Schechter function fits are given in Table~\ref{tab:dpfit}.
Likelihood ratio tests show that the double-power-law Schechter function
provides significantly better fits than the standard Schechter function 
for the combined and red galaxy samples, at least for the redder bands.
For the blue galaxies, however, the double-power-law Schechter function fits
are actually worse than the standard Schechter function fits
when taking into account the two additional degrees of freedom.

\begin{figure}
\includegraphics[width=\linewidth]{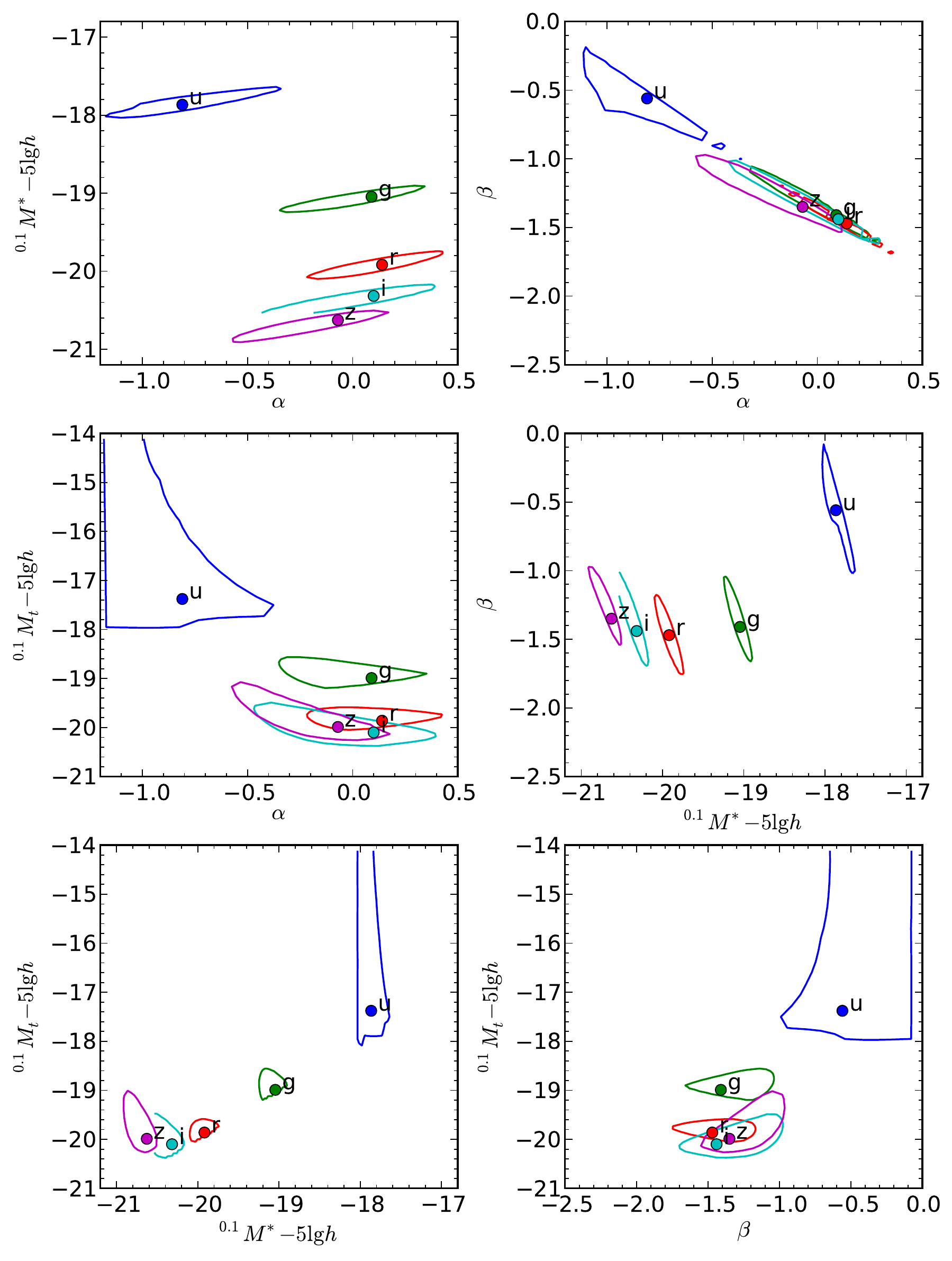}
\caption{2-$\sigma$ likelihood contours for various parameter pairs
in double power-law Schechter function fits to the combined sample
for $ugriz$ bands as labelled.
}
\label{fig:like_faint}
\end{figure}

\begin{figure}
\includegraphics[width=\linewidth]{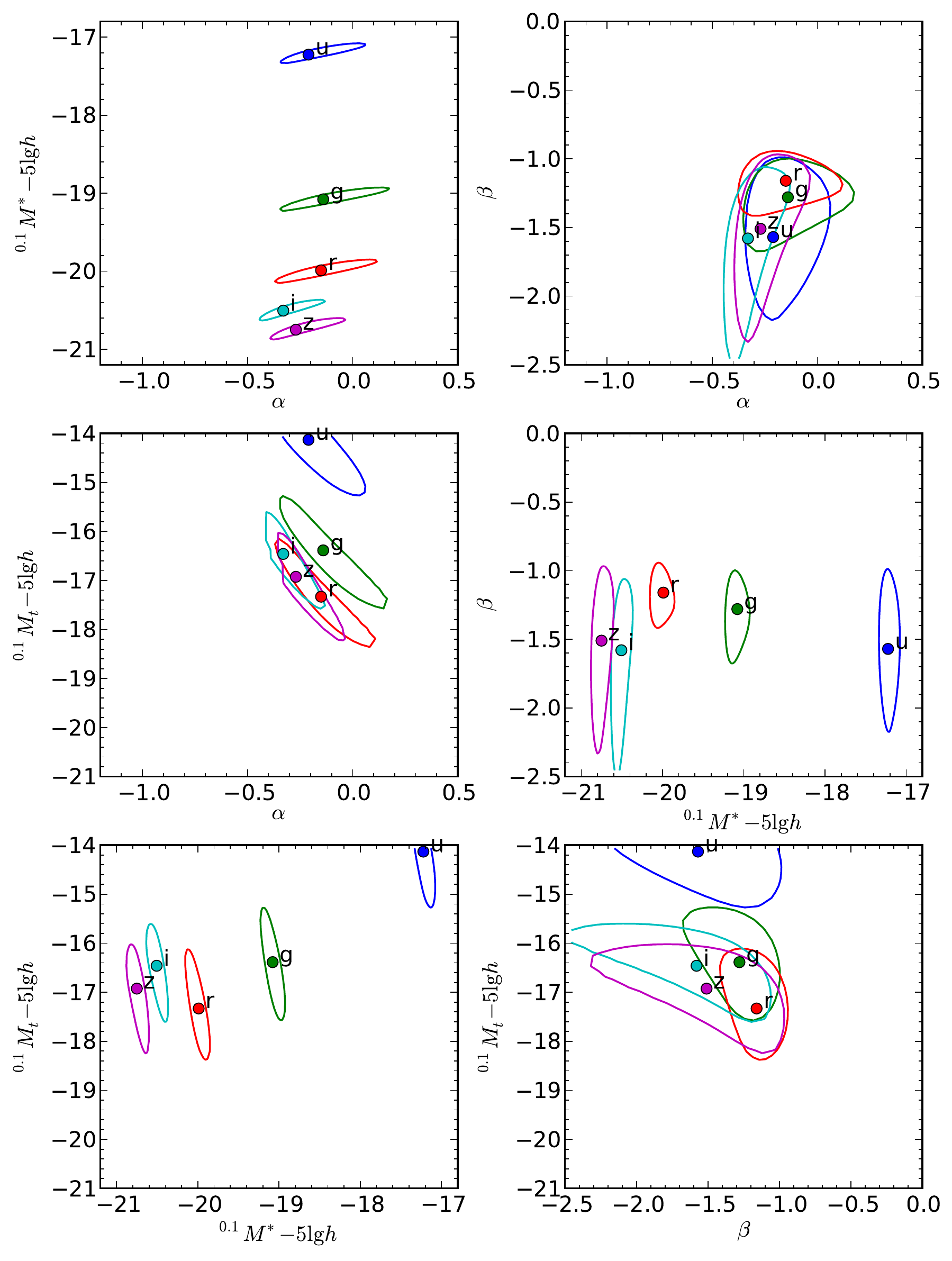}
\caption{2-$\sigma$ likelihood contours for various parameter pairs
in double power-law Schechter function fits to the red sample
for $ugriz$ bands as labelled.
}
\label{fig:like_faint_red}
\end{figure}

The quoted errors need to be treated with caution 
due to strong correlations between the parameters,
particularly in the case of the five-parameter, 
double power-law Schechter function fits.
Fig.~\ref{fig:like_faint} shows $2\sigma$ likelihood contours for each 
pair of parameters from $\alpha, M^*, \beta$ and $M_t$ for the combined sample.
We see that $\alpha$ and $\beta$ individually are very poorly constrained, 
with an uncertainty $\Delta \alpha \simeq \Delta \beta \simeq 0.5$.
However, the overall faint-end slope $\alpha + \beta$ 
is very well constrained, with a consistent value in all five passbands 
$\alpha + \beta = -1.37 \pm 0.05$ for the combined sample.
For blue galaxies, $\alpha + \beta = -1.50 \pm 0.03$ and for red galaxies,
$\alpha + \beta = -1.6 \pm 0.3$.
Consistent faint-end slopes are found for the stellar mass function
\citep{Baldry2011} .
Also from Fig.~\ref{fig:like_faint}, we see that the characteristic magnitude 
$M^*$ is positively correlated with slope $\alpha$ but negatively correlated 
with $\beta$.
The transition magnitude $M_t$
is only weakly correlated with either slope parameter $\alpha$ or $\beta$
and almost completely uncorrelated with the characteristic magnitude $M^*$.

Fig.~\ref{fig:like_faint_red} shows $2\sigma$ likelihood contours for the 
red sample.
We see that, while still uncertain, the slope parameters $\alpha$ and $\beta$ 
are only weakly correlated.
The characteristic magnitude $M^*$ is positively correlated with $\alpha$ but
almost completely uncorrelated with $\beta$.
The transition magnitude $M_t$ is strongly anti-correlated with $\alpha$ and
virtually independent of $\beta$.

Since the shape of the blue galaxy LF is reasonably well fitted by a 
standard Schechter function,
there are huge degeneracies between the double power-law Schechter function
parameters, and so the contour plots contain no useful information, 
and hence are not shown.

In summary, in analysing the faint end of the LFs, we have found that:
\begin{enumerate}
\item While a standard Schechter function provides an acceptable fit
to the blue galaxy LF in all bands, the red galaxy LF exhibits a decline
just faintwards of $M^*$ followed by a pronounced upturn at magnitude
$M_t \simeq M^* + 3$.
Such an LF is well-fitted by a double-power-law Schechter function.
\item We caution that the faint end of the red galaxy LF is possibly
dominated by dust-reddened systems, rather than by galaxies
with intrinsically red stellar populations.
\item Neither standard nor double-power-law Schechter function faint-end
slopes show any systematic dependence on passband: while strongly
colour-dependent, faint-end slopes are largely independent of passband.
\item The characteristic magnitude $M^*$ (and to a lesser extent, the
transition magnitude $M_t$) brightens systematically and significantly with
passband effective wavelength.
\end{enumerate}

\subsection{Luminosity function evolution} \label{sec:lfev}

\begin{figure*}
\includegraphics[angle=90,width=0.85\linewidth]{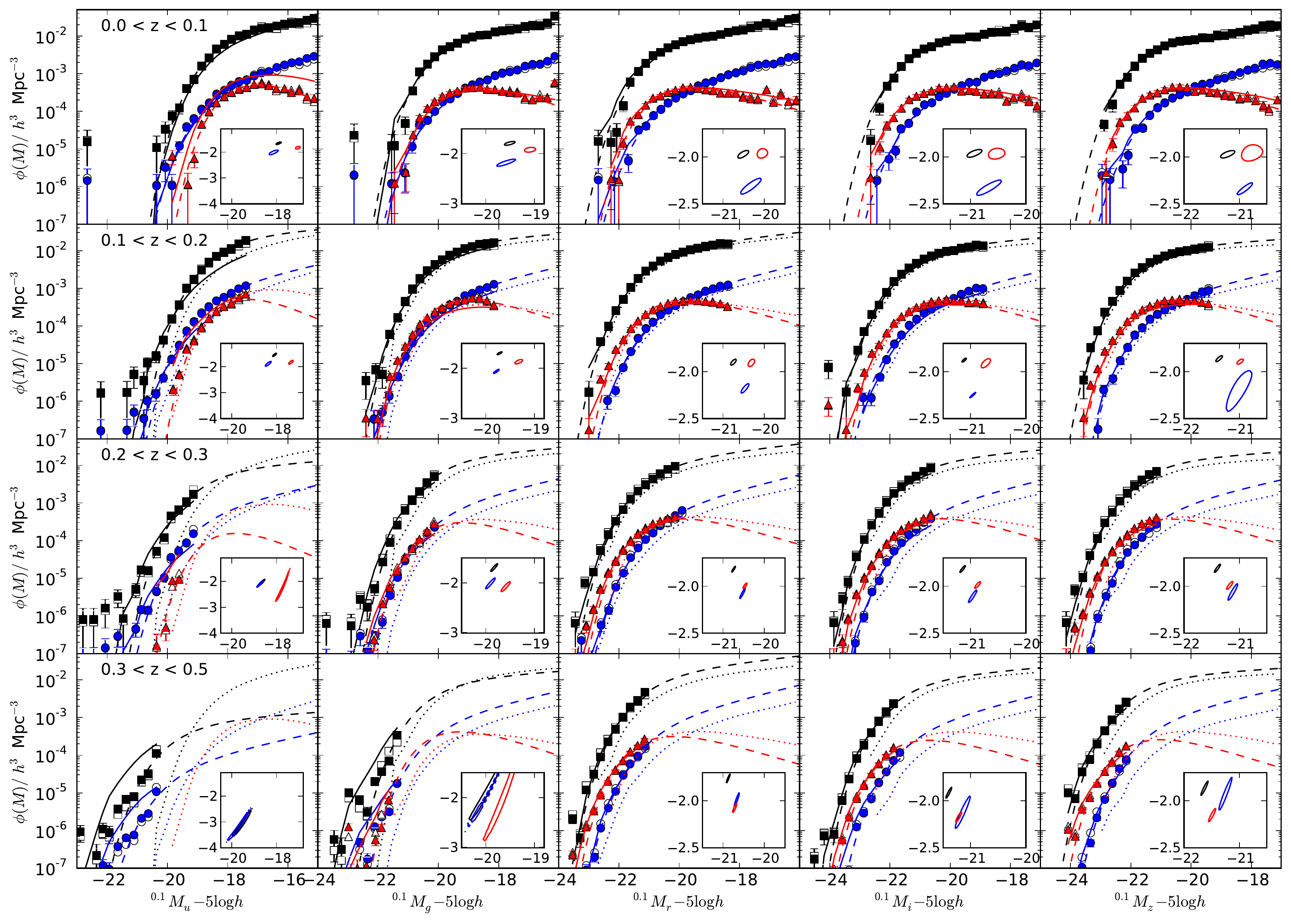}
\caption{Evolution of the $ugriz$ LFs.
The five columns show the $ugriz$ LFs respectively from left to right.
The four rows show the LFs in four redshift ranges increasing from top
to bottom as indicated in the leftmost panels.
Filled black squares show SWML estimates for combined red and blue samples, 
filled blue circles and red triangles show SWML LFs for the blue and red 
samples respectively.
Open symbols show the corresponding $1/\Vmax$ estimates --- in most cases these
are indistinguishable from the SWML estimates.
Continuous lines show the parametric evolving LF for each sample
The dotted lines reproduce the parametric LF fit for each sample 
from the lowest redshift bin.
Dashed lines show least-squares fits to the SWML estimates with $\alpha$ 
fixed at higher redshifts.
The insets show the 95 per cent likelihood contours for ($M^*$, $\lg \phi^*$) 
parameters obtained from these fits.
LFs (but not contours) for the blue and red sample have been scaled by a 
factor of 0.1 to aid legibility.
}
\label{fig:lf_ev}
\end{figure*}

We present LFs for the combined, blue and red
samples in the $ugriz$ bands in four redshift ranges in Fig.~\ref{fig:lf_ev}.
Table~\ref{tab:evfit} gives the magnitude limits (chosen to exclude the
upturn seen in the LF of red galaxies)\footnote{This choice of magnitude
limits also corresponds closely to those of \citet{blan2003L}.}, 
observed and predicted numbers of galaxies, 
and Schechter and evolution parameters in each band.
Qualitatively, the $riz$ LFs appear to be well fitted by the parametric 
evolution model, although this model is formally excluded by the 
likelihood ratio test for almost every colour, band and redshift combination.
Even by eye, we see that the evolving Schechter function fits
are in extremely poor agreement with the $u$ and $g$ band non-parametric 
(SWML and $1/\Vmax$) estimates
in the highest redshift range, in the sense that the model overpredicts the
number density of luminous galaxies by almost an
order of magnitude in the $u$ band.
Equation~\ref{eqn:evol} thus provides a poor fit to evolution of
the $u$- and $g$-band LFs beyond $z \simeq 0.2$ and $z \simeq 0.3$,
respectively.
It is very possible that the $u$-band flux of more luminous, 
higher redshift galaxies is being dominated by AGN.
We intend to investigate the LFs of AGN-dominated/starforming/quiescent 
galaxies in a future paper.

In addition to the parametric fit, we also fit Schechter functions
to the SWML estimates for each redshift slice using least-squares.
Because the LF faint end is poorly sampled at redshift $z \ga 0.1$,
we only fit for all three Schechter parameters $\alpha$,  $M^*$ and $\phi^*$
in the lowest-redshift slice.  
At higher redshifts we hold $\alpha$ fixed and allow only $M^*$ and $\phi^*$ 
to vary.
The results of these fits are shown as dashed lines in Fig.~\ref{fig:lf_ev} and
the insets show 95 per cent likelihood contours of ($M^*$, $\lg \phi^*$).
These least-squares fits are for illustration only.
In order to calculate the parameters $Q_{\rm SWML}$ and $P_{\rm SWML}$ given
in Table~\ref{tab:evfit}, and shown below in Fig.~\ref{fig:QPlikeCont},
we sub-divide into eight redshift bins, 
perform least-squares fits to the SWML estimates in each, and then fit
straight lines to $\lg \phi^*$ and $M^*$ versus redshift.
We now discuss LF evolution separately, for the $u$, $g$ and $riz$ bands.

\begin{table*}
\caption{Evolving Schechter function fits to $ugriz$ LFs.
Columns are the same as in Table~\ref{tab:faintfit}, with the addition of.
evolution parameters $Q_{\rm par}$ and $P_{\rm par}$ determined
from the parametric model and
$Q_{\rm SWML}$ and $P_{\rm SWML}$ determined from least-squares fits
to SWML estimates in eight redshift slices as described in the text.
}
\label{tab:evfit}

    \begin{math}
    \begin{array}{cccrrrrrrrrr}
    \hline
     &^{0.1}M_1 & ^{0.1}M_2 & 
    \multicolumn{1}{c}{N_{\rm gal}} & 
    \multicolumn{1}{c}{N_{\rm pred}} & 
    \multicolumn{1}{c}{\alpha} &
    \multicolumn{1}{c}{^{0.1}M^* - 5 \lg h} &
    \multicolumn{1}{c}{Q_{\rm par}} &
    \multicolumn{1}{c}{P_{\rm par}} &
    \multicolumn{1}{c}{Q_{\rm SWML}} &
    \multicolumn{1}{c}{P_{\rm SWML}} &
    \multicolumn{1}{c}{\phi^* \times 100} \\
    & \multicolumn{2}{c}{- 5 \lg h} & & & & & & & & & / \denunit\\
    \hline
    \mbox{All}\\
        u &
            -23.0 & -15.0 &
            21120 & 13616 \pm  1941 & 
             -1.10 \pm   0.08 &
            -17.98 \pm   0.08 &
              6.2 \pm   0.5 &
             -8.5 \pm   1.2 &
              4.6 \pm   0.7 &
             -1.1 \pm   0.4 &
            3.10 \pm 0.53 \\
            g &
            -24.0 & -16.0 &
            37245 & 31909 \pm  2876 & 
             -1.10 \pm   0.02 &
            -19.58 \pm   0.03 &
              2.9 \pm   0.5 &
             -1.5 \pm   1.1 &
              0.2 \pm   0.5 &
              2.1 \pm   0.4 &
            1.80 \pm 0.22 \\
            r &
            -24.0 & -16.0 &
            90554 & 87163 \pm  5494 & 
             -1.23 \pm   0.01 &
            -20.70 \pm   0.04 &
              0.7 \pm   0.2 &
              1.8 \pm   0.5 &
              0.2 \pm   0.2 &
              1.6 \pm   0.2 &
            0.94 \pm 0.10 \\
            i &
            -25.0 & -17.0 &
            66069 & 57351 \pm  3290 & 
             -1.12 \pm   0.02 &
            -20.97 \pm   0.03 &
              1.5 \pm   0.1 &
              0.0 \pm   0.4 &
              0.6 \pm   0.2 &
              1.2 \pm   0.2 &
            1.16 \pm 0.15 \\
            z &
            -25.0 & -17.0 &
            51657 & 44771 \pm  2803 & 
             -1.07 \pm   0.02 &
            -21.22 \pm   0.04 &
              1.7 \pm   0.3 &
             -0.5 \pm   0.8 &
              0.8 \pm   0.2 &
              1.3 \pm   0.2 &
            1.26 \pm 0.18 \\
            \mbox{Blue}\\
        u &
            -23.0 & -15.0 &
            15205 & 10508 \pm  1214 & 
             -1.43 \pm   0.07 &
            -18.28 \pm   0.10 &
              5.5 \pm   0.6 &
             -7.1 \pm   1.5 &
              3.5 \pm   0.8 &
             -0.2 \pm   0.6 &
            1.31 \pm 0.25 \\
            g &
            -24.0 & -16.0 &
            21035 & 16733 \pm  2637 & 
             -1.40 \pm   0.03 &
            -19.60 \pm   0.06 &
              3.1 \pm   0.7 &
             -1.2 \pm   1.5 &
              0.4 \pm   0.8 &
              2.1 \pm   0.6 &
            0.73 \pm 0.03 \\
            r &
            -24.0 & -16.0 &
            43222 & 39901 \pm  1993 & 
             -1.49 \pm   0.03 &
            -20.45 \pm   0.06 &
              0.8 \pm   0.3 &
              2.9 \pm   0.6 &
              0.6 \pm   0.3 &
              1.4 \pm   0.2 &
            0.38 \pm 0.05 \\
            i &
            -25.0 & -17.0 &
            26845 & 22313 \pm  1608 & 
             -1.45 \pm   0.02 &
            -20.76 \pm   0.06 &
              1.7 \pm   0.4 &
              1.2 \pm   0.9 &
              0.8 \pm   0.4 &
              1.5 \pm   0.3 &
            0.42 \pm 0.06 \\
            z &
            -25.0 & -17.0 &
            18588 & 17993 \pm   855 & 
             -1.45 \pm   0.03 &
            -21.03 \pm   0.04 &
              0.9 \pm   0.2 &
              3.6 \pm   0.5 &
              0.8 \pm   0.4 &
              2.0 \pm   0.4 &
            0.34 \pm 0.04 \\
            \mbox{Red}\\
        u &
            -23.0 & -15.0 &
             5915 &  9488 \pm  2179 & 
             -0.14 \pm   0.13 &
            -17.32 \pm   0.07 &
              6.4 \pm   1.4 &
             -8.1 \pm   3.4 &
              5.5 \pm   0.9 &
             -1.2 \pm   0.7 &
            4.28 \pm 1.44 \\
            g &
            -24.0 & -16.0 &
            16210 & 11685 \pm  2696 & 
             -0.43 \pm   0.05 &
            -19.30 \pm   0.06 &
              3.6 \pm   1.4 &
             -3.9 \pm   2.8 &
              2.8 \pm   0.6 &
             -0.4 \pm   0.4 &
            1.26 \pm 0.17 \\
            r &
            -24.0 & -16.0 &
            47332 & 42882 \pm  2426 & 
             -0.57 \pm   0.02 &
            -20.34 \pm   0.03 &
              1.8 \pm   0.1 &
             -1.2 \pm   0.5 &
              1.7 \pm   0.2 &
             -0.1 \pm   0.2 &
            1.11 \pm 0.15 \\
            i &
            -25.0 & -17.0 &
            39224 & 33962 \pm  1840 & 
             -0.54 \pm   0.03 &
            -20.73 \pm   0.03 &
              2.0 \pm   0.1 &
             -1.8 \pm   0.5 &
              2.3 \pm   0.2 &
             -0.5 \pm   0.2 &
            1.16 \pm 0.15 \\
            z &
            -25.0 & -17.0 &
            33069 & 27543 \pm  1701 & 
             -0.49 \pm   0.05 &
            -20.97 \pm   0.06 &
              2.4 \pm   0.3 &
             -2.7 \pm   0.7 &
              2.7 \pm   0.2 &
             -1.0 \pm   0.2 &
            1.32 \pm 0.16 \\
            
    \hline
    \end{array}
    \end{math}
    
\end{table*}

\subsubsection{$u$-band evolution}

We observe a gradual brightening of $M^*$ for all samples between 
$z \simeq 0.05$ and $z \simeq 0.15$, and already 
a bright-end excess above a Schechter function is becoming apparent.
By $z \simeq 0.25$, the evolving model provides a very poor fit
to the non-parametric estimates: the former is much steeper than the latter,
and by $z \simeq 0.4$ the parametric fit over-predicts the
number density of galaxies by almost an order of magnitude.
From the least-squares fits to the non-parametric estimates,
we see a dramatic brightening of $M^*$ in the highest redshift
range (there are too few red galaxies to obtain a sensible LF fit in this bin).
This is due to the very shallow slope at the bright end of the LF, 
leading the Schechter fit to prefer brightening $M^*$ to increasing $\phi^*$.
In reality, we suspect that this shallow slope is caused by an increasing
fraction of highly-luminous AGNs,
rather than by such strong luminosity evolution in non-active galaxies.
An alternative explanation is that photometric errors in the $u$ band
are manifesting themselves as unrealistically strong luminosity evolution.
This possibility will explored when VST KIDS data become available in
the GAMA regions.

\subsubsection{$g$-band evolution}

The parametric model provides a good fit out to redshift $z \simeq 0.2$,
beyond which a bright-end excess results in an over-prediction of the
LF relative to the non-parametric estimates.
From these latter estimates, one sees that the number density of
blue galaxies is gradually increasing with redshift, whereas red galaxies
show the opposite trend.
Photometric errors are less likely to be a problem in the $g$
than in the $u$ band, and so again we suspect that AGNs are dominating the 
bright end of the LF at higher redshifts.
This interpretation does not necessarily imply rapid evolution of the 
AGN population --- the volume sampled at low redshifts is simply too small to
detect them in significant numbers.

\subsubsection{$riz$-band evolution}

Evolution in the $r$, $i$ and $z$ bands is qualitatively very similar,
and so we discuss them together.
The parametric model provides a reasonable fit in all redshift slices,
although it should be said that the formal fit probabilities from
the likelihood ratio test are mostly below 1 per cent.
This does not necessarily mean that the model is a poor fit,
as we see from simulations (Appendix~\ref{sec:test}) that the non-parametric
LF estimates are biased when the underlying LF is evolving.
The likelihood ratio test provides improved probabilities when we consider
narrower ranges in redshift.
From the non-parametric fits, we see that $M^*$ brightens with redshift for
all samples.
At low redshifts, red galaxies have a much higher space density than blue,
but as redshift increases, the density of red galaxies drops and that of blue
galaxies increases, until blue galaxies come to dominate by redshifts
$z \simeq 0.4$.

\subsubsection{Comparison with previous results}

\begin{figure*}
\includegraphics[width=\linewidth]{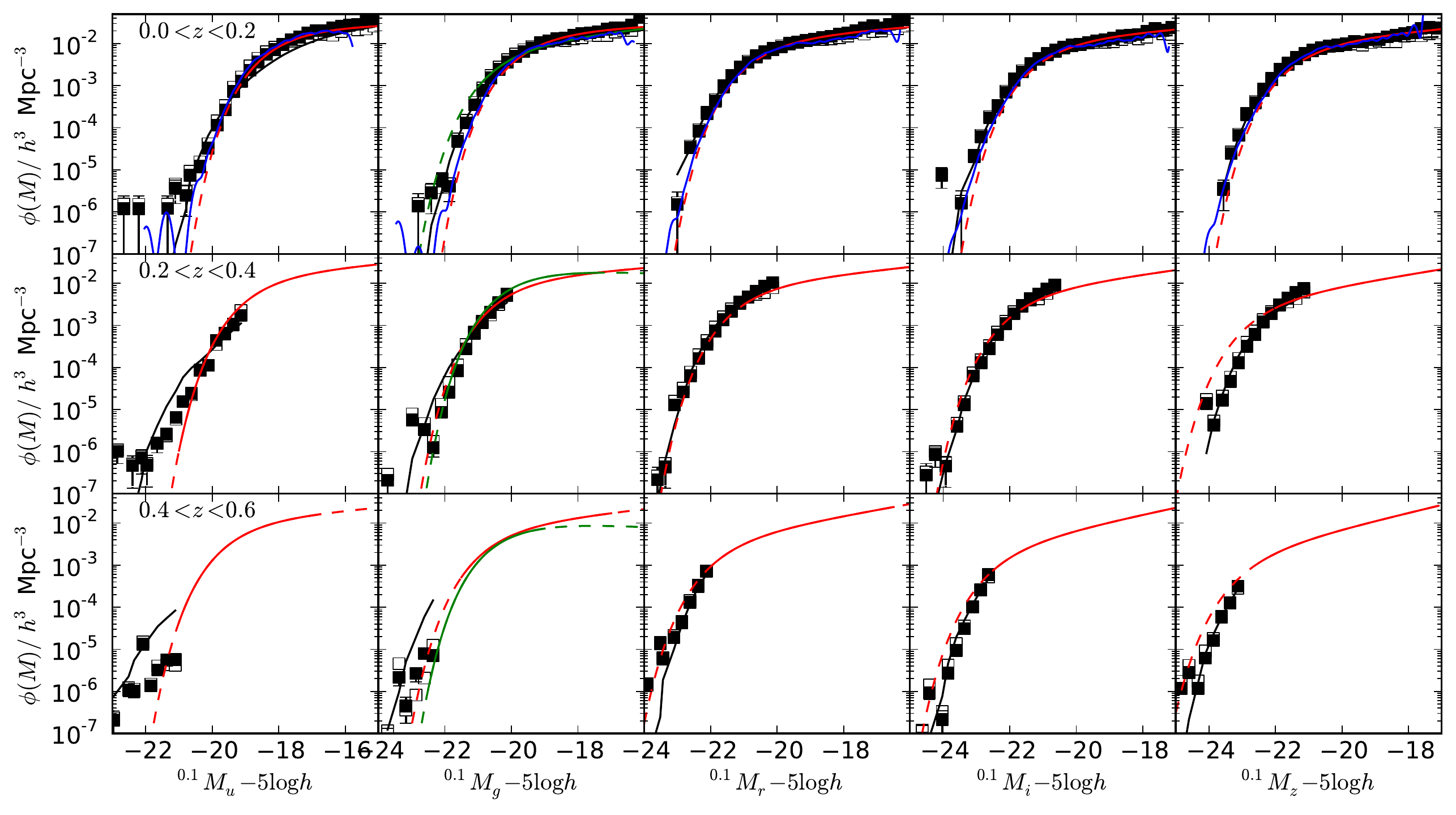}
\caption{Comparison of of our $ugriz$ LFs for the combined sample in three
redshift ranges as indicated with previous estimates.
Filled and open symbols and black lines show our SWML, $1/\Vmax$ 
and parametric fits respectively.
Blue lines (upper panels only) show the $z \simeq 0.1$ SDSS
LFs estimated by \protect\citet{blan2003L}.
Red lines show the VVDS LFs in bands $UBVRI$ respectively from
Ilbert et al. 2005.
Green lines show the zCOSMOS $B$-band LFs from Fig.~1 of Zucca et al. 2009.
The VVDS and zCOSMOS fits are shown as solid lines over the magnitude 
range actually fitted;
dashed lines show extrapolations outside the fitted magnitude range.
}
\label{fig:ev_comp}
\end{figure*}

We compare our evolving LFs with previous estimates in Fig.~\ref{fig:ev_comp}.
Here we plot the LFs in three redshift ranges (0.002, 0.2], (0.2, 0.4] and
(0.4, 0.6].
We choose these ranges to coincide with the first three redshift bins
used by \citet{2005A&A...439..863I} in their analysis of the VIMOS-VLT deep
survey (VVDS).
Following these authors, we assume approximate correspondence between
the restframe $BVRI$ and the $^{0.1}griz$ passbands, and assume that
$^{0.0}M^*_U =\ ^{0.1}M^*_u - 0.25$.
Their LFs are shown as red lines: solid over the magnitude range 
actually fitted, dashed where extrapolated.

In the low-redshift range (top row), we also show the non-parametric LF
estimates of SDSS galaxies from \citet{blan2003L} as blue lines.
Their estimates are in good agreement with our non-parametric SWML and 
$1/\Vmax$ estimates except that we see a slightly
higher density of luminous galaxies, particularly in the $g$ band
($^{0.1}M_g - 5 \lg h \la -21$ mag).
This difference is likely to be due to the greater depth of the
GAMA sample compared with SDSS: the mean redshift in this range is
$\bar{z} = 0.13$ for GAMA versus $\bar{z} = 0.10$ for SDSS.
The GAMA sample thus contains a higher fraction of more distant and hence 
more evolved galaxies.
The VVDS Schechter fits in the lowest redshift bin show a slightly
lower density of luminous galaxies than seen in GAMA and SDSS.
Note, however, that the bright end of the low-redshift LF is very
poorly constrained by Ilbert et al. due to their bright apparent
magnitude limit of $I_{AB} = 17.5$.
The zCOSMOS $B$ band LF from Fig.~1 of \citet{2009A&A...508.1217Z} 
(Zucca private communication, green line)
shows a bright-end excess relative to GAMA and to VVDS.
While reaching about 1.5 mag brighter than VVDS, the zCOSMOS low-redshift
LF still relies on extrapolation at magnitudes brighter than
$M_B - 5 \lg h = -21$ mag.

At intermediate redshifts (middle row), our LFs are in good agreement
with VVDS and zCOSMOS apart from an excess of $u$-bright galaxies in GAMA, 
and, conversely, a much higher bright-end fit by VVDS in the $I$ 
($^{0.1}z$) band.
This latter discrepancy is almost certainly due to poor coverage of the
bright end of the $I$ band LF at redshifts $0.2 < z < 0.4$ by VVDS.

In the highest redshift range (bottom row), the extrapolation of the
VVDS Schechter fit shows a higher abundance of
luminous galaxies in the redder $VRI$ bands than GAMA.
In this redshift range the comparison is not quite fair, 
since GAMA contains very few galaxies
beyond $z = 0.5$, and so much of the VVDS excess is likely to be due
to galaxies in the redshift range $0.5 < z < 0.6$.

Overall, our evolving LF estimates are in reasonable agreement with
the previous results of \citet{blan2003L}, \citet{2005A&A...439..863I}
and \citet{2009A&A...508.1217Z}.

\subsubsection{Schechter parameter likelihood contours}

\begin{figure}
\includegraphics[width=\linewidth]{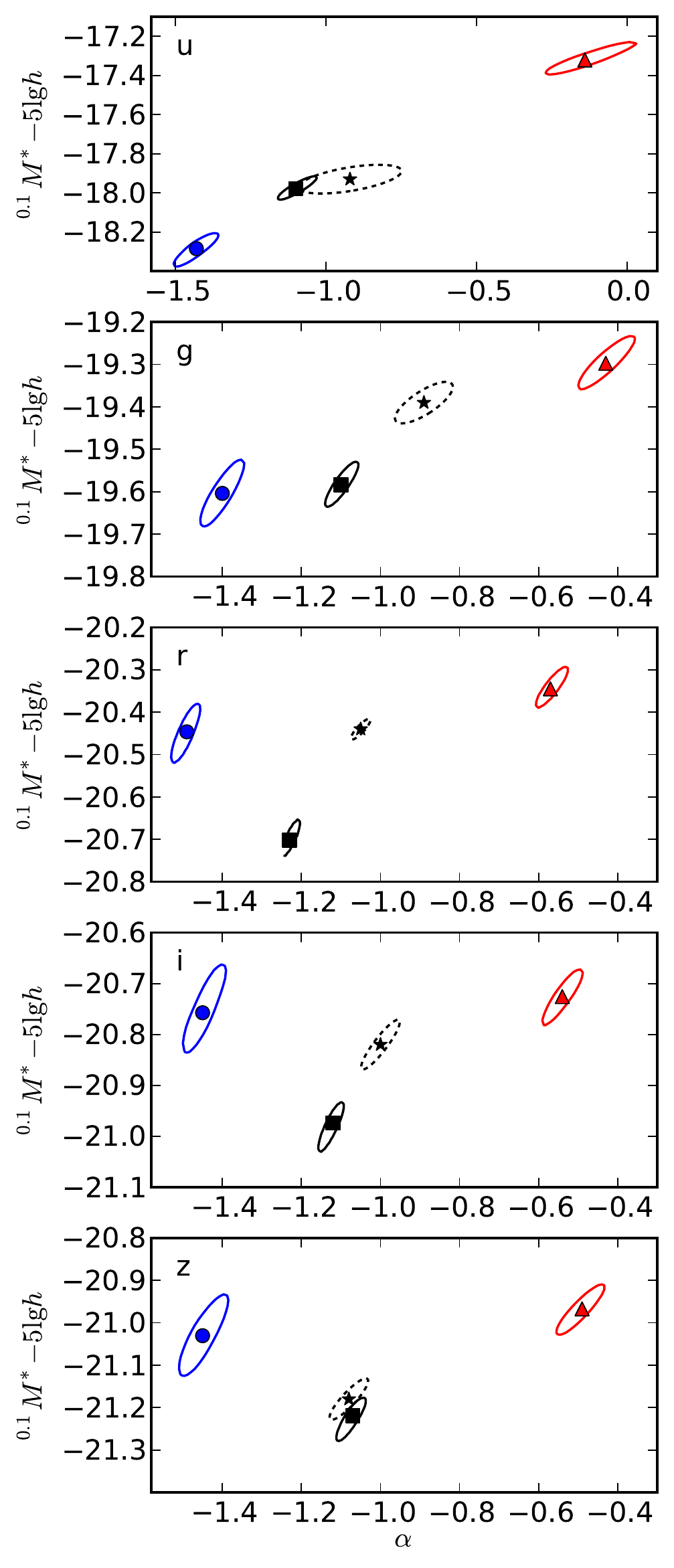}
\caption{2-$\sigma$ likelihood contours for evolving Schechter function 
parameters $\alpha$ and $M^*$ in $ugriz$ bands for 
combined, blue and red samples (black, blue and red contours respectively).
Asterisks and dotted ellipses show the best-fit values and 2-$\sigma$ error 
ellipses on the parameters reported by
\protect\cite{blan2003L} (combined colours only).
}
\label{fig:AMlikeCont}
\end{figure}

\begin{figure}
\includegraphics[width=\linewidth]{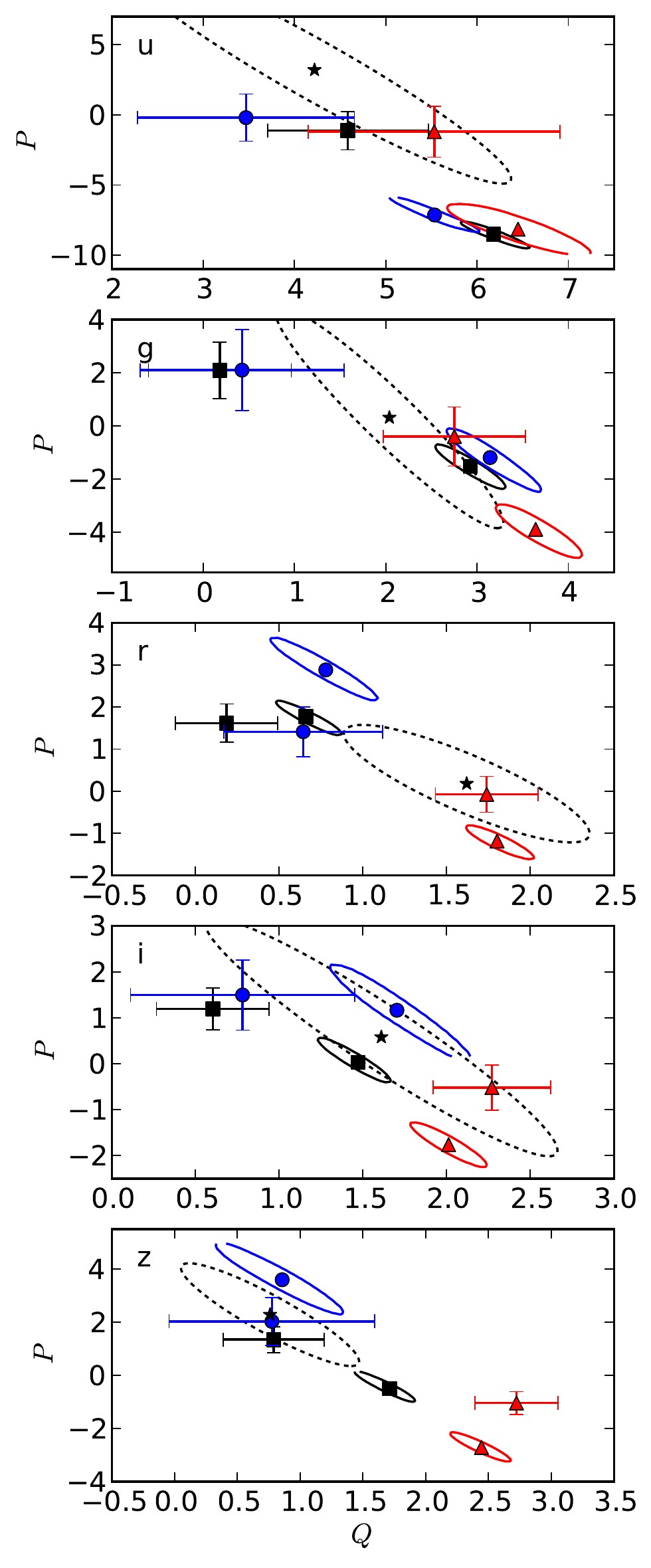}
\caption{2-$\sigma$ likelihood contours for evolving Schechter function 
parameters $Q$ and $P$ in $ugriz$ bands for 
combined, blue and red samples (black, blue and red contours respectively).
Error bars with symbols show evolution parameters and their 2-$\sigma$ 
errors determined from a least-squares fit to SWML estimates of the LF
in eight redshift ranges covering ($0.002 < z < 0.5$).
Asterisks and dotted ellipses show the best-fit values and 2-$\sigma$ error 
ellipses on the parameters reported by
\protect\cite{blan2003L} (combined colours only).
}\label{fig:QPlikeCont}
\end{figure}

$2\sigma$ likelihood contours for the Schechter parameters ($\alpha$, $M^*$),
determined using the parametric fits,
are shown in Fig.~\ref{fig:AMlikeCont}.
For the combined sample, we find roughly consistent faint-end slopes 
$\alpha = -1.1 \pm 0.1$ in all bands.
The slightly steeper slope seen in the $r$ band is most likely due to the
fact that GAMA is selected in the $r$ band, and hence we probe slightly
further down the LF in this than the surrounding bands.
Similarly, the slightly steeper slopes in our fits to the low-redshift sample
(Table~\ref{tab:faintfit}) are due to the inclusion of fainter-magnitude
galaxies in those fits.
As expected, $M^*$ increases systematically in brightness 
with increasing wavelength from $u$ to $z$.
Red galaxies have systematically shallower faint-end
slopes ($\alpha \simeq -0.5$ in $griz$) than blue galaxies 
($\alpha = -1.45 \pm 0.05$) in all bands.
The characteristic magnitudes $M^*$ are fainter for red galaxies
than for blue in the $ugr$ bands and are comparable in the $iz$ bands.

Fig.~\ref{fig:AMlikeCont} also shows Schechter function parameters
estimated from the SDSS main galaxy sample by \cite{blan2003L}.
We find systematically steeper faint-end slopes (apart from in the $z$ band)
and brighter characteristic magnitudes (apart from the $u$ band).
Since our non-parametric estimates are in good agreement 
(Fig.~\ref{fig:ev_comp}),
these differences most likely arise due to strong degeneracies 
between the parameters
$\alpha$, $M^*$ and $Q$:
($\alpha$, $M^*$), ($\alpha$, $Q$) and ($M^*$, $Q$) are all 
positively correlated.

\subsubsection{Evolution parameter likelihood contours}

In Fig.~\ref{fig:QPlikeCont} we show $2\sigma$ likelihood contours 
for the luminosity evolution parameters $Q$ and $P$ from our parametric fits,
along with estimates of these quantities and their errors from least-squares 
fits of Schechter functions to the SWML estimates made in eight
redshift ranges.
The differences between the estimates of these parameters are frequently
larger than the formal errors associated with each method.
This indicates that our assumption of linear evolution of $M^*$ and
$\lg \phi^*$ with redshift is only approximate.

For the combined sample, luminosity evolution is least in the $r$ band 
($Q_{\rm par} \simeq 0.7$), 
increasing to $Q_{\rm par} \simeq 1.6$ in the $i$ and $z$ bands.
Luminosity evolution is even more pronounced in the $g$ and $u$ bands
($Q_{\rm par} \simeq 2.9$ and 6.2, respectively), although,
as previously noted, the parametric model performs very poorly in these 
bands, and so these values are unreliable at best.
Luminosity evolution is more pronounced for the red galaxy population
than the blue.

Blue galaxies exhibit positive density evolution,
($P_{\rm par} > 0$ in all bands apart from $u$ and $g$,
$P_{\rm SWML} > 0$ in all bands apart from $u$),
whereas red galaxies show negative density evolution,
both $P_{\rm par}$ and $P_{\rm SWML}$ are negative in all bands.
This observation is in good qualitative agreement with an analysis 
of the zCOSMOS survey by \citet{2009A&A...508.1217Z}, who find that both
early- and late-spectroscopic-type galaxies brighten in $M^*$ by $\simeq 0.5$
mag over the redshift range $z \simeq 0.2$ to $z \simeq 0.9$, but that 
$\phi^*$ for early types decreases by a factor $\simeq 1.7$ over the same
redshift range; for late types $\phi^*$ increases by a factor $\simeq 1.8$.

Density evolution for the combined sample is positive
in the $r$ band; in other bands $P$ is either negative or consistent with zero,
compensating for the stronger luminosity evolution in these bands.
Thus the contrary density evolution of blue and red galaxies largely cancels 
out in the combined sample.
For all bands and samples, the evolution parameters ($Q$, $P$) 
are strongly anticorrelated.
We remind the reader that the maximum-likelihood luminosity evolution 
$Q_{\rm par}$ is determined along with
$\alpha$ and $M^*$, independently of the normalisation of the LF.
Density evolution $P_{\rm par}$ does depend on the fitted value of 
$Q_{\rm par}$, as well as the Schechter parameters,
resulting in the observed anti-correlation between $Q$ and $P$.
In the redder bands, $riz$, the combined LD evolution $P+Q$ is 
stronger for blue galaxies, $(P+Q)_{\rm par} \simeq 3.7 \pm 0.8$, than for red,
$(P+Q)_{\rm par} \simeq 0.2 \pm 0.5$.

Fig.~\ref{fig:QPlikeCont} also shows evolution parameters determined 
from the SDSS main galaxy sample by \cite{blan2003L}.
The $2\sigma$ likelihood contours intersect in $gri$, and narrowly miss
in $u$ and $z$.
In the $r$ band we find weaker luminosity evolution and
a compensating stronger density evolution, vice versa in $z$.
Our least-squares fits to the SWML estimates in the $u$ band yield comparable
$Q$ estimates to \cite{blan2003L}.
Their density evolution, unlike ours, is positive, but has a very large error.

Although sampling a smaller volume, the GAMA data analysed here
have a mean redshift $\bar z \simeq 0.2$ compared with $\bar z \simeq 0.1$ 
for the data analysed by Blanton et al.
We thus have a longer redshift baseline over which to measure evolution.

\citet{2004ApJ...615..209H}, in an analysis combining constraints 
from the star formation rate density of the Universe and 1.4-GHz
radio source counts, found $Q = 2.70 \pm 0.60$, $P=0.15 \pm 0.60$
for the star forming galaxy population\footnote{Note that Hopkins actually
models evolution as $L \propto (1+z)^Q$ and $\phi \propto (1+z)^P$.}.
This measurement, sensitive to the star-forming population up to $z \simeq 1$,
is consistent with our parametric fit results for blue galaxies 
in the $g$ band at the low-redshift end of this range.
However, given the very large discrepancy between $Q_{\rm par}$ and 
$Q_{\rm SWML}$ for blue galaxies in the $g$ band, the apparent agreement may
be fortuitous.

For red galaxies, \citet{2007ApJ...654..858B} find that $M^*$ in the
$B$ band brightens by $\simeq -0.7$ mag from redshift $z = 0.2$ to $z = 1$ 
while $\phi^*$ declines by about 25 per cent, in qualitative agreement 
with our results

\subsubsection{Evolution summary}

To summarize our findings regarding evolution of the LF:
\begin{enumerate}
\item The evolutionary model (equation~\ref{eqn:evol}) provides a reasonable
fit in the redder bands, $riz$, but performs poorly in the $u$ and $g$ bands, 
overpredicting the LF of luminous galaxies at high redshift.
This is possibly due to a significant contribution from AGNs.
\item Our non-parametric LF estimates are in good agreement with SDSS
measurements at low redshift and with results from the
VVDS and zCOSMOS surveys at higher redshifts, 
over magnitude ranges where our LF estimates overlap.
\item There is a strong degeneracy between the luminosity and density evolution
parameters $Q$ and $P$.
One should be wary in using them in isolation, e.g. using the $Q$ parameter 
to apply evolutionary corrections.
\item Nevertheless, red galaxies in all bands show evidence for 
positive luminosity evolution ($Q > 0$) and negative density evolution 
($P < 0$).
\item Blue galaxies show less luminosity evolution but show evidence
for positive density evolution.
\item The observation of decreasing number density of blue galaxies 
but increasing number density of red galaxies with cosmic time implies that the
transition from blue cloud to red sequence is an important and ongoing
phenomenon since redshifts $z \simeq 0.5$.
\item The combined luminosity plus density evolution is stronger for blue 
than for red galaxies.
\end{enumerate}

\subsection{Luminosity density evolution} \label{sec:lumdens}

\begin{figure}
\includegraphics[width=0.9\linewidth]{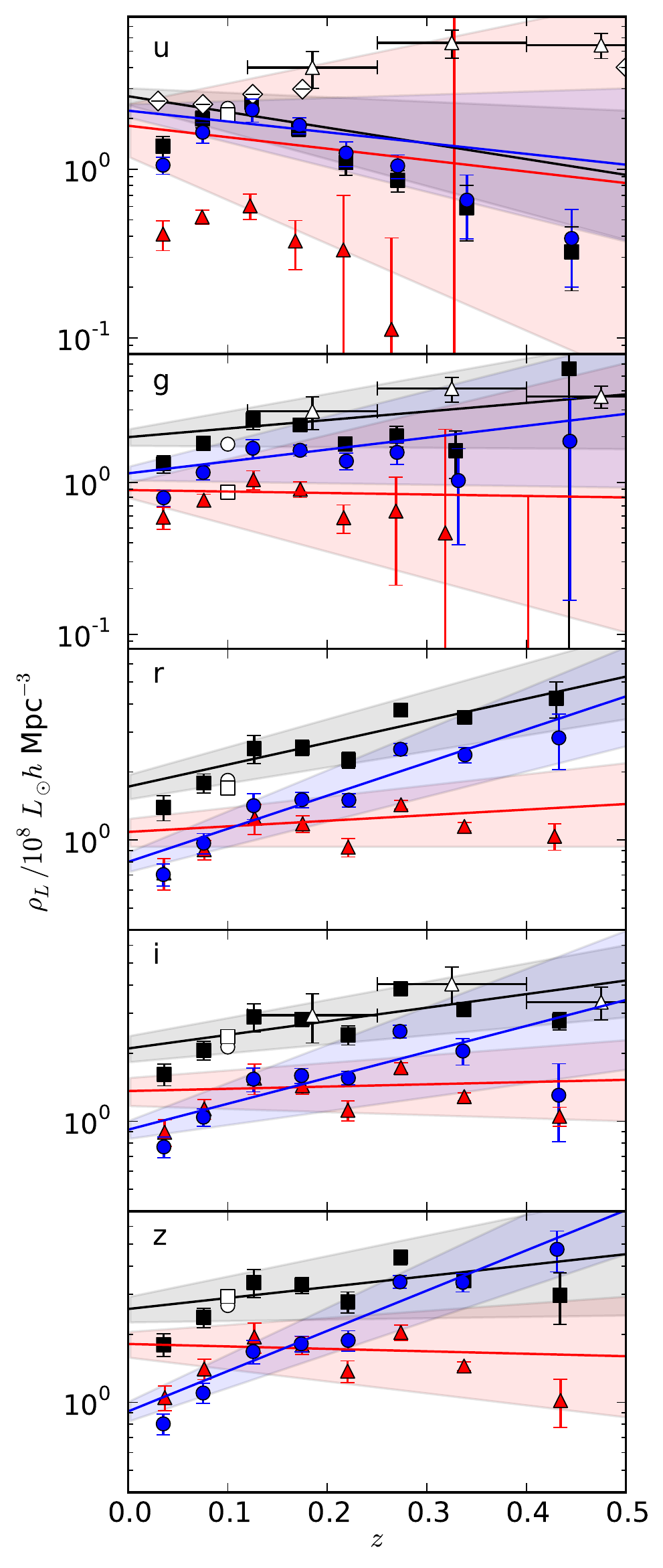}
\caption{Luminosity density in the five bands $ugriz$ as labelled as a function
of redshift.
Points with error bars show the LD estimated by summing
galaxies in each luminosity range (equation~\ref{eqn:ldsum}).
Lines show the predictions from the parametric fits 
(equation~\ref{eqn:ldfit})
and shaded regions show confidence limits obtained by combining the
lower and upper 1-$\sigma$ limits on ${\rho_L}(0)$, $Q$ and $P$.
Black squares, blue circles and red triangles show results for the combined, 
blue and red samples respectively.
Open circles and squares show the LD estimated at
redshift $z \simeq 0.1$
from SDSS data by \citet{blan2003L} and \citet{2009MNRAS.399.1106M}
respectively.
Open triangles show LD estimates from CNOC2 (Table~3
of Lin et al. 1999) in $U$, $B$ and $R$ bands corresponding roughly
to $^{0.1}u - 0.25$, $^{0.1}g$ and $^{0.1}i$ respectively.
Diamonds show $u$-band LD estimates from 
\citet{2009MNRAS.397...90P}.
}
\label{fig:lumdens}
\end{figure}

\begin{table*}
\caption{Luminosity density evolution.
The column labelled `Fit' gives the redshift-zero LD from the
parametric fit (equation~\ref{eqn:ld0}), subsequent columns show
the LD obtained from summing over galaxies 
(equation~\ref{eqn:ldsum}) in the indicated redshift ranges.
Units are $10^8 \ldenunit$.}
\label{tab:lumdens}

    \begin{math}
    \begin{array}{cccccccccc}
    \hline
    \mbox{Redshift} & \mbox{Fit} & \mbox{0.0 -- 0.05} & \mbox{0.05 -- 0.1} &
    \mbox{0.1 -- 0.15} & \mbox{0.15 -- 0.2} &
    \mbox{0.2 -- 0.25} & \mbox{0.25 -- 0.3} &
    \mbox{0.3 -- 0.4} & \mbox{0.4 -- 0.5}\\
    \hline
    \mbox{All}\\
        u & 2.70 \pm 0.31 & 1.37 \pm 0.19 & 1.98 \pm 0.22 & 2.49 \pm 0.38 & 1.73 \pm 0.17 & 1.09 \pm 0.17 & 0.86 \pm 0.12 & 0.59 \pm 0.21 & 0.32 \pm 0.13\\g & 1.99 \pm 0.25 & 1.33 \pm 0.17 & 1.81 \pm 0.17 & 2.58 \pm 0.36 & 2.39 \pm 0.21 & 1.79 \pm 0.16 & 2.02 \pm 0.32 & 1.62 \pm 0.56 & 5.61 \pm 8.27\\r & 1.72 \pm 0.21 & 1.40 \pm 0.18 & 1.79 \pm 0.17 & 2.54 \pm 0.36 & 2.56 \pm 0.20 & 2.26 \pm 0.19 & 3.75 \pm 0.17 & 3.49 \pm 0.14 & 4.23 \pm 0.76\\i & 2.10 \pm 0.28 & 1.61 \pm 0.18 & 2.06 \pm 0.20 & 2.89 \pm 0.41 & 2.82 \pm 0.20 & 2.40 \pm 0.23 & 3.86 \pm 0.18 & 3.12 \pm 0.15 & 2.78 \pm 0.24\\z & 2.59 \pm 0.33 & 1.80 \pm 0.21 & 2.37 \pm 0.24 & 3.39 \pm 0.48 & 3.30 \pm 0.27 & 2.78 \pm 0.30 & 4.38 \pm 0.33 & 3.45 \pm 0.18 & 2.98 \pm 0.76\\\mbox{Blue}\\
        u & 2.22 \pm 0.20 & 1.05 \pm 0.12 & 1.65 \pm 0.23 & 2.25 \pm 0.35 & 1.82 \pm 0.20 & 1.24 \pm 0.21 & 1.04 \pm 0.17 & 0.65 \pm 0.27 & 0.39 \pm 0.19\\g & 1.15 \pm 0.12 & 0.79 \pm 0.10 & 1.16 \pm 0.12 & 1.68 \pm 0.24 & 1.63 \pm 0.14 & 1.38 \pm 0.17 & 1.58 \pm 0.26 & 1.03 \pm 0.64 & 1.87 \pm 1.70\\r & 0.80 \pm 0.08 & 0.70 \pm 0.08 & 0.97 \pm 0.10 & 1.42 \pm 0.19 & 1.51 \pm 0.12 & 1.50 \pm 0.10 & 2.52 \pm 0.15 & 2.38 \pm 0.19 & 2.83 \pm 0.78\\i & 0.92 \pm 0.08 & 0.77 \pm 0.08 & 1.04 \pm 0.09 & 1.53 \pm 0.19 & 1.59 \pm 0.12 & 1.55 \pm 0.11 & 2.50 \pm 0.16 & 2.04 \pm 0.27 & 1.30 \pm 0.49\\z & 0.91 \pm 0.09 & 0.80 \pm 0.09 & 1.10 \pm 0.11 & 1.68 \pm 0.20 & 1.82 \pm 0.15 & 1.88 \pm 0.19 & 3.41 \pm 0.22 & 3.39 \pm 0.32 & 4.75 \pm 0.97\\\mbox{Red}\\
        u & 1.81 \pm 0.62 & 0.41 \pm 0.08 & 0.52 \pm 0.05 & 0.61 \pm 0.10 & 0.37 \pm 0.12 & 0.33 \pm 0.37 & 0.11 \pm 0.28 & \mbox{---} & \mbox{---}\\g & 0.89 \pm 0.10 & 0.59 \pm 0.10 & 0.76 \pm 0.07 & 1.04 \pm 0.15 & 0.90 \pm 0.11 & 0.59 \pm 0.13 & 0.65 \pm 0.44 & 0.46 \pm 1.77 & 0.05 \pm 0.77\\r & 1.09 \pm 0.15 & 0.72 \pm 0.11 & 0.91 \pm 0.09 & 1.25 \pm 0.19 & 1.18 \pm 0.10 & 0.93 \pm 0.09 & 1.43 \pm 0.06 & 1.15 \pm 0.05 & 1.04 \pm 0.14\\i & 1.36 \pm 0.19 & 0.89 \pm 0.12 & 1.13 \pm 0.11 & 1.55 \pm 0.24 & 1.44 \pm 0.12 & 1.12 \pm 0.11 & 1.73 \pm 0.09 & 1.28 \pm 0.05 & 1.05 \pm 0.10\\z & 1.81 \pm 0.23 & 1.05 \pm 0.13 & 1.41 \pm 0.14 & 1.95 \pm 0.30 & 1.79 \pm 0.16 & 1.38 \pm 0.15 & 2.04 \pm 0.16 & 1.45 \pm 0.06 & 1.02 \pm 0.25\\
    \hline
    \end{array}
    \end{math}
    
\end{table*}

As we have seen in the previous section, while it can be difficult to
isolate the effects of luminosity and density evolution,
evolution in LD is better constrained.
Fig.~\ref{fig:lumdens} shows the LD ${\rho_L}_{\rm sum}$ 
measured in eight
redshift bins up to $z=0.5$, according to equation~\ref{eqn:ldsum},
along with the prediction ${\rho_L}_{\rm fit}$ of the parametric model 
(equation~\ref{eqn:ldfit}).
These results are tabulated in Table~\ref{tab:lumdens}.
For the combined sample, in all bands other than $u$, we see LD increasing with
redshift, steeply between redshifts $z = 0$ and $z \simeq 0.15$, slightly more
gradually thereafter.
The blue galaxy LD
increases more steeply with redshift than the combined sample.
The LD of red galaxies barely evolves with redshift beyond $z \simeq 0.15$, 
thus the relative contribution to LD from blue galaxies comes to dominate
by redshifts $z \simeq 0.2$.
Given our choice of colour cut (equation~\ref{eqn:colourcut}),
red and blue galaxies contribute roughly equally to the LD 
in the $r$ and $i$ bands at low redshifts ($z \la 0.15$),
red galaxies are slightly dominant in the $z$ band but under-represented in
$u$ and $g$ bands.

Fig.~\ref{fig:lumdens} also shows the LD estimated
from the SDSS by \citet{blan2003L} and \citet{2009MNRAS.399.1106M}
at a mean redshift of $z \simeq 0.1$.
Our results are in excellent agreement with those of \citet{blan2003L}.
\citet{2009MNRAS.399.1106M} appear to have significantly underestimated the
$g$-band LD, although their estimates in other bands
are in agreement with ours and those of \citet{blan2003L}.
Open triangles show LD estimates from CNOC2.
We have taken the $q_0 = 0.1$ `Total' values from table~3
of Lin et al. 1999 in $U$, $B$ and $R$ bands, corresponding roughly
to $^{0.1}u - 0.25$, $^{0.1}g$ and $^{0.1}i$ respectively.
We convert the CNOC2 luminosity densities from physical units of 
$h$ W Hz$^{-1}$ Mpc$^{-3}$ to AB magnitudes using
$M_{AB} = 34.1 - 2.5 \lg \rho_L$ and then convert into Solar luminosities
using the assumed absolute magnitudes of the Sun quoted in 
Section~\ref{sec:ld}.
The $B$- and $R$-band LD estimates are in reasonable agreement with our
$g$-and $i$-band estimates, while the CNOC2 $U$-band LD is many times
larger than our $u$-band LD at redshifts $z \ga 0.15$.
Open diamonds show $u$-band LD estimates from 
\citet{2009MNRAS.397...90P}.

Note that our $u$-and $g$-band LD estimates will be adversely affected
by the poor fit of the parametric model for LF evolution.
The selection function in the denominator of equation~(\ref{eqn:ldsum}) will be
overestimated at high redshifts in these bands, and hence the summed LD itself
will be underestimated, leading to the decline in LD with redshift seen for
the $u$ band in Fig.~\ref{fig:lumdens}.

Previous studies of LD evolution (e.g.
\citealt{1996ApJ...460L...1L,lin99,2004ApJ...608..752B,baldry2005,2006ApJ...647..853W,2007ApJ...665..265F,2009MNRAS.397...90P}) 
have found that for blue and non-colour-selected galaxies, 
${\rho_L}$ increases monotonically
with redshift, while for red galaxies, it is approximately constant, with
a possible decline beyond redshift $z \simeq 1$.
We have presented the most detailed investigation to date of LD
evolution since redshift $z \simeq 0.5$, finding consistent results
in the $gri$ and $z$ bands with previous analyses which have focused
primarily on evolution beyond redshifts $z \simeq 0.5$.

\section{Conclusions} \label{sec:concs}

We have presented the first measurements of the $ugriz$ galaxy 
LFs from the GAMA survey, 
after correcting for imaging, target and spectroscopic incompleteness.
At low redshift ($z < 0.1$), 
the shapes of the blue galaxy LFs are reasonably matched, 
albeit not in detail, by standard Schechter functions.
LFs for red galaxies show a noticeable dip at intermediate magnitudes,
requiring double power-law Schechter functions to obtain an adequate fit.
One should be cautious in interpreting this as the upturn predicted by
halo occupation distribution models
\citep[e.g.][]{2008ApJ...682..937B} and the 
\citet{2010ApJ...721..193P} quenching model, since the faint end of
our red galaxy LF contains a significant fraction of 
edge-on disc systems, which are likely to be dust-reddened.
We find consistent faint-end slopes in all bands,
$\alpha + \beta = -1.35 \pm 0.05$ for the combined sample.

In order to determine evolution of the LF, we employ the parametric model of 
\citet{lin99} in which characteristic magnitude $M^*$ and log density
$\lg \phi^*$ are allowed to vary linearly with redshift.
We test the parametric model by comparing with estimates using the
$1/\Vmax$ and SWML estimates.
We find that the $r$, $i$ and $z$ bands are qualitatively well fitted 
by this model, although the model provides poor likelihood fits
compared with SWML.
The model predicts an excessively high number density in the $u$ and $g$
bands at high redshift, most likely due to 
QSO/Seyfert contamination \citep{2009MNRAS.399.1106M}.
With this caveat in mind, we find positive (i.e. increasing with redshift)
luminosity evolution in all bands and for all colour samples.
Luminosity evolution is stronger for red than for blue galaxies,
with blue galaxies brightening by $\simeq 1$--1.5 mag per unit redshift,
red galaxies brightening by $\simeq 2$--2.5 mag per unit redshift.

Number density evolution for blue galaxies is positive in the redder 
$riz$ bands in which it can be reliably measured, 
while red galaxies exhibit negative density evolution.
This observation of decreasing number density of blue galaxies 
but increasing number density of red galaxies with cosmic time implies that the
transition from blue cloud to red sequence is an important and ongoing
phenomenon since redshifts $z \simeq 0.5$.
Investigation of the mechanism that causes this transition will be 
the subject of future work,
but it appears unlikely that mergers play a dominant role at these moderate
redshifts, given the low merger fraction ($\sim 5$ per cent or less) observed at
low redshift by e.g. \citet{2009MNRAS.394.1956C} and \citet{arXiv:1108.2508v1}.

Luminosity density increases from redshift zero until $z \simeq 0.15$, 
beyond which redshift it increases more gradually for the combined sample.
The LD of red galaxies is roughly constant beyond $z \simeq 0.15$, whereas
that for blue galaxies keeps on increasing, leading to blue galaxies dominating
the LD at higher redshifts.

In this paper, we have not considered the effects of internal dust extinction
on the LF, nor have we considered the effects of using total as opposed to
Petrosian magnitudes \citep{2005AJ....130.1535G,2011MNRAS.412..765H}.
These extensions to the analysis will be considered in a future paper,
along with a measurement of the galaxy LF for AGN-dominated, star-forming
and quiescent galaxies which, it is hoped, will resolve the problems 
encountered while attempting to fit an evolutionary model in the $u$ and $g$ bands.

\section*{Acknowledgements}

JL acknowledges support from the Science and Technology Facilities Council
(grant numbers ST/F002858/1 and  ST/I000976/1).
PN acknowledges financial support from a Royal Society URF and an ERC StG grant
(DEGAS-259586).
We thank Elena Zucca for providing her Schechter function fits to
the zCOSMOS $B$ band LF \citep[][Fig.~1]{2009A&A...508.1217Z}
and Alister Graham, Simon Lilly and the referee for useful comments.

GAMA is a joint European-Australasian project based around a
spectroscopic campaign using the Anglo-Australian Telescope. The GAMA
input catalogue is based on data taken from the Sloan Digital Sky
Survey and the UKIRT Infrared Deep Sky Survey. Complementary imaging
of the GAMA regions is being obtained by a number of independent
survey programs including GALEX MIS, VST KIDS, VISTA VIKING, WISE,
Herschel-ATLAS, GMRT and ASKAP providing UV to radio coverage. GAMA is
funded by the STFC (UK), the ARC (Australia), the AAO, and the
participating institutions. The GAMA website is:
\url{http://www.gama-survey.org/}.

Funding for the SDSS and SDSS-II has been provided by the Alfred
P. Sloan Foundation, the Participating Institutions, the National
Science Foundation, the U.S. Department of Energy, the National
Aeronautics and Space Administration, the Japanese Monbukagakusho, the
Max Planck Society, and the Higher Education Funding Council for
England. The SDSS Web Site is \url{http://www.sdss.org/}.
The SDSS is managed by the Astrophysical Research Consortium for the
Participating Institutions. The Participating Institutions are the
American Museum of Natural History, Astrophysical Institute Potsdam,
University of Basel, University of Cambridge, Case Western Reserve
University, University of Chicago, Drexel University, Fermilab, the
Institute for Advanced Study, the Japan Participation Group, Johns
Hopkins University, the Joint Institute for Nuclear Astrophysics, the
Kavli Institute for Particle Astrophysics and Cosmology, the Korean
Scientist Group, the Chinese Academy of Sciences (LAMOST), Los Alamos
National Laboratory, the Max-Planck-Institute for Astronomy (MPIA),
the Max-Planck-Institute for Astrophysics (MPA), New Mexico State
University, Ohio State University, University of Pittsburgh,
University of Portsmouth, Princeton University, the United States
Naval Observatory, and the University of Washington.

{}

\appendix

\section{Testing the methods} \label{sec:test}

We here test our methods using simulated mock catalogues.
We generate clustered distributions of points with known evolving
LF and then apply the GAMA selection effects to create 
a set of mock catalogues.

\subsection{Clustered simulations}

We use a \citet{sp78} type simulation to generate a clustered distribution 
of points in a cube 1200 \hMpc\ on a side with periodic boundary conditions.
The parameters in the simulation are chosen to yield
similar clustering properties to those
measured by \cite{zehavi2005} for a flux-limited sample of SDSS galaxies, 
namely a correlation function with power-law slope $\gamma \simeq 1.7$ and
correlation length $r_0 \simeq 5.2 \hMpc$.
This allows us to investigate the effects of large-scale structure
on our LF estimates.

With the observer located at one corner of the cube,
each galaxy in the simulation was assigned an absolute magnitude $M$ within
the range $-24 \le M_r \le -10$ mag, 
drawn at random from a LF with parameters specified
in Table~\ref{tab:sims}.
Since these are static simulations, we assume a linear redshift--distance
relation, with $r = cz/H_0$ and volume element $dV \propto r^2$.
Apparent magnitudes $m$ are calculated using $m = M + 25 + 5\lg r(1 + z)$
with no $K$-correction.
Strictly, of course, there is no factor $(1+z)$ in luminosity distance
in Euclidean space.
We choose, however, to include this factor in the simulations to make them
more realistic --- without it one predicts far too many galaxies at higher
redshifts.
The number of observable galaxies $N_{\rm obs}$ to redshift $z_{\rm max} = 0.4$
in each simulation cube was determined by
integrating the model LF $\phi(L,z)$ over luminosity and redshift:
\begin{equation}
N_{\rm obs} = \int_0^{z_{\rm max}} \int_{L_{\rm min}(z, m_{\rm lim})}^\infty
\phi(L,z) dL \frac{dV}{dz} dz,
\end{equation}
with $m_{\rm lim} = 19.8$ mag.

We crudely mimic the GAMA survey geometry by selecting 
galaxies within each of three $12 \times 4$ deg$^2$ regions
(bounded by latitude $0^\circ < \phi < 4^\circ$ and longitudes 
$0^\circ < \theta < 12^\circ$, 
$36^\circ < \theta < 48^\circ$ and $78^\circ < \theta < 90^\circ$).

Imaging completeness is determined 
by linear interpolation of the curve in Fig.\ref{fig:imcomp}.
$r$-band half-light surface brightness, $\mu_{50,r}$, for each simulated galaxy
is assigned according to the empirically observed relation
between $\mu_{50,r}$ and $r$-band absolute magnitude $^{0.1}M_r$ 
for GAMA galaxies,
$\mu_{50,r} = 22.42 + 0.029\, ^{0.1}M_r$ with 1-sigma scatter of 0.76 mag.
Target completeness is obtained using the empirically observed completeness
shown in Fig.~\ref{fig:targcomp}.
Note that we do not attempt to follow the dependence of target completeness
on sky coordinates, and so therefore any dependence of target completeness
on target density in the real data will not be simulated.
Given the better than 98 per cent completeness of GAMA spectroscopy, this should not
be a significant issue.
Finally, spectroscopic completeness is obtained by generating a fibre magnitude
for each simulated galaxy according to the empirically observed relation
between Petrosian and fibre magnitudes for GAMA galaxies,
$r_{\rm fib} = 5.84 + 0.747 r_{\rm Petro}$ with 1-sigma scatter of 0.31 mag.
Redshift completeness is then obtained using the sigmoid function fit
to redshift success shown in Fig.~\ref{fig:speccomp}.

Considering each form of completeness in turn, galaxies are selected at 
random with a probability equal to the completeness.
Weights for the simulated galaxies that survive this culling process
are assigned in the same way as for 
observed galaxies (equation~\ref{eqn:weight}).

\subsection{Simulation results}

\begin{table*}
 \caption{LF parameters estimated from both non-evolving
and evolving simulated data.
The recovered values show the mean and standard deviation from
eight mock catalogues.
}
 \label{tab:sims}
 \begin{math}
 \begin{array}{rccccc}
 \hline
 & \alpha & M^* - 5 \lg h & \phi^*/\denunit & Q & P\\
 \hline
\mbox{True} & -1.20  & -20.80 & 0.0100 & 0.00 & 0.00\\
\mbox{Recovered} & -1.19 \pm 0.01 & -20.78 \pm 0.02 & 0.0101 \pm 0.0007 & 0.04 \pm 0.06 & -0.03 \pm 0.30\\
  \hline
\mbox{True} & -1.20  & -20.80 & 0.0100 & 2.00 & 0.00\\
\mbox{Recovered} & -1.20 \pm 0.01 & -20.79 \pm 0.02 & 0.0101 \pm 0.0004 & 2.05 \pm 0.05 & -0.07 \pm 0.13\\
  \hline
 \end{array}
 \end{math}
\end{table*}

\begin{figure}
\includegraphics[width=\linewidth]{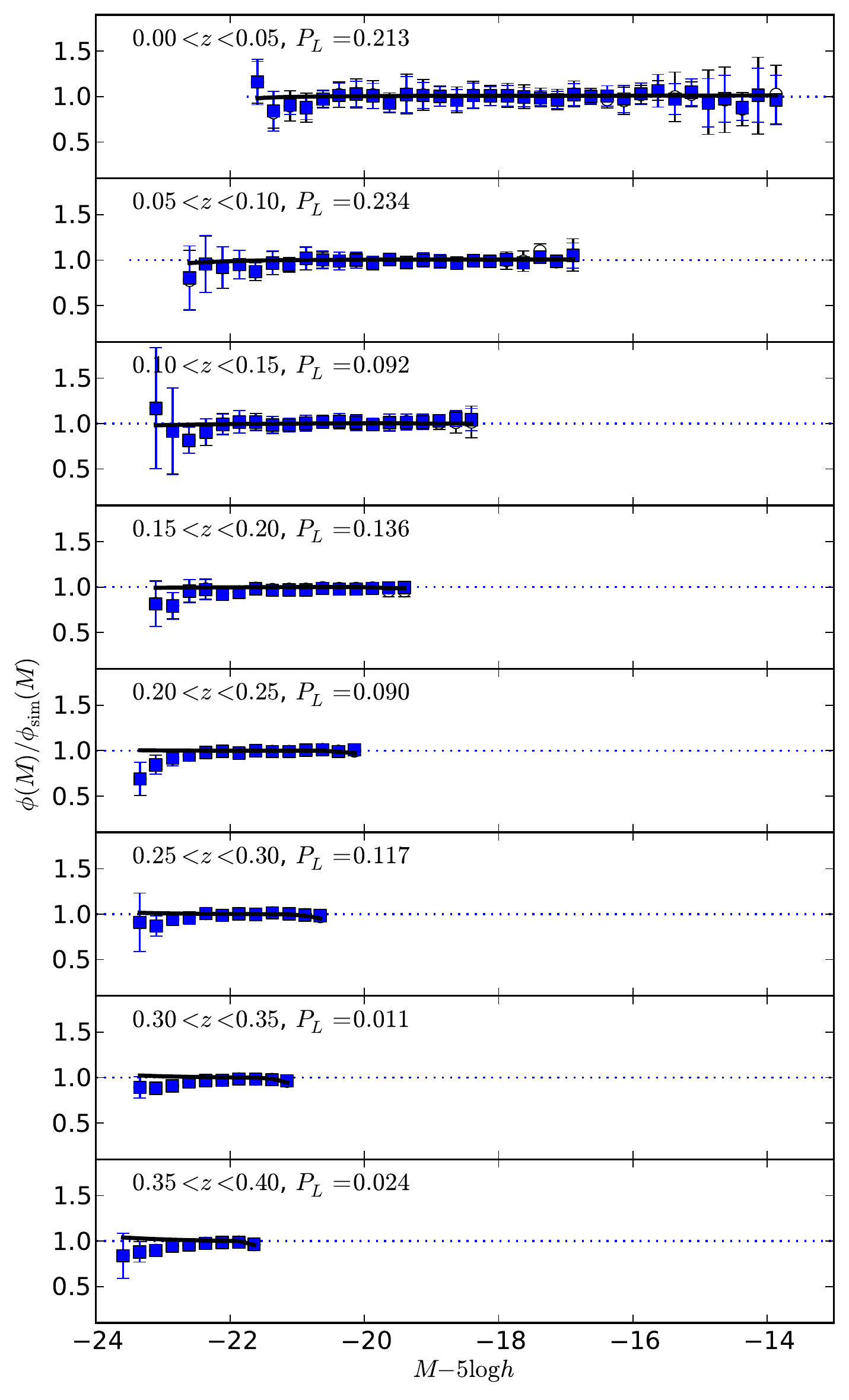}
\caption{Simulated evolving LF estimates in eight
redshift slices as indicated.
All estimates have been normalised by the 'true' underlying LF
of the evolving simulations;
a perfect estimate would like along the horizontal dotted line.
Open symbols show the mean from eight mock catalogues
determined using the $1/V_{\rm max}$ estimator;
filled symbols are from SWML estimator.
Error bars for each come from the scatter between the eight mocks.
The continuous lines show the best-fit evolving parametric fit,
as given by equation~\ref{eq:stybin}.
$P_L$ gives the likelihood ratio 
determination of how well the the parametric fit describes the SWML
observations in that range.
}
\label{fig:lfsim}
\end{figure}

\begin{figure}
\includegraphics[width=\linewidth]{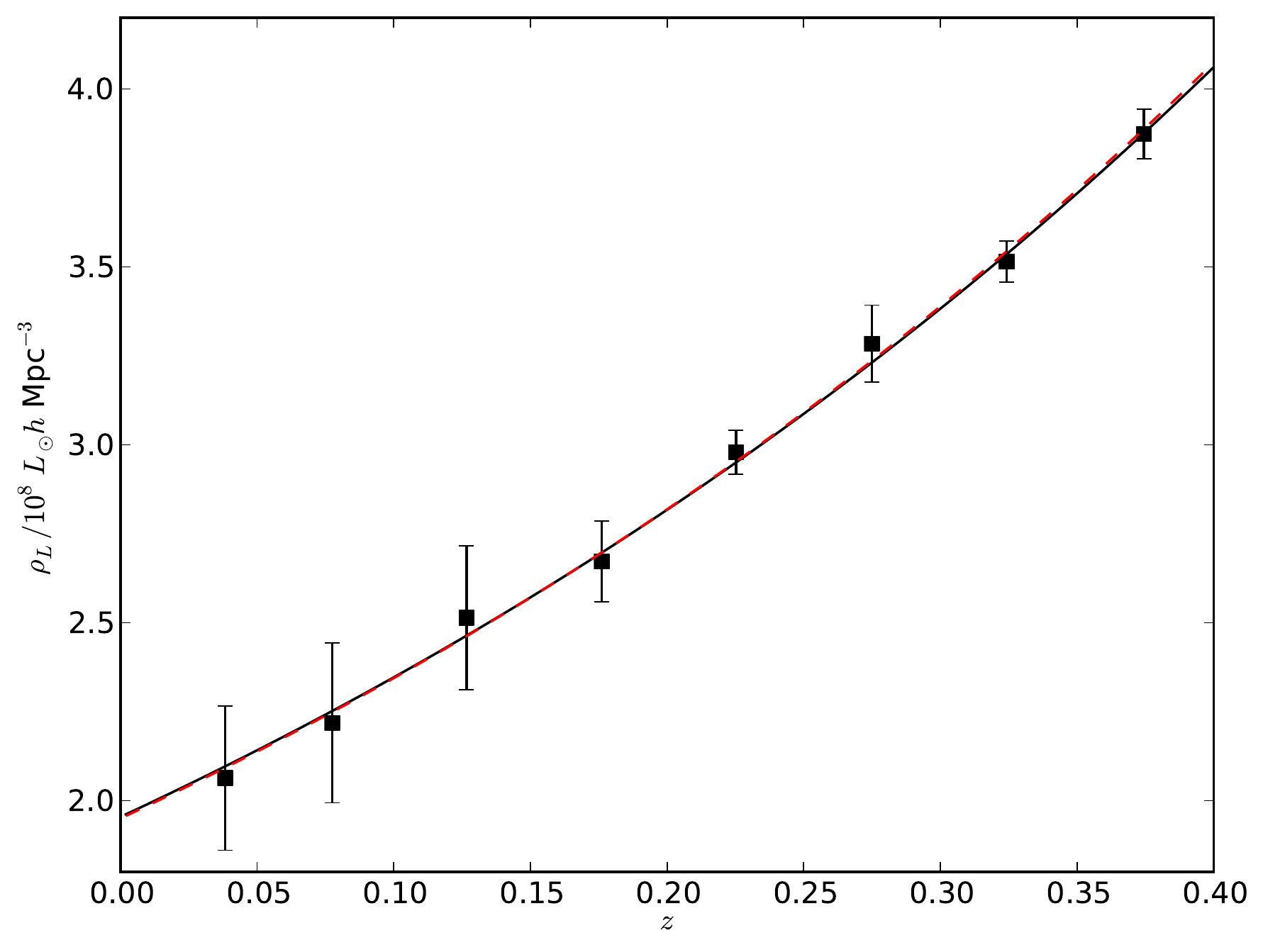}
\caption{Luminosity density estimated from simulated data in eight
redshift slices.
Symbols with error bars show the mean and rms from eight mock catalogues
determined using equation~(\ref{eqn:ldsum}).
The continuous line shows the mean parametric prediction of luminosity 
density evolution (equation~\ref{eqn:ldfit}) and the red dashed line
shows the expected evolution, given the simulation parameters.
}
\label{fig:ldsim}
\end{figure}

Eight independent mock catalogues were generated for each of two
different input LFs, as described above, and used as input catalogues
for our LF estimation code.
Naturally, when analysing these simulations, we assume a consistent
cosmology in calculating distances, apparent magnitudes and volumes.

The Schechter function parameters recovered by the parametric LF estimator,
for both non-evolving and pure luminosity evolution 
simulations, are given in Table~\ref{tab:sims}.
In both cases we recover the true LF parameters with minimal bias,
$\sim 1 \sigma$ at worst.

Fig.~\ref{fig:lfsim} shows the LF recovered in eight redshift
slices from our evolving mock catalogues.
In order to amplify any discrepancies,
all estimates have been normalised by the true LF, obtained by substituting
the input LF parameters into equation~(\ref{eq:stybin}).
We only plot binned estimates when there is at least one galaxy in that
magnitude bin in all eight realisations in order to avoid biasing the mean high
if only realisations with one or more galaxies are included or biasing it low
if all realisations are used.
This eliminates bins fainter than $M = -13$ mag, for which galaxies are only
found in a subset of the simulations.

For the low redshift ($z < 0.05$) slice, all estimates are in good agreement
with the true LF.
Faintwards of $M_r \simeq -16$ mag, 
the scatter in $1/V_{\rm max}$ estimates starts
to increase due to density variations induced by large-scale structure.
The SWML estimator, being insensitive to density variations, 
has smaller error bars at the faint end.

For the higher redshift slices, we see that $1/V_{\rm max}$ 
and SWML estimators give essentially identical results.
There is a tendency for both of these binned estimates 
to slightly underestimate the bright end of the LF, 
a consequence of that fact that both binned estimates,
unlike our parametric fit, 
make the assumption that the LF is independent of redshift.
Even dividing the simulation into eight redshift slices of width
$\Delta z = 0.05$, there is a systematic change of $\Delta M^* = 0.1$
across each slice in these $Q=2$ simulations.
With broader redshift slices the discrepancy worsens.
For example, with $\Delta z = 0.1$, the likelihood ratio probabilities
are below 0.001 for the three higher redshift slices.
It is likely that our binned estimates of the GAMA LF 
(Section~\ref{sec:results}) are biased in a similar fashion.
Note that \citet{2011MNRAS.416..739C} 
has recently proposed a method for estimating
binned LFs whilst simultaneously fitting for luminosity and density
evolution.

Fig.~\ref{fig:ldsim} shows the LD estimated from 
the evolving simulations.
The recovered LD, both in redshift bins, 
equation~(\ref{eqn:ldsum}), and as predicted
by the parametric fit, equation~(\ref{eqn:ldfit}),
is in excellent agreement with the prediction given
the simulation parameters.
The decreasing errors at higher redshift indicate that sample variance is the
largest contributing factor to errors in LD for these 
simulations.
This is not the case with the observed LD 
(Section~\ref{sec:lumdens}), where the dominant source of error,
particularly for the $u$ and $g$ bands, is the applicability of
the evolution model (equation~\ref{eqn:evol}).

\end{document}